\PassOptionsToPackage{hyphens}{url}
\documentclass[sigplan,screen,nonacm]{acmart}

\usepackage{amsmath}
\usepackage{adjustbox}	
\usepackage{placeins}
\usepackage{array}

\makeatletter
\@ifpackageloaded{subfig}
{}
{\usepackage[labelformat=simple,subrefformat=parens,caption=false]{subfig}
 
 }
\makeatother
\usepackage{graphicx}

\usepackage{color}
\usepackage{listings}
\usepackage{relsize}
\usepackage{multirow}
\usepackage[normalem]{ulem} 

\usepackage{xspace}

\newcommand{\originalgrumbler}[2]{\begin{quote}\textcolor{blue}{\sl{\bf #1 says:} #2}\end{quote}}
\newcommand{\grumbler}[2]{\originalgrumbler{#1}{#2}}

\newcommand{\mike}[1]{\grumbler{Mike}{#1}}

\newcommand{\kaan}[1]{\grumbler{Kaan}{#1}}

\newcommand{\harry}[1]{\grumbler{Harry}{#1}}

\definecolor{darkgreen}{rgb}{0,0.4,0}

\newcommand{\sfsmaller}{}

\newcommand{\bench}[1]{\textsf{\sfsmaller#1}}
\newcommand{\code}[1]{\textsf{\sfsmaller#1}}

\newcommand{\eg}{e.g.\xspace}
\newcommand{\ie}{i.e.\xspace}
\newcommand{\cf}{cf.\xspace}
\newcommand{\etal}{et al.\xspace}

\newcommand\naive{na\"{\i}ve\xspace}

\newcommand{\thr}[1]{\textsf{\sfsmaller#1}}

\makeatletter
\@ifundefined{state}{%
}{}
\makeatother

\usepackage{etoolbox}

\usepackage{tikz}
\usetikzlibrary{automata, positioning, arrows, fit}

\usepackage{enumitem}

\usepackage{algorithm}
\usepackage{algpseudocode}
\algnewcommand{\LineComment}[1]{\State \(\triangleright\) #1}
\algnewcommand{\LineCommentx}[1]{\Statex \(\triangleright\) #1}
\algblockdefx{BeginRegion}{EndRegion}{\textbf{begin\_persistent\_region}}{\textbf{end\_persistent\_region}}

\newcommand{\Crafty}{\crafty}
\newcommand{\crafty}{Crafty\xspace}
\newcommand\ndundologging{nondestructive undo logging\xspace}
\newcommand\Ndundologging{Nondestructive undo logging\xspace}

\newcommand\phase[1]{\textsc{#1}}
\newcommand\logp{\phase{Log}\xspace}
\newcommand\Logp{\logp}
\newcommand\redo{\phase{Redo}\xspace}
\newcommand\Redo{\redo}
\newcommand\validate{\phase{Validate}\xspace}
\newcommand\Validate{\validate}

\newcommand\LOGGED{\code{LOGGED}\xspace}
\newcommand\COMMITTED{\code{COMMITTED}\xspace}
\newcommand\HTMonly{Non-durable\xspace}
\newcommand\htmonly{\HTMonly}
\newcommand\bankfee{bank\xspace}

\newcommand\logged{\textcolor{red}{logged}\xspace}
\newcommand\committed{\textcolor{red}{committed}\xspace}

\newcommand\undoentry[2]{\ensuremath{\langle \code{#1},\allowbreak\code{#2} \rangle}}
\newcommand\ts{\ensuremath{\mathit{ts}}\xspace}

\newcommand\notes[1]{\begin{quote}\textcolor{darkgreen}{\textbackslash \textbf{notes\{}} #1 \textcolor{darkgreen}{\}}\end{quote}}
\newcommand\later[1]{\begin{quote}\textcolor{darkgreen}{\textbackslash \textbf{later\{}} #1 \textcolor{darkgreen}{\}}\end{quote}}

\renewenvironment{acks}{%
  \makeatletter\if@ACM@anonymous\makeatother
  \else\makeatother 
  \begingroup
  \section*{Acknowledgments}
  \phantomsection\addcontentsline{toc}{section}{Acknowledgments}
}{%
  \endgroup
  \fi
}

\renewcommand\paragraph[1]{\subsubsection*{\em #1}}

\newtoggle{extended-version}

\renewcommand{\grumbler}[2]{}
\renewcommand{\notes}[1]{}
\renewcommand{\later}[1]{}

\toggletrue{extended-version}

\begin{document}


\newcommand\mytitle[1]{\Crafty: Efficient, HTM-Compatible {#1}Persistent Transactions}


\title{\mytitle{\\}\title{\mytitle{}}}

\iftoggle{extended-version}{
  \begin{teaserfigure}
  \begin{center}
  \vspace*{-0.5em}
  \framebox{\Large This extended arXiv version of our PLDI 2020 paper adds an appendix with additional results}
  \bigskip
  \end{center}
  \end{teaserfigure}
}{}



\author{Kaan Gen{\c{c}}}
\affiliation{
  \institution{Ohio State University \small (USA)}
}
\email{genc.5@osu.edu}

\author{Michael D. Bond}
\affiliation{
  \institution{Ohio State University \small (USA)}
}
\email{mikebond@cse.ohio-state.edu}

\author{Guoqing Harry Xu}
\affiliation{
  \institution{UCLA \small (USA)}
}
\email{harryxu@cs.ucla.edu}

\setcopyright{acmlicensed}
\acmPrice{15.00}
\acmDOI{10.1145/3385412.3385991}
\acmYear{2020}
\copyrightyear{2020}
\acmSubmissionID{pldi20main-p221-p}
\acmISBN{978-1-4503-7613-6/20/06}
\acmConference[PLDI '20]{Proceedings of the 41st ACM SIGPLAN International Conference on Programming Language Design and Implementation}{June 15--20, 2020}{London, UK}
\acmBooktitle{Proceedings of the 41st ACM SIGPLAN International Conference on Programming Language Design and Implementation (PLDI '20), June 15--20, 2020, London, UK}



\begin{abstract}
Byte-addressable persistent memory, such as Intel/Micron 3D XPoint, is
an emerging technology that bridges the gap between volatile memory
and persistent storage.
Data in
persistent memory survives crashes and restarts; however, it is
challenging to ensure that this data is consistent after failures. Existing
approaches incur significant performance costs to ensure crash consistency.

This paper introduces \emph{\crafty}, a new approach for ensuring consistency and
atomicity on persistent memory operations using \emph{commodity hardware}
with existing hardware transactional memory (HTM) capabilities, while incurring low
overhead. \Crafty employs a novel technique called \emph{\ndundologging}
that leverages commodity HTM to control persist ordering.
Our evaluation shows that \crafty outperforms state-of-the-art prior work
under low contention, and performs competitively under high
contention.
\end{abstract}

\maketitle

\section{Introduction}
\label{sec:intro}

Non-volatile memory (NVM) technologies, such as phase change memory
(PCM)~\cite{pcm-nvm,pcm@ieee,Yoon@taco}, resistive
random-access memory (RRAM)~\cite{rram@ieee}, spin-transfer torque
memory (STT-MRAM)~\cite{STT-RAM@ispass2013}, and 3D
XPoint~\cite{3dxpoint}, are designed to combine DRAM's
byte-addressability and storage's durability: A
program's updates to data structures residing in persistent
memory can persist across failures such as a
program crash or power interruption.
As a result, NVM has the
potential to fundamentally change the dichotomy between DRAM and
durable storage in many important domains such as storage
systems~\cite{strata-sosp17, nova-fast16, nova-fortis-sosp17,
  mojim-asplos15, log-structured-nvmm}, databases~\cite{huang-vldb14,
  bztree-vldb18, logging-vldb16, zuo-osdi18}, and big data
analytics~\cite{panthera-pldi19}.


\paragraph{State of the art.}

As with any storage system, the first challenge in effectively using
NVM is to provide \emph{crash consistency}~\cite{memory-persistency,
mnemosyne}, which allows a program to correctly recover from
persistent data upon a failure. Crash consistency is often achieved by
leveraging \emph{transactional support} in a high-level programming
model. The developer specifies \emph{persistent transactions}, in which
updates to persistent memory appear to be one atomic unit---upon a
program crash, either all or none of these updates are committed,
ensuring that important data structures are always left in a
consistent state.

However, prior work's mechanisms for persistent transactions have two
main drawbacks. First, all of the mechanisms---undo
logging~\cite{NV-heaps,transactions-for-pm}, redo
logging~\cite{mnemosyne}, and
copy-on-write~\cite{kamino-tx,dudetm,nv-htm,persistent-htm-giles-2017,pmthreads}---incur
performance costs such as persist latency on each write, lookups at
program reads, maintenance of shadow memory, and poor multithreaded
scalability.

Second, while commodity hardware transactional memory (HTM) such as
Intel's transactional synchronization extensions
(TSX)~\cite{yoo-rtm-2013,tm-book,tm-herlihy-moss} is an appealing
mechanism for supplementing persistent transactions to achieve full
ACID transactions, persistent transaction mechanisms are
\emph{incompatible} with commodity HTM because of the following
dilemma: \emph{To ensure correct recovery, log entries must be persisted before a transaction
  commits, yet the nature of transactions dictates
  that executing transactions cannot perform persist operations.}
Although some recent work shows how to make hardware transactions
persistent~\cite{dudetm,nv-htm,persistent-htm-giles-2017,htpm,phtm,ptm,phytm},
it has major drawbacks such as requiring log lookups at reads, using
shadow memory, incurring scalability bottlenecks, or relying on
nontrivial hardware changes (Section~\ref{sec:background}).

\paragraph{Contributions.}

This paper addresses \emph{both} major limitations of prior work by
\emph{leveraging} commodity HTM to control persist ordering.  We
introduce a new kind of persistent transaction mechanism,
\emph{\ndundologging}, that exploits commodity HTM to populate and
persist undo logs before making persistent writes visible.  Key to
\ndundologging---which runs a persistent transaction's code in a
hardware transaction and logs persistent writes in an undo log---is
that the hardware transaction \emph{rolls back} its persistent writes
prior to committing, effectively creating its undo log entries without
performing actual persistent writes.  This behavior breaks HTM's
persist--commit dependence cycle mentioned above, by decoupling the
undo log updates from the persistent writes.  After committing its
hardware transaction that computes the undo log entries, a persistent
transaction can perform its persistent writes---albeit in a way
that is consistent with the persisted undo log entries and with other
threads' persistent transactions.

We apply \ndundologging in introducing \emph{\crafty},
a novel and general approach for
correct and efficient persistent transactions using \emph{unmodified
commodity HTM}.
For each persistent transaction, \Crafty first uses \ndundologging to compute and persist undo log entries.
It then performs the transaction's writes---by performing the logged writes directly if contention is low,
or by repeating the transaction's execution while validating its consistency with the persisted undo log entry
if contention is high.
\Crafty can operate in a \emph{thread-unsafe} mode
that provides only failure atomicity (relying on some other mechanisms
such as locks for thread atomicity), or it can operate in a
\emph{thread-safe} mode that provides both failure and thread
atomicity (\ie, full ACID transactions).

\notes{
\mike{I excluded the following \textbackslash paragraph for now,
since I think the qualitative comparison belongs elsewhere
(and we already covered NV-HTM and DudeTM earlier in this section),
and the quantitative comparison should be in the Results \textbackslash paragraph.}
\paragraph{\Crafty vs.\ DudeTM and NV-HTM.}
\mike{I think we don't want to focus on the comparison with NV-HTM and DudeTM so much in the intro.
But we do compare against them quantatively, so perhaps we should incorporate the following content into the results summary.}
One clear advantage \crafty has over the copy-on-write techniques~\cite{kamino-tx,dudetm,nv-htm,pmthreads} is
that \crafty does not need a shadow volatile memory, while these
existing techniques maintain two copies of the program state,
requiring DRAM to be at least as large as NVM. 
Since in practice NVM is often orders of magnitude larger than DRAM,
these techniques rely heavily on paging, incurring significant overhead. 
\mike{I don't think that's a big deal since NVM can be used as (essentially volatile) shadow memory.}
Furthermore, in order to ensure consistency between writes to volatile
memory and writes to NVM, DudeTM and NV-HTM incur scalability
bottlenecks that impact performance at high thread counts.
\Crafty works well in these scenarious, particularly when there are
many threads but few conflicts.
}


We implemented \crafty by extending the publicly available
implementation of \emph{NV-HTM}~\cite{nv-htm}, which also implements
\emph{DudeTM}~\cite{dudetm}; both approaches support persistent
transactions with HTM using shadow-memory-based copy-on-write
mechanisms.  Our evaluation uses several programs with varying levels of thread contention:
persistent transaction microbenchmarks and transactional benchmarks.
Our results demonstrate that \crafty outperforms the two
state-of-the-art HTM-based persistent transaction implementations
NV-HTM and DudeTM, especially under low thread contention.
Furthermore, \crafty usually adds low run-time overhead over
\emph{non-durable} transactions, and its overhead is largely thread local
and thus scales well with additional threads.

These results suggest that \ndundologging and \crafty are promising approaches
for providing efficient, HTM-compatible persistent transactions.

\later{
\harry{I felt like the introduction is too short and doesn't convey
  much important information. For readers who don't have much
  background, the introduction would make no sense and they can't
  learn anything but a bunch of terms. For example, what does this new
  thing ``non-destructive undo logging'' do exactly? How is it fundamentally
  different than existing approaches? Again, this
  is probably my style of reading/writing papers --- the reader should
  already have most of his/her questions answered and want to accept
  the paper after finishing the introduction :)
\mike{I fleshed out the design of \ndundologging and \crafty somewhat more in this section.
We could potentially do more.}}
}

\section{Background and Motivation}
\label{sec:background}

\begin{figure*}
\centering
\small
\begin{tabular}{@{}l@{\;\;}|@{\;\;}l@{\;\;}|@{\;\;}l@{\;\;}|@{\;\;}l@{}}
\begin{minipage}{0.15\linewidth}
\begin{scriptsize}
\begin{lstlisting}
failure_atomic {
  *p = 1;
  ... = *q;
  *r = 2;
}






%**)
\end{lstlisting}
\end{scriptsize}
\end{minipage}
&
\begin{minipage}{0.25\linewidth}
\begin{scriptsize}
\begin{lstlisting}
undoLog.append(p, *p);
flush(last log entry);
drain;
*p = 1;
... = *q;
undoLog.append(r, *r);
flush(last log entry);
drain;
*r = 2;
undoLog.append(COMMITTED);

%**)
\end{lstlisting}
\end{scriptsize}
\end{minipage}
&
\begin{minipage}{0.24\linewidth}
\begin{scriptsize}
\begin{lstlisting}
redoLog.put(p, 1);
... = redoLog.lookup(q);
redoLog.put(r, 2);
redoLog.append(COMMITTED);
foreach entry in redoLog
  flush(entry);
drain;
foreach <ptr,val> in redoLog
  *ptr = val;


%**)
\end{lstlisting}
\end{scriptsize}
\end{minipage}
&
\begin{minipage}{0.25\linewidth}
\begin{scriptsize}
\begin{lstlisting}
*p = 1; //writes shadow mem
redoLog.append(p, 1);
... = *q; // reads shadow mem
*r = 2; //writes shadow mem
redoLog.append(r, 2);
redoLog.append(COMMITTED);
foreach entry in redoLog
  flush(entry); 
drain;
foreach <ptr,val> in redoLog
  // Write to persistent addr
  *getPersistAddr(ptr) = val;
\end{lstlisting}
\end{scriptsize}
\end{minipage}\\
(a) A persistent transaction & (b) Undo logging applied to (a) & (c) Redo logging applied to (a) & (d) Copy-on-write applied to (a) \\
\end{tabular}

\caption{To providing failure atomicity for the persistent transaction in (a),
a system uses one of the following crash-consistency mechanisms:
(b) undo logging, (c) redo logging, or (d) copy-on-write. Initial values for all locations are \code{0}.
\label{fig:failure-atomicity}}
\vspace{-1em}
\end{figure*}


This section covers background on persistent memory programming models
and motivates the need for better mechanisms for persistent transactions.

\subsection{Persistent Memory Programming Model}

The key challenge of supporting persistent memory is ensuring that if
a failure occurs, a \emph{recovery observer} can restore persistent
memory to a state that is usable by the restarted program.  This
property is often provided through \emph{failure atomicity}: in the
event of a crash or power failure, persistent memory state can be
restored so that each \emph{persistent transaction} appears to have
executed fully or not executed at
all~\cite{atlas,nvthreads,atlas-follow-up,persistency-sfr,NV-heaps,
  persistent-memory-transactions,justdo-logging,nv-htm,dudetm}.

In addition, a multithreaded program generally needs \emph{thread
  atomicity}---persistent transactions execute atomically with respect
to other threads---and state reconstructed by the recovery observer
should be consistent with the commit order of persistent transactions.
A program with persistent transactions can provide thread atomicity by
using locks~\cite{atlas,nvthreads,persistency-sfr}, or by using transactional
memory~\cite{nv-htm,dudetm,mnemosyne,NV-heaps,persistent-htm-giles-2017}---in
which case the transactions have full ACID properties.

\paragraph{Requirements.}

\notes{
\mike{The following gets too specific too quickly.
\Crafty can provide failure atomicity of any region, whether explicitly or implicity designated.}
Following the persistent memory literature~\cite{dudetm, nv-htm, atlas, nvthreads}, we assume that programmers annotate persistent transactions,
and all writes to persistent memory locations are in persistent transactions.
}

An implementation of persistent transactions must ensure that, after a
crash, the recovery observer can restore the program's persistent
state so that it corresponds to a serialization of persistent
transactions consistent with the program's multithreaded
execution. For example, if transaction $A$ \emph{happened before}
transaction $B$, the recovered state must correspond to one of the
following three execution scenarios: (1) $B$ executed after $A$, (2)
only $A$ executed, or (3) no transaction executed at all.

Furthermore, recovered state should correspond to a point in time
that is not too ``far back'' from the crash time. Otherwise, the amount of
work that needs to be re-executed may be too large to be practical.
\notes{Furthermore, irrevocable, externally visible operations such as
network communication should be considered as ``persist points'' at
which all prior persistent transactions must be fully persisted so
that upon a failure the recovered persistent state will be consistent
with external environments.
\mike{This is something we only handle in the implementation.}}%


\notes{
Finally, we assume that the program provides code that can
continue the execution from the persistent state
(after the recovery observer has made the persistent state consistent as described above).
}

\subsection{Persistent Transaction Mechanisms}
\label{subsec:background-mechanisms}

Upon a crash or power failure,
the recovery observer must reconstruct a state in which transactions
appear to have executed fully or not at all.
This challenge is exacerbated by the fact that stores do not reach
persistent memory in their issuing order.  This is because processor
caches effectively buffer writes until eviction or explicit write-back
of the dirty line to persistent memory.


To ensure that stores reach persistent memory in order,
one can use \emph{persist} operations.
A persist operation consists of one or more \code{flush} operations that write back
specified cache lines to persistent memory, followed by a \code{drain} operation
that waits until the flush operations have completed.
On x86, \code{flush} can be implemented with the \code{CLWB} (cache line write-back) instruction,
and \code{drain} can be implemented with the \code{SFENCE} (store fence)
instruction~\cite{persistent-memory-programming}.
A persist operation is
expensive because it incurs the roundtrip write latency of NVM, which
is expected to be several hundreds of nanoseconds~\cite{pcm-nvm,
  pcommit}.  Even if the NVM controller buffers persistent stores and
includes the buffer as part of the persistence
domain~\cite{persistent-memory-programming}, persist latency (\ie,
the time for roundtrip communication with the NVM controller) is still
significant.  If a commodity approach can be
developed that amortizes persistent latency effectively across many
persistent writes, one can make a case for removing the buffer from
the persistence domain, simplifying future hardware designs.

Persistent transactions generally use one of the following three
mechanisms to provide crash consistency: \emph{undo
  logging}~\cite{NV-heaps,transactions-for-pm}, \emph{redo
  logging}~\cite{mnemosyne}, and
\emph{copy-on-write}~\cite{kamino-tx,dudetm,nv-htm,pmthreads}.  Marathe
\etal\ compared these mechanisms quantitatively and found that no
mechanism is a clear winner in all situations (\eg, across thread
counts or transaction sizes)~\cite{persistent-memory-transactions}.
Here we describe each mechanism and its drawbacks.  We use
Figure~\ref{fig:failure-atomicity}(a) as a simple example persistent
transaction.
\notes{Note that in real programs, the write set of a
transaction would be obscured from static analysis by control and data
dependencies (\eg, indirect stores via pointers).}%


\paragraph{Undo logging.}

In undo logging, a persistent transaction logs the old value of a
persistent location in a persistent undo log before the location is
updated by a memory store.
Undo logging enables fast read accesses:
Since each store performs an \emph{in-place} memory update, any
memory load can directly read the latest value from persistent memory
without being remapped to a different address. 
However, to ensure correct rollback after a crash,
the implementation must persist (\ie, flush and drain) \emph{each update to the undo log}
before writing to the corresponding persistent memory location,
incurring a high write latency for each NVM write.

Figure~\ref{fig:failure-atomicity}(b) shows
how undo logging works for the persistent transaction in
Figure~\ref{fig:failure-atomicity}(a).
To signal the end of a persistent transaction's log entries, the
implementation appends a \COMMITTED entry to the undo log.  A
multithreaded implementation can include a timestamp (not shown in the
figure) with the \COMMITTED entry to enable the recovery observer
to reconstruct a state corresponding to some globally consistent point
in time.

\paragraph{Redo logging.}  

In contrast, instead of performing in-place updates to
persistent memory, redo logging \emph{buffers} all
persistent writes and
performs them together at the end of the transaction. By
buffering writes, redo logging pays the cost of persist ordering
\emph{once} only at the end of each transaction, effectively
amortizing the latency across all of the writes.
However,
it adds an overhead for
each persistent \emph{read} because the read needs to find the latest
value in a set of buffered writes. Since reads often significantly
outnumber writes, redo logging can also incur significant overhead.

Figure~\ref{fig:failure-atomicity}(c) illustrates how redo logging
works.
Writes and reads to persistent memory are replaced with updates and lookups, respectively,
to a map-based log so that reads of persistent memory correctly read from
any preceding writes.

\paragraph{Copy-on-write.}

Recent work proposes \emph{copy-on-write} mechanisms that maintain a
\emph{volatile shadow} for each persistent page to be
modified~\cite{kamino-tx,dudetm,nv-htm,pmthreads}.  We focus
on copy-on-write mechanisms that use shadow paging because it allows
efficient in-place writes.  Other copy-on-write mechanisms use
indirection to copy an object upon the first write in a transaction,
incurring costs similar to redo
logging~\cite{persistent-memory-transactions,persistent-htm-giles-2017}.
Persistent transactions perform reads and writes normally, since
virtual addresses are mapped to physical volatile shadow memory
addresses.  At the end of the transaction, changes to each shadow page
are persisted to its corresponding non-volatile page.
Figure~\ref{fig:failure-atomicity}(d) shows how this mechanism works.

Although copy-on-write techniques enjoy the performance benefits of
undo and redo logging---and can be made compatible with commodity
HTM as described shortly---shadowing the entire NVM is
expensive and impractical.  Most significantly, copy-on-write
mechanisms must ensure consistency between the updates to volatile and
non-volatile pages, leading to scalability bottlenecks, as detailed
below.
\notes{\textcolor{red}{Furthermore, in a realistic setting, the size of DRAM can be
much (\eg, an order of magnitude) smaller than the size of NVM and,
hence, memory shadowing leads to frequent swapping and poor
performance.}
\mike{I don't think this is significant concern (beyond the general concern of wasting memory):
There's no reason the shadow pages couldn't use non-volatile memory (treating it like volatile memory).}}%

\subsection{Transactional Memory}
\label{subsec:background-tm}

A natural way to implement persistent transactions that provide full
ACID properties is to leverage \emph{transactional memory}~\cite{tm-herlihy-moss,tm-book}.
Much of the existing work on
persistent transactions extends \emph{software} transactional memory
(STM)~\cite{lightweight-transact}, which incurs a high
overhead in detecting and resolving conflicts between concurrent
transactions.

\emph{Hardware} transactional memory (HTM), which detects and resolves
conflicts at the hardware level, is an appealing technique for implementing efficient persistent transactions.
However,
\emph{commodity HTM} implementations including  Intel's \emph{restricted
TM} (RTM)~\cite{yoo-rtm-2013} are fundamentally
incompatible with persistency. Because log updates must occur before
memory updates, there is an obvious dilemma: On the one hand, undo or
redo log entries must be persisted before the hardware transaction
commits the actual memory updates (to ensure crash consistency), while
on the other hand, the nature of the transaction dictates that these
log entries cannot be persisted before the transaction
commits---otherwise they cannot be revoked upon an abort.  The updates
to the log entries and the actual memory updates depend on each
others, forming a dependence cycle that seemingly thwarts the use of HTM for
persistent transactions.

\notes{
\mike{Covered in Related Work:}
Researchers have proposed nontrivial
hardware extensions to commodity HTM to support persistent
transactions~\cite{htpm,phtm,ptm,phytm}. However, it is unclear
when, if ever, these extensions might make it into commercial
systems.
}

Recent approaches use commodity HTM for persistent
transactions, by \emph{decoupling} persistence from HTM's
concurrency control. \emph{DudeTM}~\cite{dudetm} and
\emph{NV-HTM}~\cite{nv-htm} show how copy-on-write mechanisms
can support HTM-based persistent transactions.
\later{\textbackslash footnote\{\emph{cc-HTM} uses copy-on-write with
  HTM, but it uses explicit lookups instead of shadow paging~\cite{persistent-htm-giles-2017}, slowing
  execution more than DudeTM and NV-HTM do, so we focus
  on the comparison with DudeTM and
  NV-HTM.\}}%
Hardware transactions perform \emph{in-place} reads and writes to shadow memory.
After a transaction commits, redo log entries can be persisted before copying the transaction's writes
to persistent memory.
In addition, by writing redo log entries and program writes to persistent memory asynchronously,
writes to the same persistent locations can be combined.

The drawback, though, is that DudeTM and NV-HTM have significant
disadvantages in maintaining shadow state and keeping updates to persistent
memory consistent with the order of transactions writing to the volatile
shadow state.
First, these approaches add space overhead by maintaining two copies
of program state, as discussed above.  Second, they must ensure consistency between the
order of the writes to DRAM inside a transaction and that to NVM at
the end of the transaction. DudeTM computes timestamps by incrementing
a global variable in commodity HTM, making it effectively incompatible
with commodity HTM~\cite{dudetm}.

NV-HTM, on the other hand, works with unmodified commodity HTM, but it
has two major scalability bottlenecks that limit performance at higher
thread counts~\cite{nv-htm}. First, each persistent transaction cannot
complete until \emph{every other ongoing transaction} completes.  In
particular, each transaction cannot write a COMMIT entry to its redo
log until it ensures that no ongoing transaction may still write a
COMMIT entry for an earlier transaction, since redo logs are used by
the recovery observer to roll the persistent state forward after a
crash.  Waiting ensures that if the recovery observer sees a COMMIT
entry for a transaction, it sees COMMIT entries for all earlier
transactions. Of course, this incurs overhead.
\notes{\mike{Is this needed even if we don't care about providing immediate persistence after a transaction completes?
We should be comparing NV-HTM and \crafty on the same terms.
\mike{Update: We discussed at it seems like NV-HTM needs this even if it doesn't
care about providing immediate persistence after a transaction completes.}}}%

Second, threads that persist logs and program writes to persistent
memory must do it in a serialized manner.  In NV-HTM, an asynchronous
background thread applies transactions' writes (based on their redo
log entries) to persistent memory locations in timestamp order.
This serialization
of writes to persistent memory is inherent in the fact that
transactions record a timestamp (for efficiency), from which only a
global transaction order can be inferred.



The DudeTM paper surmises that decoupling persistence from HTM may be
``the best (and possibly the only) way to avoid the drawbacks of both
undo and redo logging and reduce the performance
penalty''~\cite{dudetm}.  Our work seeks to \emph{counter} that
supposition and overcome the performance disadvantages of existing
persistent transaction mechanisms.

\section{\Crafty Overview}
\label{sec:design-overview}

As Section~\ref{subsec:background-tm} explained, the main obstacle that precludes
efficient use of commodity HTM in implementing persistent transactions
is the dependence cycle that results from the tight coupling of log
entry updates and program memory updates: If a hardware transaction
contains a mix of these two types of updates, it can neither commit
before persisting, nor persist before committing.

To address this problem, we introduce a new persist transaction design called \emph{\crafty}
that leverages a new logging mechanism called \emph{\ndundologging}.
Key to \Crafty's success is breaking the persist--commit dependence cycle by
executing the log entry updates and the program memory updates in
\emph{separate hardware transactions}, effectively decoupling these two types
of updates.  In \ndundologging, a hardware transaction performs a
\emph{\logp} phase that executes a persistent transaction in a way
that updates only undo log entries, not the program's persistent data. These log
entries are persisted after the hardware transaction commits. Next,
\Crafty executes the program writes using another hardware
transaction. These writes are performed in a way that is consistent
with the updates of the log entries and also with other threads'
executed transactions.

\paragraph{Challenges and insights.}

Achieving a correct and efficient design 
presents three major challenges. The \emph{first challenge} is how to
make the \logp phase only update undo or redo log entries without
modifying program memory locations. To overcome this challenge, \crafty
uses undo logging when executing the \logp phase: Before each write to
a persistent memory location, the old value in the location is
recorded in a \emph{thread-local undo log}. At the end of the
transaction, \crafty \emph{rolls back} all of these writes by applying
the entries of the undo log in a \emph{reverse order}, effectively
setting the modified values back to the their original values
before the transaction executed.
Furthermore, during this rollback process, when both the
old and new values are visible, the hardware transaction builds a
\emph{redo} log for these locations.  After the \logp phase commits,
all of the undo log entries are persisted into persistent memory.
Figure~\ref{fig:failure-atomicity-crafty} shows how the \logp phase uses \ndundologging to
construct an undo log for the persistent transaction from Figure~\ref{fig:failure-atomicity}.

\begin{figure}
\small
\begin{lstlisting}
HTM_BEGIN;
undoLog.append(p, *p);
*p = 1;
... = *q;
undoLog.append(r, *r);
*r = 2;
foreach entry <ptr, oldVal> in undoLog in reverse
  redoLog.append(ptr, *ptr);
  *ptr = oldVal; // undo each write
HTM_END;
foreach entry in undoLog
  flush(entry);
drain; // persist the undo log entries
/*...%*\textcolor{darkgreen}{\em Transaction's}*) writes can now be performed here...*/
undoLog.append(COMMITTED);
\end{lstlisting}

\caption{How \crafty's crash-consistency mechanism, \ndundologging,
provides failure atomicity for the persistent transaction in Figure~\ref{fig:failure-atomicity}(a).
\label{fig:failure-atomicity-crafty}}
\end{figure}

The \emph{second challenge} is how to execute the program's memory
updates in the same order as the updates to log entries.  To do this,
\crafty starts the second phase, which updates program memory
locations. In theory, all we need is a \emph{\redo} phase that applies
the redo log constructed at the end of the \logp phase.  This \naive
approach would work if persistent transactions were protected by a
pessimistic technique such as locking, because transactions executed
by different threads would conflict with each other.
However, if persistent transactions can
conflict, then a thread's \logp and \redo phases---which executed in two
separate hardware transactions---may not execute together
\emph{atomically}. This can potentially lead to inconsistencies
between the log entries and the contents in their corresponding memory
locations.  To solve this problem, \crafty lets the \redo phase check
a conservative \emph{conflict constraint} based on timestamps. Failure
of this check is a \emph{necessary but insufficient} condition for a
transaction conflict. To guarantee safety, \crafty aborts the HTM
transaction that executes this \redo phase.

The \emph{third challenge} is what to do if and when \redo aborts.
Due to the conservative nature of our conflict constraint, a \redo
abort does not necessarily indicate a real conflict. Hence, if and only if \redo
aborts, \Crafty executes a \emph{\validate} phase, which
\emph{re-executes the persistent transaction} to check the validity of
the undo log entries that were persisted in the \logp phase. If all of
the undo log entries are still valid, the transaction succeeds,
allowing the memory writes to be committed and visible to other
threads. Any mismatch between the values in a log entry and its
corresponding memory location makes \validate abort, indicating that
another thread has committed new, conflicting writes after the current
thread's \logp phase finished.  The aborted thread handles this case
by starting over---by re-executing the \logp phase and constructing new
undo and redo logs.

\notes{

\harry{To me, all of those discussed above should belong to a discussion section \emph{after} Crafty's major design points. If I were a reader, why would I want to know this before understanding the three major transactions performed by Crafty?}

\paragraph{Supporting crash consistency.}

\Crafty executes persistent transactions in two steps:
It first logs old values for a transaction,
at which point the transaction has been \emph{logged}.
Second, \crafty writes new values for a transaction,
at which point the transaction has been \emph{committed}.

\harry{We need be more precise --- e.g., what do we mean by ``old values'' and ``new values''? without seeing the entire design, it'd be hard to understand/appreciate what ``logged'' and ``committed'' mean.}

\Crafty supports crash consistency by maintaining per-thread persistent undo logs.
Each entry in the undo log has one of the following forms:
\begin{itemize}[leftmargin=*]
\item An \undoentry{addr}{oldValue} entry
records a persistent address and its original value before a \logged transaction.
\item A \undoentry{\LOGGED}{\ts} entry
records a timestamp corresponding to a \logged transaction.
\item A \undoentry{\COMMITTED}{\ts} entry
records a timestamp corresponding to a \committed transaction.
\end{itemize}
Each \logged or \committed persistent transaction consists of any number of \undoentry{addr}{oldValue} entries,
followed by a \LOGGED entry and possibly a \COMMITTED entry,
represented by the following regular expression:
\[
(\undoentry{addr}{oldValue}^* \; \undoentry{\LOGGED}{\ts} \; \undoentry{\COMMITTED}{\ts}?)*
\]

\mike{We should also define what a timestamp represents:
I guess it's basically a Lamport clock, \ie, two values are ordered if the transactions are ordered.}
\harry{This is low-level information. Why should it appear here? Why do we show the format of our logs before even discussing the major design points?}

\noindent
If the program crashes, \crafty can recover persistent state
by rolling back threads' \logged and \committed persistent transactions with respect to a global timestamp.
}

\paragraph{Outline.}

Section~\ref{Sec:executing-transactions} describes how
\crafty executes persistent transactions to provide atomicity at run time and support recovery on a crash.
Section~\ref{sec:recovery} describes how recovery restores
persistent state after a crash.

\section{How \Crafty Executes Transactions}
\label{Sec:executing-transactions}

This section describes how \crafty leverages \ndundologging
to execute persistent transactions.

\paragraph{Execution modes.}

\Crafty can operate in either of two modes.  In its \emph{thread-safe
  mode} (this paper's focus), programmers specify persistent
transaction boundaries, and \crafty provides both atomicity and
durability (\ie, all ACID properties) for persistent transactions.

\Crafty's \emph{thread-unsafe mode} is appropriate when locks or
another mechanism already provides atomicity, so \crafty only needs to
provide durability.  In this mode, programmers can specify transaction
boundaries explicitly or inform \crafty to treat all critical
sections~\cite{atlas,nvthreads,atlas-follow-up,justdo-logging,ido} or
synchronization-free regions~\cite{persistency-sfr} as persistent
transactions.

\begin{figure}[t]
\small
\centering
\tikzset{
  ->, 
  node distance=1.8cm, 
  >=stealth, 
  every state/.style={rectangle, thick, fill=gray!20}, 
  }%
  \begin{tikzpicture}[align=center]
    \node[] (START) {Start persistent transaction};
    \node[state, below of=START] (LOG) {\textbf{\Logp} phase}; 
    \node[state, below of=LOG] (WRITE) {\textbf{\Redo} phase};
    \node[state, below right of=WRITE, yshift=-1cm] (VALIDATE) {\textbf{\Validate} phase};
    \node[state, right of=LOG, xshift=2cm] (SGL) {SGL-based execution \\ in \emph{thread-unsafe mode}};
    \node[below of=VALIDATE] (FINISH) {End persistent transaction};
    \path (START) edge node{} (LOG);
    \path (LOG) edge[left] node{success} (WRITE);
    \path (LOG) edge[above] node{fallback} (SGL);
    \path (LOG) edge[bend right=75, right] node{read-only} (FINISH);
    \path (WRITE) edge[right] node{failure} (VALIDATE);
    \path (WRITE) edge[bend right=40, left] node{success} (FINISH);
    \path (VALIDATE) edge[bend right=40, above right] node{validation \\ failure} (LOG);
    \path (VALIDATE) edge[right] node{success} (FINISH);
    \path (VALIDATE) edge[right] node{fallback} (SGL);
    \path (SGL) edge[bend left, right] node{} (FINISH);
  \end{tikzpicture}
\caption{\Crafty's phases in \emph{thread-safe} mode.\label{fig:crafty-thread-safe}}
\bigskip
  \centering
  \small
  \tikzset{
  ->, 
  node distance=1.8cm, 
   >=stealth, 
  every state/.style={rectangle, thick, fill=gray!20}, 
   }%
  \makebox[0.5\linewidth]{%
  \begin{tikzpicture}[align=left]
    \node[] (START) {Start persistent transaction; \\ $k \gets \code{MAX\_WRITES}$};
    \node[state, below of=START] (BEFOREWRITE) {Execute until persistent \\ write or transaction end};
    \node[state, below of=BEFOREWRITE] (LOG) {\textbf{\Logp} phase for \\ up to $k$ writes};
    \node[state, left of=LOG, xshift=-1.2cm] (HANDLEFAIL) {Return to last \\ \logp phase's \\ start; $k \gets \frac{k}{2}$};
    \node[state, below of=LOG] (REDO) {\textbf{\Redo} phase for \\ \logp phase's writes};
    \node[state, below right of=BEFOREWRITE, xshift=1.75cm] (SINGLEWRITE) {Persist undo log entry \\ before persistent write};
    \node[state, right of=REDO, xshift=1.05cm] (FINISH) {End persistent \\ transaction};
    \node[left of=HANDLEFAIL, xshift=0.6cm] (INT) {...};
    \path (START) edge node{} (BEFOREWRITE);
    \path (BEFOREWRITE) edge[left] node{persistent write \\ \;\;\;\qquad(if $k>1$)} (LOG);
    \path (BEFOREWRITE) edge[right, bend left] node{persistent write \\ ~~~~(if $k=1$)} (SINGLEWRITE);
    \path (BEFOREWRITE) edge[right] node{\\\\\\\\\qquad transaction end} (FINISH);
    \path (LOG) edge[right] node{success} (REDO);
    \path (LOG) edge[above] node{failure} (HANDLEFAIL);
    \path (REDO) edge[above right] node{failure} (HANDLEFAIL);
    \path (HANDLEFAIL) edge[bend left] node{} (BEFOREWRITE);
    \path (REDO) edge[above right, out=180, in=270] node{success} (INT);
    \path (INT) edge[out=90, in=180] node{} (BEFOREWRITE);
    \path (SINGLEWRITE) edge[] node{} (BEFOREWRITE);
  \end{tikzpicture}
  }%
\caption{\Crafty's phases in \emph{thread-unsafe} mode.\label{fig:crafty-thread-unsafe}}
\end{figure}

\begin{figure}
\centering
\small
\lstset{
  basicstyle=\footnotesize\sffamily,
  morekeywords={atomic_and_durable},
  linewidth={18cm},
}
\newcommand\undoLog[1]{\ensuremath{\code{undoLog}_{\code{T#1}}}}
\newcommand\redoLog[1]{\ensuremath{\code{redoLog}_{\code{T#1}}}}
\newcommand\lastLog[1]{\ensuremath{\code{lastTS}_{\code{T#1}}}}
\begin{tabular}{@{}l@{\qquad\qquad}l@{}}
Thread 1       & Thread 2        \\\hline
\begin{lstlisting}
atomic_and_durable {
  *p = *q;
  *r = 1;
}
\end{lstlisting} &
\begin{lstlisting}
atomic_and_durable {
  *q = 2;
  *s = 3;
}
\end{lstlisting} \\
\end{tabular}\\
(a) Two persistent transactions.
\bigskip\\
\setlength{\tabcolsep}{1em}
\hspace*{-1em}
\begin{tabular}{@{}l@{\qquad\qquad\quad}l@{}}
Thread 1       & Thread 2        \\\hline
\begin{lstlisting}[boxpos=t]
%*\textcolor{blue}{\Logp phase:}*)
HTM_BEGIN                   
%*\undoLog{1}*).add(p, 0)         
*p = *q                      
%*\undoLog{1}*).add(r, 0)         
*r = 1                       
// Start roll back:
%*\redoLog{1}*).add(r, 1)         
*r = 0 // from undo log      
%*\redoLog{1}*).add(p, 0)         
*p = 0 // from undo log      
%*\lastLog{1}*) = ts()           
%*\undoLog{1}*).add(LOGGED,
               %*\lastLog{1}*))
HTM_END                     






%*\textcolor{blue}{\Redo phase:}*)
HTM_BEGIN
check gLastRedoTS < %*\lastLog{1}*)
*p = 0 // from redo log       
*r = 1 // from redo log       
gLastRedoTS = ts()        
%*\undoLog{1}*).add(COMMITTED,
               gLastRedoTS)
HTM_END
\end{lstlisting}
&
\begin{lstlisting}[boxpos=t]












%*\textcolor{blue}{\Logp phase:}*)
HTM_BEGIN              
%*\undoLog{2}*).add(q, 0)    
*q = 2                  
%*\undoLog{2}*).add(s, 0)    
*s = 3
// Start roll back:
%*\redoLog{2}*).add(s, 3)
*s = 0 // from undo log
%*\redoLog{2}*).add(q, 2)  
*q = 0 // from undo log
%*\lastLog{2}*) = ts()  
%*\undoLog{2}*).add(LOGGED,
               %*\lastLog{2}*))
HTM_END

%*\textcolor{blue}{\Redo phase:}*)
HTM_BEGIN // redo
check gLastRedoTS < %*\lastLog{2}*)
ABORT // check failed

%*\textcolor{blue}{\Validate phase:}*)
HTM_BEGIN
check *q == 0 // from undo log
*q = 2
check *s == 0 // from undo log
*s = 3
check # writes == # log entries
gLastRedoTS = ts()
%*\undoLog{2}*).add(COMMITTED,
               gLastRedoTS)
HTM_END
\end{lstlisting}
\end{tabular} \\
(b) A possible execution of the persistent transactions in (a) using \crafty in its thread-safe mode,
which provides all ACID properties.

\caption{An example of \Crafty's thread-safe mode
executing persistent transactions.
Initial values of \code{*p}, \code{*q}, \code{*r}, and \code{*s} are 0.
The example omits \code{flush} and \code{drain} instructions.}
\label{fig:crafty-exec-example}
\end{figure}


Figures~\ref{fig:crafty-thread-safe} and
\ref{fig:crafty-thread-unsafe} show how \crafty operates in its
thread-safe and thread-unsafe modes
(Section~\ref{sec:design-overview}), respectively.
Sections~\ref{subsec:log-phase}--\ref{subsec:validate-phase} provide
a detailed description of the \logp, \redo, and \validate phases
in the context of \crafty's thread-safe mode.
In thread-safe mode,
repeated aborts cause \crafty to transition to thread-unsafe mode 
while holding a \emph{single
global lock} (SGL), as Figure~\ref{fig:crafty-thread-safe} shows and
Section~\ref{subsec:sgl-fallback} describes.






The rest of this section uses Figure~\ref{fig:crafty-exec-example} as an example
to show how \crafty works.

\subsection{\Logp Phase}
\label{subsec:log-phase}

\Crafty's \logp phase generates undo log entries for an executed persistent
transaction and then persists these entries.  The key treatment here
is that the \logp phase does \emph{not} commit any program writes to
persistent memory.  The \logp phase achieves this outcome by allowing the persistent transaction to
perform writes normally during its execution, but rolling back the
writes before the hardware transaction commits.

Algorithm~\ref{alg:log-phase} shows the details of the \logp phase,
which executes the persistent transaction body in a hardware
transaction.  Before each persistent write, the \logp phase records
the old value of the written-to address in the executing thread's
persistent undo log.
For example, in Figure~\ref{fig:crafty-exec-example},
each persistent transaction's \logp phase adds old values
to the undo log before each write.

\begin{algorithm}[t]
\caption{\hfill \Logp phase}
\small
\begin{algorithmic}[1]
\State \code{HTM\_BEGIN}
\LineCommentx{Start of persistent transaction}
\smallskip
\Statex \dots
\smallskip
\LineCommentx{Program write to persistent variable:}
\State Add \undoentry{addr}{oldValue} to \code{T.undoLog} \Comment{\thr{T} is current thread}
\State \code{*addr = newValue} \Comment{Original program write}%
\smallskip
\Statex \dots
\smallskip
\LineCommentx{End of persistent transaction}
\State Roll back transaction's writes using \code{T.undoLog}, and populate local redo log from \code{T.undoLog}
\State Add \undoentry{\LOGGED}{getTimestamp()} to \code{T.undoLog}
\State \code{HTM\_END}
\State \code{flush}(\code{T.undoLog} entries for this transaction)
\end{algorithmic}
\label{alg:log-phase}
\end{algorithm}

At the transaction end, the \logp phase uses the undo log entries to
roll back the transaction's writes, by applying the undo log entries'
old values in the reverse order.  When rolling back the writes,
\crafty simultaneously builds a volatile \emph{redo log} for the
transaction, which can be used by the subsequent \redo phase to
perform program writes.
For example, in Figure~\ref{fig:crafty-exec-example}, starting from the
``Start roll back:'' comment, the persistent transaction's \logp phase
rolls back the writes by applying the values from the undo log.
Before committing the hardware transaction,
the \logp phase adds a \LOGGED entry with a Lamport timestamp\footnote{If
two events are ordered by happens-before,
their logical times are correspondingly ordered~\cite{happens-before}.}
denoting the
current logical time (which is equivalent to the logical time at the
beginning of the hardware transaction since HTM ensures atomicity).
The timestamps will be used by recovery to order undo log entries by different threads.
In Figure~\ref{fig:crafty-exec-example}, each \logp phase concludes by inserting
a \LOGGED entry into the undo log before committing the hardware transaction.

After committing the transaction, the \logp phase flushes the
transaction's undo log entries to persistent memory.
The algorithm flushes the transaction's undo log entries
but does not wait for them to be written back to persistent memory
(\ie, \code{flush} but no \code{drain})
because the program writes will be committed by the \redo or \validate phase inside
of a hardware transaction,
which has \code{drain} semantics (\eg, an RTM transaction has \code{SFENCE} semantics).

Once the \logp phase completes, undo log
entries for the transaction have been persisted, but the transaction
has effectively \emph{not} executed from the perspective of other
threads and memory because none of the memory updates have been
performed.
In Figure~\ref{fig:crafty-exec-example}, after the \logp phase completes,
\code{*p}, \code{*r}, \code{*q}, and \code{*s} still have their initial values (\code{0}).
To make these updates visible to other threads and
persistent memory, \crafty uses the \redo or, if needed, the \validate
phase.



A read-only transaction need not add a \LOGGED
entry to the undo log or perform any persist operations, and it can skip the \redo and \validate
phases, as shown in Figure~\ref{fig:crafty-thread-safe}.

\subsection{\Redo Phase}

The \redo phase applies the writes from the redo log
(in the reverse order of how they were recorded in the \logp phase),
as illustrated in Algorithm~\ref{alg:redo-phase}.

\begin{algorithm}[t]
\caption{\hfill \Redo phase}
\small
\begin{algorithmic}[1]
\Statex \textbf{Thread-safe \redo phase:}
\State \code{HTM\_BEGIN}
\If {$\code{gLastRedoTS} < \textnormal{ \LOGGED timestamp from \logp phase}$}
  \State $\code{gLastRedoTS} \gets \code{getTimestamp()}$
  \State Perform thread-unsafe \redo phase
\Else
  \State Abort transaction and fail \redo phase\label{line:redo-abort}
\EndIf
\State \code{HTM\_END}
\State \code{flush}(written-to addresses)
\bigskip
\Statex \textbf{Thread-unsafe \redo phase:}
\State Perform writes from redo log
\If {not in hardware transaction}
  \State \code{flush}(written-to addresses)
\EndIf
\State Add \undoentry{\COMMITTED}{getTimestamp()} to \code{T.undoLog}
\end{algorithmic}
\label{alg:redo-phase}
\end{algorithm}

If the program is single-threaded or no other threads access
persistent memory, it is safe to execute the \redo phase
\emph{unconditionally}. However, if multiple threads are executing
persistent transactions, atomicity can be violated. For
example, thread B's \redo phase can occur between thread A's \logp and
\redo phases.  Hence, it is important to ensure that \emph{A's \redo
  phase completes only if it executes atomically with its preceding
  \logp phase}.

To this end, \crafty uses a global variable \code{gLastRedoTS} that
represents the timestamp of the last writes committed by any
thread.
Figure~\ref{fig:crafty-exec-example} demonstrates how \code{gLastRedoTS} is updated and used.
\Crafty checks \code{gLastRedoTS} at the start of the \redo phase in Thread~1.
The check succeeds because no thread has committed writes since Thread~1's \logp phase.
The \redo phase then performs the writes from the redo log.
Thread~1 completes the \redo phase by updating \code{gLastRedoTS} and
adding a timestamped \COMMITTED entry to the log.
This timestamp represents the time at which the transaction's
writes happened in relation to other threads' transactions.


A failed check indicates a potential atomicity
violation.
In Figure~\ref{fig:crafty-exec-example}, Thread~2's check of
\code{gLastRedoTS} fails because Thread~1 updated \code{gLastRedoTS}
to reflect that it committed writes (in its \redo phase) \emph{after}
Thread~2's \logp phase. Thread~2's \redo phase thus fails, and
\crafty tries the \validate phase. (Alternatively, under different timing,
Thread~2's \redo phase could start and
complete before Thread~1's \redo phase read \code{gLastRedoTS},
allowing Thread~2 to commit its writes with the \redo phase.
Thread~1 would fail the \redo phase and try the \validate phase,
re-execuing the transaction and writing an updated value of
\code{*q}, \code{2}, to \code{*p}.)

If successful, the \redo phase concludes by flushing the transaction's
writes to persistent memory, but does not wait for the write-backs to
finish (\ie, \code{flush} but no \code{drain}).  The next persistent
transaction's \logp phase will perform a hardware transaction, which
has \code{drain} semantics, and the recovery algorithm always rolls
back each thread's latest transaction in case its writes had not fully
persisted (Section~\ref{sec:recovery}).

The \redo phases of all transactions are effectively serialized.
This does not necessarily cause a bottleneck in performance because the \redo phase
is often short and can execute concurrently with \logp and \validate phases.

\subsection{\Validate Phase}
\label{subsec:validate-phase}

The goal of the \validate phase is to execute a persistent transaction
that is consistent with the persisted undo log entries.
The \validate phase checks consistency by comparing the old values
recorded in the undo log with the current values of the same locations,
as Algorithm~\ref{alg:validate-phase} illustrates.

\begin{algorithm}[t]
\caption{\hfill \Validate phase}
\small
\begin{algorithmic}[1]
\State Reset \code{T.undoLog} to the beginning of the transaction
\State \code{HTM\_BEGIN}
\LineCommentx{Start of the persistent transaction}
\smallskip
\Statex \dots
\smallskip
\LineCommentx{Program write to a persistent variable:}
\State \textbf{let} \undoentry{expectedAddr}{expectedValue} be next entry in \code{T.undoLog}
\If {$\code{addr} \ne \code{expectedAddr} \lor \code{*addr} \ne \code{expectedValue}$}
  \State Abort transaction and fail validation
\EndIf
\State $\code{*addr} \gets \code{newValue}$ \Comment{Original program write}%
\smallskip
\Statex \dots
\smallskip
\LineCommentx{End of the persistent transaction}
\State Check that the next entry in \code{T.undoLog} is a \LOGGED entry (if not, abort the transaction and fail the validation)\label{line:validate-abort}
\State $\code{gLastRedoTS} \gets \code{getTimestamp()}$
\State Add \undoentry{\COMMITTED}{\code{gLastRedoTS}} to \code{T.undoLog} after the \LOGGED entry
\State \code{HTM\_END}
\State \code{flush}(written-to addresses)
\end{algorithmic}
\label{alg:validate-phase}
\end{algorithm}

The \validate phase checks whether, for each program write, its
corresponding entry in the undo log matches the write's address and
the value at the address. If it does, this implies the validity of the
undo log entry.  For example, in Figure~\ref{fig:crafty-exec-example},
Thread~2's \validate phase checks that both writes to \code{q} and
\code{s} match the original addresses and old values in the undo log,
and that there are no new writes.  At the end of the persistent
transaction, the hardware transaction is committed, and the writes are
persisted and made visible to other threads.  Note that it is
important to re-execute the transaction---by validating the undo log
entries rather than just performing the writes---to ensure
(implicitly) that values \emph{read} by the transaction are still
consistent with the undo log entries.  Like the \redo phase, after
performing the writes, \crafty adds a
\undoentry{\COMMITTED}{getTimestamp()} entry to the log, which
represents the time at which the transaction's writes happened in
relation to other threads.

Note that the \validate phase executes only if the \redo phase fails,
Every persistent transaction commits its writes exactly once, with
either the \redo or \validate phase.

\subsection{Single Global Lock Fallback\label{subsec:sgl-fallback}}

A hardware transaction may abort for a variety of reasons including a
conflict with other threads, cache capacity overflow, or an
unsupported event such as an interrupt~\cite{yoo-rtm-2013}.  Since
commodity HTM generally provides no progress guarantee, special care
must be taken to ensure that an execution makes progress.  As
Figure~\ref{fig:crafty-thread-safe} shows, \crafty's thread-safe mode
retries an aborted transaction several times; if no \redo or \validate
phase commits successfully, it falls back to acquiring a \emph{single
  global lock (SGL)} to provide progress guarantees.  The SGL serves
two purposes. First, it eliminates conflicts among different threads.
Second, it allows \crafty to execute in
\emph{thread-unsafe mode}, where \crafty can execute shorter hardware
transactions (fewer instructions) or without hardware transactions to ensure
progress.

The SGL is a global variable that a thread acquires by updating it
atomically from 0 to 1, and releases by setting it to 0 (with proper
memory fencing).  To ensure consistency with respect to other threads
executing hardware transactions in thread-safe mode, each hardware
transaction in thread-safe mode must check whether the SGL is 0 at the
beginning of the transaction; if the SGL is 1, the transaction must
abort (not shown in the algorithms). This handling ensures consistency
with an ongoing SGL section or with an SGL section that starts while
the transaction is executing (since the transaction's read set
contains the SGL).  This fallback method is referred to as
\emph{speculative lock elision} in the literature and has been widely
studied~\cite{yoo-rtm-2013,sle,aahtm}.

After acquiring the SGL, a thread can safely execute in \crafty's
thread-unsafe mode, as illustrated in
Figure~\ref{fig:crafty-thread-unsafe}. In this mode, the SGL ensures
atomicity, so HTM serves solely to implement \ndundologging (\ie, to
prevent updated cache lines from being written back to persistent
memory prematurely), \emph{not} for thread atomicity.

As a result, thread-unsafe mode uses hardware transactions for the
\logp phase only, which can wait to start a hardware transaction until
the first persistent write of the persistent transaction.
Thread-unsafe mode does not use HTM for the \redo phase because no
other threads can update the global timestamp \code{gLastRedoTS}, and
hence this phase always succeeds. The \validate phase is not needed at
all.

\paragraph{Ensuring progress.}

Even without contention from other threads, a hardware transaction may
still abort for cache capacity or other reasons.
The \logp phase in thread-unsafe mode can ensure progress by breaking
a persistent transaction into smaller hardware transactions, each
executing at most $k$ persistent writes. After executing $k$
persistent writes (or fewer, if the persistent transaction ends before
$k$ is reached), the \logp phase completes normally, rolling back
writes and persisting undo log entries including a \LOGGED entry.  The
\redo phase then performs the $k$ (or fewer) persistent writes, except
that it does \emph{not} add a \code{\COMMITTED} entry, which should
only be used to indicate the end of the (SGL-based) transaction.  The
\logp and \redo phases continue executing the persistent transaction,
in chunks of up to $k$ writes.  If $k=1$, the \logp phase writes and
persists an undo log entry \emph{before} performing the write to
memory, without using any hardware transaction.

When entering thread-unsafe mode for a persistent transaction, \crafty
begins with a (relatively large) value of $k$ (\eg, 64) with the goal
to amortize persist latency across multiple writes.  After each
transactional abort in thread-unsafe mode, \crafty decreases $k$
geometrically for the next hardware transaction. When the value of $k$
drops to 1, thread-unsafe mode is guaranteed to make progress.
Figure~\ref{fig:crafty-thread-unsafe} illustrates the logic for
thread-unsafe mode, which executes the transaction in $k$-write chunks
until completion.

Before releasing the SGL, \crafty adds a \COMMITTED entry to the
persistent undo log.  All of the SGL section's
hardware transaction's \LOGGED and \COMMITTED entries
use the \emph{same} timestamp (from the first call to \code{getTimestamp()})
to ensure that the recovery algorithm either rolls back all or none of
the SGL section's writes.

\Crafty thus adaptively adjusts transaction sizes to provide a
tradeoff between persist latency and the risk of aborting.  Prior work
in other contexts splits transactions to balance between
per-transaction costs and aborts
costs~\cite{legato-cgo-2017,block-chop}.

\section{How \Crafty Recovers After a Crash}
\label{sec:recovery}

This section presents \crafty's recovery logic. We first describe how
recovery is done under the assumptions that \emph{an infinite log is
  initially zeroed (\ie, no wraparound or reuse)} and \emph{log
  entries are persisted atomically}. Then we show how
  to handle logs without these simplifying assumptions.

\subsection{Basic Recovery Logic}

Under the above-mentioned assumptions, the recovery observer can
detect \emph{persisted entries}, which are entries with a nonzero \code{addr} field
(which is either an address or a \LOGGED or \COMMITTED tag).
A \emph{fully persisted sequence} is a consecutive sequence of persisted \undoentry{addr}{oldValue}
entries preceded by a (persisted) \LOGGED or \COMMITTED entry and concluded by a persisted \LOGGED entry.

The recovery observer needs to roll back the last \emph{fully
  persisted sequence of each thread} because some of the corresponding
writes may have persisted, but not all of them have definitely
persisted.
\notes{\sout{In addition, if there are no persisted entries after a thread's last fully
persisted sequence, then the actual writes of the previous sequence
may not have persisted either. As such, the recovery observer
needs to roll back its writes as well.}
 \mike{Really? I don't think that's correct.
 \mike{Update: Kaan says he agrees it's not correct.}}}%
Let \undoentry{\LOGGED}{\ts} be the sequence's concluding entry.
Then we define a sequence's timestamp to be
\notes{$\ts-1$
\mike{Not sure about the minus one.
\kaan{Assume there are 2 threads, T1 and T2. Say that in their logs, T1's final sequence ends with \LOGGED while T2's ends with \COMMITTED. Note that T1 can not have another sequence, even partial, following this sequnce otherwise it would have had a \COMMITTED entry.
If T2's timestamp is equal to or greater than T1's, it will be rolled back whether we do minus one or not.
If T2's timestamp is smaller than T1's by more than 1, it will not be rolled back (assuming it has other fully persisted sequences following it).
The only case where adding the minus one changes anything is if T2's timestamp was exactly 1 less than T1's timestamp. However even in that case, I don't see why T2's transaction should be rolled back, it is no different from any other transaction that has a smaller timestamp than T1's transaction.
As such, I'm removing this distinction from the paper. Please feel free to add it back if I'm missing something.}}%
if the sequence
is succeeded by a \undoentry{\COMMITTED}{\ts} entry, or
\ts otherwise.}%
\ts. To arrive at a globally consistent snapshot,
the recovery observer must also
roll back every sequence that has a timestamp \emph{later than or
  equal to} the timestamp of any sequence being rolled back.
Any persisted entries outside a fully persisted sequence must not be rolled back because their corresponding writes definitely have not persisted. 

To roll back a sequence, the recovery observer applies the
\undoentry{addr}{oldValue} entries in reverse order, performing
\code{*addr = oldValue} for each entry.  The recovery observer rolls
back the fully persisted sequences in the reverse timestamp order.


\subsection{Handling High-Performance Logs}
\label{sec:circular-logs}

Next, we discuss how \crafty provides correct
recovery in the absence of simplifying assumptions
about the logs.
The following design handles circular logs that reuse entries,
and it does not require a log entry to be persisted atomically.
The design also
addresses a limitation of \crafty's design as presented so far:
Because the recovery observer rolls back at least each thread's
last transaction, a rolled-back transaction can be arbitrarily far back in time if a
thread has not executed a transaction in a while.


The design assumes that
the system provides persistence at
\emph{word} (or coarser) granularity.
The design relies on each thread's circular log being large enough to hold log entries
for at least two persisted sequences (which are bounded due to HTM bounding constraints).

\paragraph{Distinguishing reused entries.}

To handle reuse of undo log entries (\eg, via a circular log),
the recovery observer needs to be
able to tell whether an entry \undoentry{addr}{oldValue} is from the
latest transaction or the last wraparound of the log.  Inspired by
prior mechanisms~\cite{logging-algorithms-oopsla17}, \crafty's
execution of transactions maintains a per-thread \emph{wraparound bit}
that is encoded in each word and flips each time the log wraps around.
This wraparound bit allows recovery to differentiate words
written \emph{after} versus \emph{before} the latest log wraparound.
Because logged transactions occur more often than wraparound, recovery
can only observe log entries that are \emph{after} the next-to-latest
wraparound and hence a single wraparound bit suffices.

We further assume that all addresses are word (4- or 8-byte)
aligned. This allows us to steal two or three bits of the \code{addr}
word in each \undoentry{addr}{value} log entry. One of them is used as
the wraparound bit. The \LOGGED and \COMMITTED tags are each represented as a
reserved, aligned address.

Because NVM is only guaranteed to provide persistence at word granularity, the
\code{value} word in a \undoentry{addr}{value} log entry will also
need a wraparound bit. However, \code{value} needs all of its bits for
program values.  We thus steal another bit in each \code{addr} word to
store a bit (\eg, the lowest bit) of the \code{value} word, allowing that same bit
of the \code{value} word to be replaced with the wraparound bit.

\paragraph{Discarding entries and bounding rollback severity.}

In order for \Crafty's \logp phase to reuse log entries (\eg, via a
circular log), we must be able to discard some entries that would no
longer be needed by the recovery observer.  Since \crafty must not
discard entries for a logged transaction that needs to be rolled back,
we need to ensure that the earliest possible rollback timestamp \ts is
greater than the timestamp of a logged transaction that \crafty is
ready to discard.  A related issue we address here is \emph{bounding}
how far back in time the recovery observer must roll back to. This
distance can be quite far if a thread has not executed a persistent
transaction for a while.

\newcommand\tslb{\code{tsLow\-erBound}\xspace}
\newcommand\curr{\code{currentTS()}\xspace}

The logging algorithm maintains a global timestamp \tslb that is a
lower bound on the earliest possible timestamp $r$ that recovery might
need to roll back to.  It is a lower
bound because, for performance reasons, it is kept up to date
\emph{lazily}.  When adding a \LOGGED entry to its undo log, a thread
checks that
\[
\curr < \tslb + \code{MAX\_LAG}
\]
where \code{MAX\_LAG} represents a customizable maximum time duration for which recovery might need to roll back,
and \curr is a timestamp representing the current time.
Likewise, whenever a thread \thr{T} gets halfway through its circular log, it first checks if overwriting the next half of the log
will violate
\[
\code{T.log.earliestTSToBeOverwritten} > \tslb
\]
If either of these conditions fails, thread \thr{T} performs further inspection,
by checking the following two conditions for every other thread \thr{U}:
\[
\curr < \code{U.lastCommittedTxn.ts} + \code{MAX\_LAG}
\]
\[
\code{T.log.earliestTSToBeOverwritten} < \code{U.lastCommittedTxn.ts}
\]
\thr{T} can perform these checks safely (atomically) by executing them in a hardware transaction,
performing read-only accesses to \code{U.logStart} and \thr{U}'s log.
If either condition fails on \thr{U}, then \thr{T} forces \thr{U} to append an
by committing a \undoentry{\LOGGED}{getTimestamp()} entry to \thr{U}'s log
(representing an empty completed transaction).
\thr{T} can accomplish this by using a transaction to
safely update \thr{U}'s log (because we need to be careful about interference with \thr{U}, especially its
non-transactional state manipulations).
\notes{or using a per-thread interrupt (\code{pkill\_thread}) to make
\thr{U} add the entry to its own log.}%

After \thr{T} makes each delinquent thread \thr{U} commit a more up-to-date transaction, it sets
\[
\tslb \gets \min_{\thr{U}} \; \code{U.lastCommittedTxn.ts}
\]
Note that most transactions only need to read a global shared variable (\tslb) that is mostly read-only,
resorting to more expensive operations only when they are halfway through the circular undo log.
The frequency of expensive operations can be reduced by increasing the size of
each thread's circular log.

\paragraph{Providing immediate persistence.}

Some persistent transaction systems, including
DudeTM~\cite{dudetm} and NV-HTM~\cite{nv-htm},
guarantee that if a persistent transaction completes and the thread continues execution,
then the recovered state will include the completed transaction's state.
This ``immediate persistence'' property ensures that the persistent state is consistent with
any externally visible, irrevocable actions between transactions such as system calls.
However, \crafty does \emph{not} provide ``immediate persistence''
because it does not ensure that all writes have been persisted
when a transaction completes
(which is why recovery rolls back each thread's last logged transaction).
Some prior work including PMThreads~\cite{pmthreads} also does not provide immediate persistence.

Instead of providing immediate persistence,
\crafty can provide a method for ``on-demand'' immediate persistence
(to be invoked before performing externally visible, irrevocable actions).
\Crafty can implement on-demand immediate persistence
by adding an \undoentry{LOGGED}{getTimestamp()} entry to each thread's,
similar to the approach described above
for reusing log entries and bounding rollback severity.
Modifying other threads' logs can be performed safely by
executing in a hardware transaction.
Our prototype implementation does not suppport on-demand immediate persistence.

\section{Implementation}
\label{sec:impl}

Our \crafty implementation,
which we have made publicly available,\footnote{\url{https://github.com/PLaSSticity/Crafty}}
extends the publicly available NV-HTM
implementation~\cite{nv-htm}.\footnote{\url{https://bitbucket.org/dfscastro/nvhtm-selfcontained/src/master/}}
The NV-HTM implementation also provides a configuration
that represents the prior work DudeTM~\cite{dudetm}.
It also includes an \emph{\HTMonly} configuration
that simply executes each persistent transaction in a hardware transaction
and thus does not provide any crash-consistency guarantees.
\notes{We extended the implementation to allow nested transactions by flattening them,
which was necessary to evaluate the benchmark \bench{memcached}.
\mike{I don't think we need to mention this.}}%

The NV-HTM implementation
\emph{emulates} non-volatile memory in volatile memory by performing 300 ns of busy waiting at \code{drain} operations
(emulating the roundtrip latency of each \code{SFENCE} instruction that follows one or more \code{CLWB} instrutions).
This methodology is consistent with the evaluations of prior work including DudeTM and NV-HTM~\cite{dudetm,nv-htm}.

\paragraph{\Crafty logging details.}

Each thread has an undo and a redo log.  Undo logs are in non-volatile
memory and are circular.  Redo logs are in volatile memory and 
not needed after a persistent transaction completes, so the next
persistent transaction can reuse the redo log from the beginning.

\notes{
\Crafty requires word-sized writes to be aligned. Smaller writes may be unaligned as long as they do not cross aligned word boundaries.
\kaan{Revised and moved the sentence, since we list out other requirements here, and this is part of our design.
\mike{I think this is an implementation issue. The program can do unaligned writes as long as the compiler
or run-time system converts them to multiple, smaller aligned writes.}}
}

Each undo log entry \undoentry{addr}{oldValue} contains two 8-byte
words: the written-to address and the old value.
Each \code{addr} value is 8-byte-aligned because all writes
are expressed as 8-byte, aligned stores.
The implementation merges the \LOGGED and \COMMITTED
entries into a single entry,
overwriting the entry's timestamp on commit.
\notes{When the algorithm would insert a \LOGGED entry,
the implementation inserts the entry with the current timestamp.
When the algorithm would insert a \COMMITTED entry,
the implementation updates the timestamp to the current timestamp.}%
This optimization is safe as the recovery observer does not need to
differentiate between \LOGGED and \COMMITTED entries
when deciding what sequences to roll back.
The recovery observer can check if each log entry
has persisted using the wraparound bit.
Timestamps come from \code{RDTSC}.

The implementation performs the work needed to support rollback
(\ie, the wraparound bit and the log checks in Section~\ref{sec:circular-logs}).
However, we have \emph{not} implemented the actual recovery logic,
leaving it and its evaluation to future work.

\paragraph{Mixed-mode accesses.}

The implementation requires that all writes to persistent memory happen in persistent transactions.
\notes{(However, non-transactional persistent writes could be supported in the same way
that thread-unsafe mode handles writes when $k=1$; Section~\ref{subsec:sgl-fallback}).}%
\Crafty can support writes to volatile memory in transactions by ensuring
transactions are idempotent with respect to volatile memory accesses.
Our implementation requires manual transformation of transactions to
make them idempotent with respect to \emph{function-local} variables.
It does not allow other volatile memory writes in transactions,
but could do so by adding undo logs for volatile accesses.

\notes{
\Crafty may execute the code inside a persistent transaction twice:
first in the \logp phase and again in the \validate phase.  Without
tracking volatile memory accesses, \crafty cannot roll back
changes to volatile memory as it does with non-volatile memory. As a
result, \Crafty requires all persistent transactions to be
\emph{idempotent} in terms of volatile memory accesses.
Transactions may read and write volatile memory,
but must avoid read-modify-write patterns such as incrementing a counter that outlives the transactions.
Most of the persistent transactions in our benchmarks are idempotent in
terms of volatile memory. For the few transactions that are not, we modified
their code to save and restore changed function local variables when
running under Crafty. This could be done automatically by
a compiler pass that transforms code in a similar manner.
More generally, Crafty could use volatile undo and redo logs to roll back and redo
writes to volatile memory that violate idempotency.
\mike{The above is good, but I think we should mention that volatile memory accesses
(or volatile memory accesses in transactions) will be to (function-)local variables,
allowing such a compiler pass to be successful.
Hmm, I guess this is why you said that transactons shouldn't access shared volatile memory?
I think maybe transactions shouldn't access volatile memory except for function-local variables
(so even thread-local pointer-based data structures would be unacceptable,
unless we also instrument writes to volatile data in order to roll it back).
\kaan{Noted that our changes to programs were only for function-local variables, and that volatile logs could be used to circumvent the issue.}}
}

The same (volatile or persistent) variable can be accessed both in and out of transactions,
subject to the aforementioned constraints.
Programmers must be careful to synchronize such accesses correctly:
Although Intel's RTM provides strong atomicity~\cite{tm-book},
\crafty may fall back to using a global lock for providing thread atomicity.
Programs thus must ensure transactional data race freedom~\cite{strong-atomicity-weak-idea-transact09}.



\paragraph{Memory management.}

Because the \logp and \validate phases execute the same code,
the implementation must handle side effects from \code{malloc} and \code{free}
to avoid leaking memory and failing checks in the \validate phase.
The implementation thus logs allocations during the \logp phase
and reuses the allocated memory at corresponding \code{malloc} calls during the \validate phase.
Similarly, the \logp phase logs \code{free} calls during the \logp phase,
and either performs the logged \code{free}s after completing the \redo phase
or allows the \validate phase to performed \code{free} calls and then discards logged \code{free}s.

\notes{
\mike{Here's older, longer motivation:}
Besides writes rolled back in the \logp phase,
other side effects could imperil
correctness or reduce the chances of successful validation in the \validate phase.
Most side effects (\eg, system calls) are already disallowed by
commodity HTM, but one side effect that our implementation must handle
is allocation and deallocation via \code{malloc} and \code{free}.
Crafty does not require any modifications to the allocator,
but performs some extra work for allocations as explained below.

We first describe the problems created by not handling allocation.
If a transaction calls \code{malloc}, the \logp phase will allocate memory, and
the \validate phase, if performed, will execute \code{malloc} again, potentially creating duplicate objects.
A potential memory leak can be avoided easily by freeing
objects allocated by the \logp phase.  However, a trickier issue here
is that the \validate phase, if performed, may allocate objects
at different memory locations than the \logp phase when it re-executes the transaction's code.
These differences mean that writes to these locations
will be deemed different than those in \logp, causing a validation failure.

Our implementation addresses this issue by logging allocations during
the \logp phase and reusing the allocated memory during the \validate
phase.  To do this, \Crafty intercepts calls to \code{malloc} in
persistent regions.  When the \validate phase runs, it simply checks
whether each object allocated by the \logp phase is large enough using
\code{malloc\_usable\_size}. If not, it triggers a validation failure.
Upon failure of the \validate phase, \Crafty frees the logged
objects before returning to the \logp phase or falling back to
SGL-based execution.

Deallocation is easier to deal with.  A region should free allocated
memory only for successfully completed regions, and it should only do
so exactly once (\ie, no double \code{free}s).  To this end,
the \logp phase logs calls to \code{free} without executing
them.  If the \redo phase is successful, these logged \code{free}s are
executed.  Otherwise, they are discarded: The \validate phase or
SGL-based execution will perform \code{free}s normally.
}

\section{Evaluation}

This section evaluates the performance of \crafty by comparing it with
prior work's HTM-compatible persistent transactions and with
non-durable transactions.

\subsection{Methodology}

\paragraph{Configurations.}

Our experiments run \crafty in thread-safe mode to provide full ACID transactions.
NV-HTM and Dude\-TM are run under their standard
configurations.  As a baseline, we run the implementation's
\emph{\htmonly} configuration that does not provide any guarantees on
crash consistency.

In addition to evaluating \crafty's full-blown version (in thread-safe
mode), we evaluate two variants of \crafty that exclude the \validate
and \redo phases, referred to as \emph{Crafty-NoRedo} and
\emph{Crafty-NoValidate}, respectively, in the rest of the section.
These configurations help tease out the performance effects of
\crafty's components.  Note that these configurations are also fully
functioning configurations that provide the same guarantees as
\crafty.

\paragraph{Evaluated programs.}

We use two microbenchmarks and a set of transactional memory
benchmarks.

The \emph{\bankfee} microbenchmark is
from the publicly available NV-HTM implementation~\cite{nv-htm} that
performs random transfers between accounts.  We configure the
benchmark to run five transfers (ten persistent writes) per
transaction with three levels of contention: \emph{high},
\emph{medium}, and \emph{no conflict}.  The difference in conflict
rates is achieved by varying the number of accounts---the medium- and
high-conflict configurations operate on 4,096 and 1,024
cache-line-aligned accounts, respectively.  The no-conflict
configuration avoids all conflicts by partitioning the accounts among
threads.

The other microbenchmark performs operations on a B+ tree and is
adapted from the implementation of Zardoshti
\etal~\cite{zardoshti-pact2019} by annotating writes to shared memory.
The benchmark provides two variants: one performs only insertions on
the tree, and the other performs a mixture of lookups, insertions, and
removals.

As a standard benchmark suite for transaction memory research, our
experiments use the transactional \emph{STAMP
  benchmarks}~\cite{stamp}.  In particular, we consider each
transaction to be a persistent transaction, and treat all
shared-memory accesses in transactions as accesses to persistent
memory. This same methodology was used in the evaluation of prior
work~\cite{nv-htm}. We exclude the benchmark \bench{yada} as it fails
to run with \HTMonly and NV-HTM due to a pointer corruption, and
\bench{bayes} because around half of the transactions fall back to the
SGL mode due to HTM incompatible instructions which makes the results
not meaningful.

\paragraph{Experimental setup.}

We run the experiments on a machine with a quiet 16-core Intel Skylake
processor with hyperthreading disabled.  The implementation uses
native hardware transactions~\cite{ritson-rtm-2013} and emulates
non-volatile memory as described in Section~\ref{sec:impl}.
Each reported result is the arithmetic mean of five trials.  The
throughput results are normalized to the throughput of the
single-thread, non-durable configuration of the same benchmark.  We
define throughput as the inverse of the execution's wall-clock time.

\subsection{Performance Results}

This subsection presents our main results: performance and scalability
for the evaluated programs across thread counts and persistent
transaction implementations.  We also perform additional measurements
that help explain the performance including (1) breakdowns of
persistent transactions by \crafty phases and (2) hardware transaction
commit and abort counts and abort causes.  These additional
measurements, as well as performance results that emulate 100 ns
(instead of 300 ns) write latency, can be found in
\iftoggle{extended-version}{Appendix~\ref{sec:extended-eval}}{an
  extended arXiv version~\cite{crafty-extended-arxiv}}.

\paragraph{Bank microbenchmark.}
\label{sec:bank-fee}



\begin{figure}
    \vspace*{-1.5em}
    \centering
    \captionsetup[subfloat]{farskip=2pt,captionskip=1pt}
    \subfloat[High contention]{
        \includegraphics[width=1.1\linewidth]{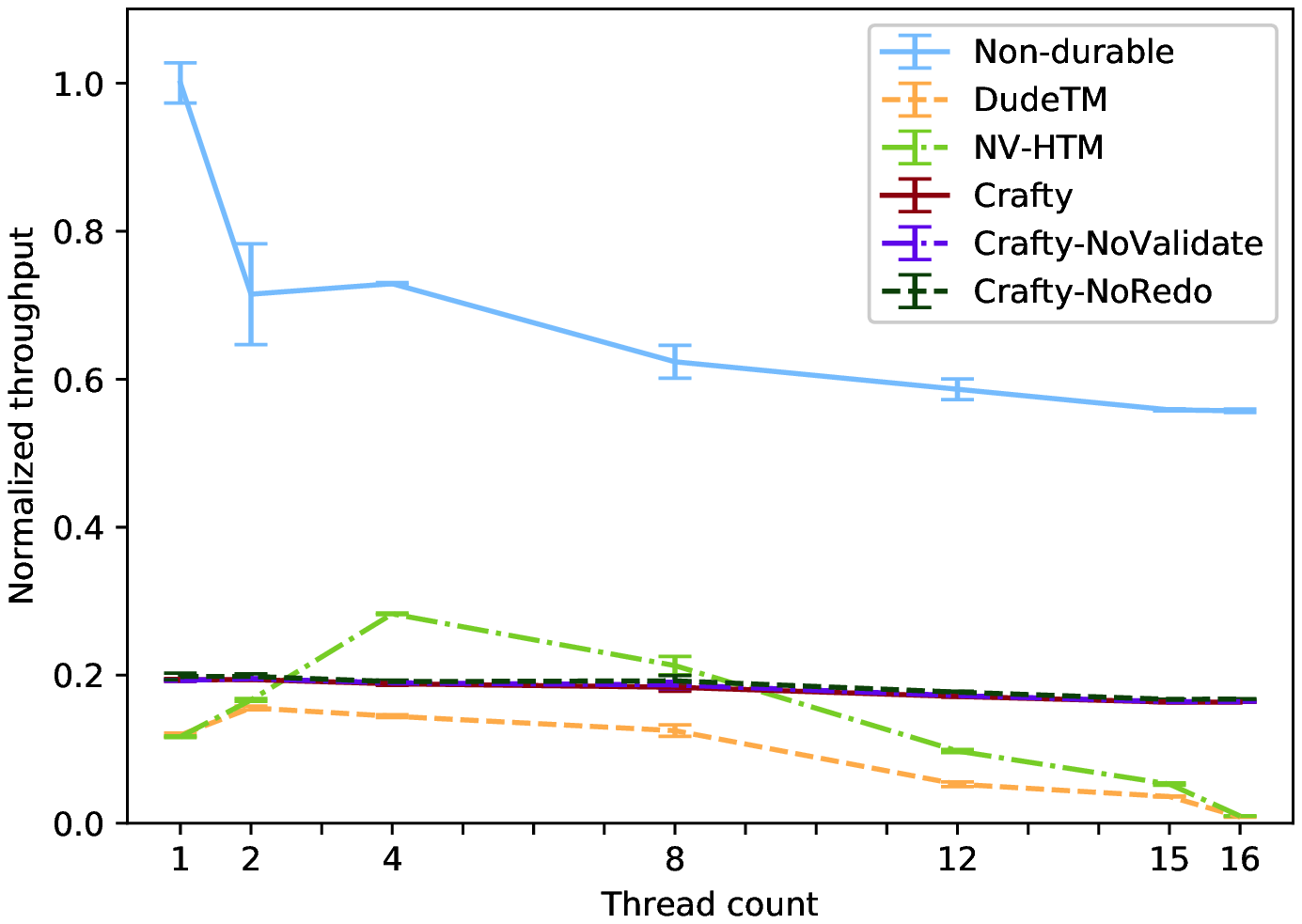}
        \label{fig:throughput:bank-fee-hc}
    }
    \\
    \subfloat[Medium contention]{
        \includegraphics[width=1.1\linewidth]{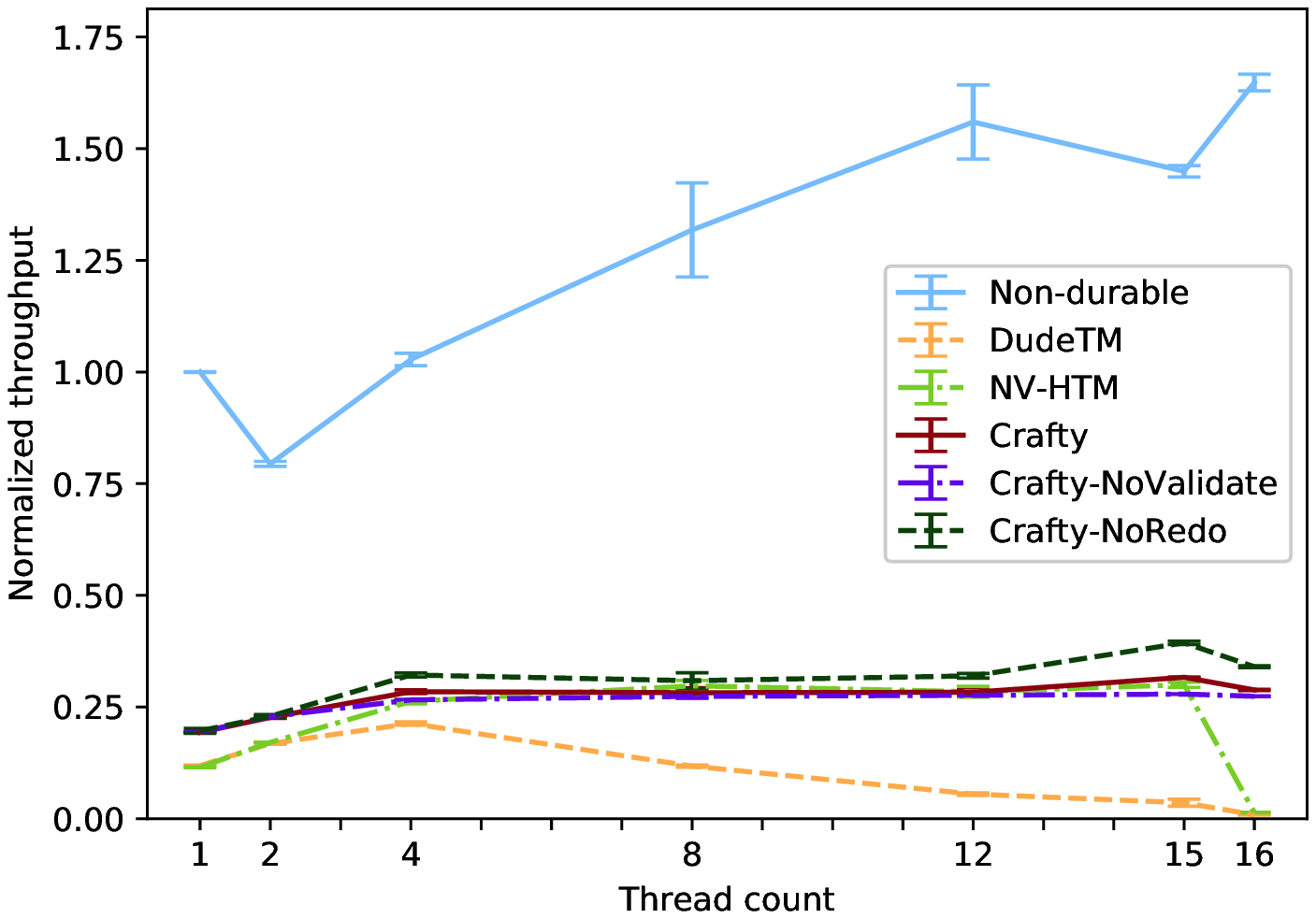}
        \label{fig:throughput:bank-fee}
    }
    \\
    \subfloat[No contention]{
        \includegraphics[width=1.1\linewidth]{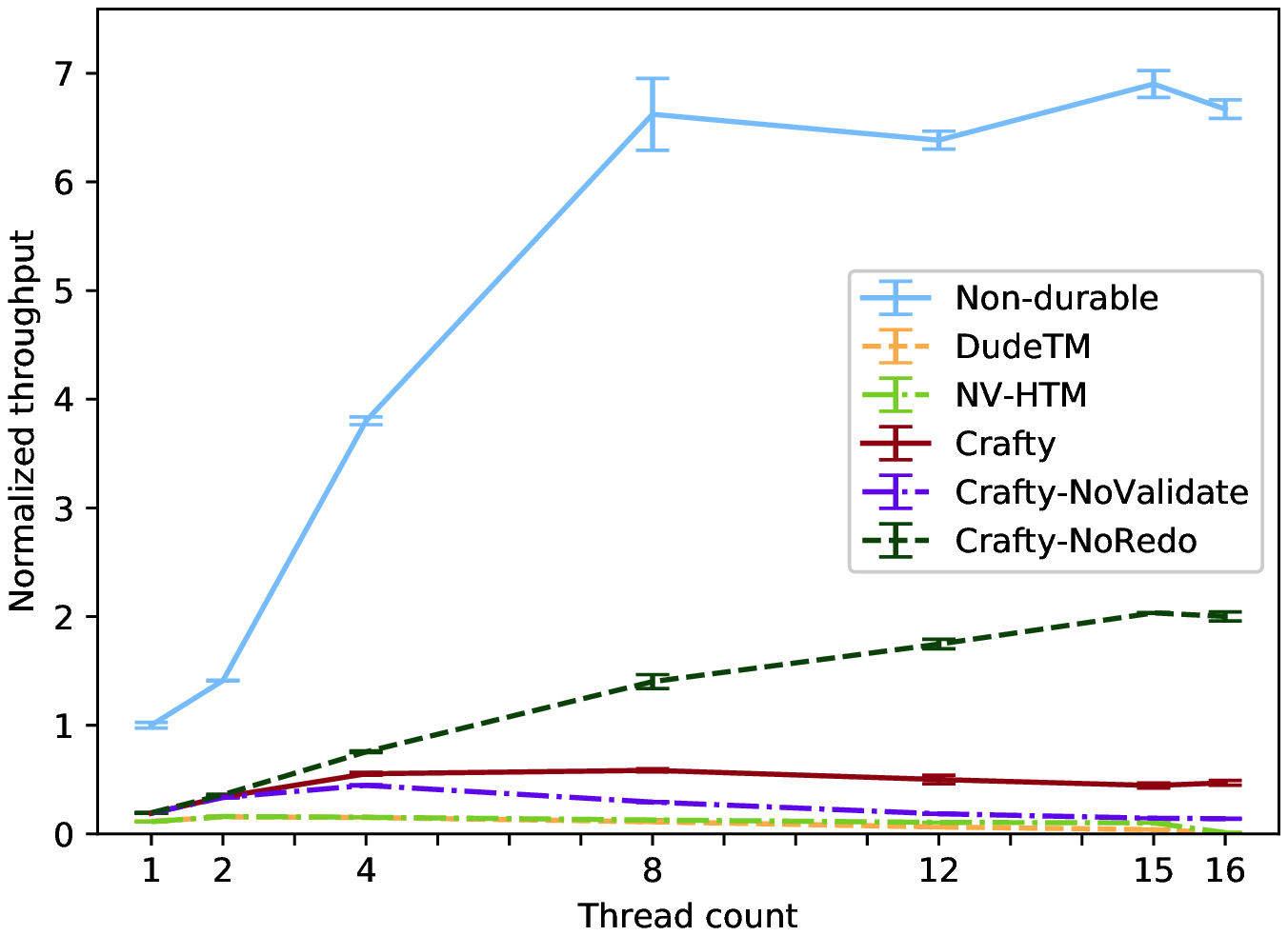}
        \label{fig:throughput:bank-fee-nc}
    }
    \caption{Throughput of Crafty and competing approaches,
    using the \bankfee microbenchmark at three contention levels.
    Crafty generally outperforms NV-HTM and DudeTM, especially under low contention and at low thread counts.}
    \label{fig:throughput:bank}
\end{figure}

Figure~\ref{fig:throughput:bank} compares Crafty and other persistent
transaction implementations, under different contention levels.  The
general trend behind these results is that Crafty outperforms NV-HTM
and DudeTM under low-contention settings, when there are few threads
or few conflicting transactions.  For example, under all contention
levels Crafty outperforms NV-HTM and DudeTM for 1--2 threads.

Under high contention, \crafty scales poorly because it Crafty amplifies the transactional conflicts by executing
persistent transactions using more hardware transactions than other approaches.
While NV-HTM scales well up to 4 threads,
its scalability limitations (Section~\ref{sec:background}) cause it to
anti-scales above 4 threads and underperforms Crafty above 8 threads.
Crafty outperforms or performs as well as the competing approaches except for NV-HTM on the high-conflict configuration at 4 threads.
\notes{the results in the supplementary material show that
Crafty incurs significantly more conflict aborts than NV-HTM.
\kaan{That's no longer true. The supplementary materials don't really explain why NV-HTM is slower than Crafty.}}

Note that NV-HTM's and DudeTM's throughput drops dramatically at 16
threads.  NV-HTM and DudeTM use an extra thread that performs writes
to persistent memory. Because 16 program threads are running
on 16 cores, the extra thread is scheduled on the same core as a
program thread, causing frequent context switches because of the
producer--consumer relationship between the two threads.


\later{Figure~\ref{fig:time:bank-fee-hc} shows why the redo phase is useful.
Except for single threaded execution, Crafty-NoRedo is slightly slower than regular Crafty.
This is because the high rate of conflicts means that many transactions do conflict with each other.
Since the validation phase is much slower than the redo phase,
\mike{If that's true, then why does Crafty-NoRedo have similar performance to Crafty and Crafty-NoValidate for a single thread?
\kaan{I'm not sure, should we just take this paragraph off? It's mostly speculation because the results in supplementary material don't tell us much about this either.}}
it is likelier to conflict with another transaction.
On the other hand, the redo phase executes much faster and is likelier to be able to finish before another thread starts executing a conflicting transaction.}

Figure~\ref{fig:throughput:bank-fee-nc} motivates the \validate phase.
When the number of threads is above 4, Crafty-NoValidate is slower than Crafty
because \redo fails due to timestamp checks, but \validate succeeds
since there is no true contention.  Results in
\iftoggle{extended-version}
         {Figure~\ref{fig:stats:bank-fee-nc} in Section~\ref{sec:extended-eval}}
         {our extended arXiv version~\cite{crafty-extended-arxiv}}
support this conclusion:
Crafty-NoValidate incurs many
explicit aborts at thread counts above 4, caused by failed
timestamp checks.

\begin{figure}
    \vspace*{-1.5em}
    \centering
    \captionsetup[subfloat]{farskip=2pt,captionskip=1pt}
    \subfloat[Insert operations only]{
        \includegraphics[width=\linewidth]{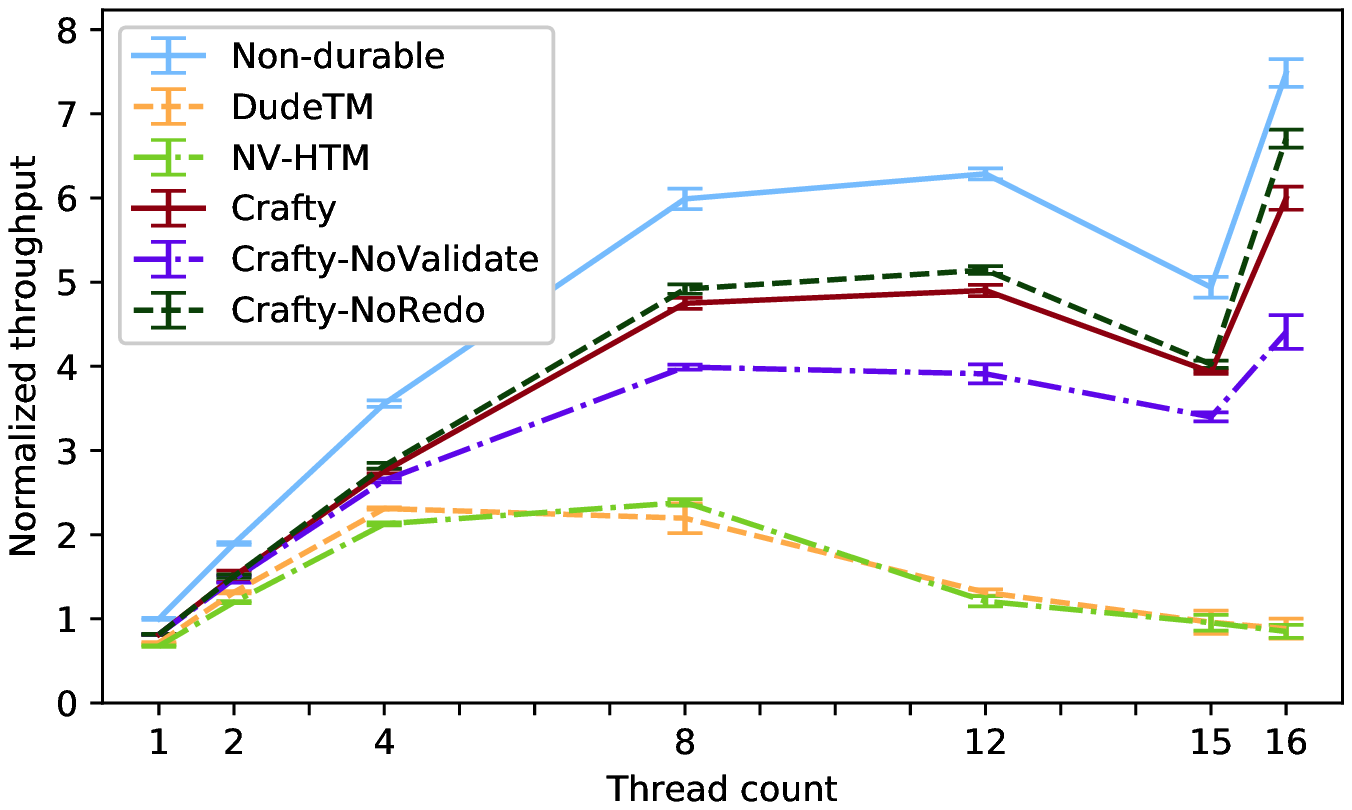}
        \label{fig:throughput:bplustree-insert}
    }
    \\
    \subfloat[Lookup, insert, and remove operations]{
        \includegraphics[width=\linewidth]{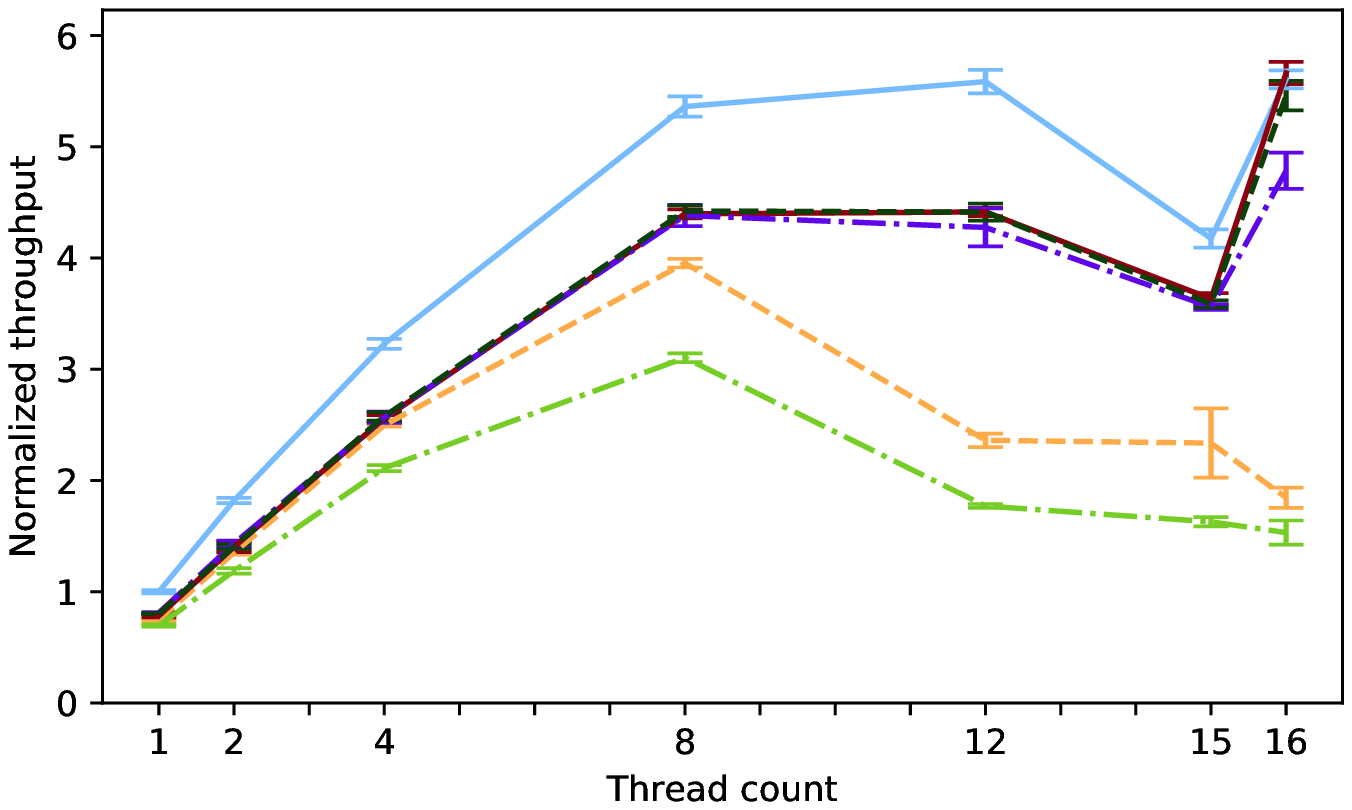}
        \label{fig:throughput:bplustree-mixed}
    }
    \caption{Throughput of \crafty and competing approaches,
    on the B+ tree microbenchmark, for mixed operations and insert only.
    \Crafty scales better than NV-HTM and DudeTM and has low overhead compared with \htmonly.}
    \label{fig:throughput:bplustree}
\end{figure}

\paragraph{B+ tree microbenchmark.}

Figure~\ref{fig:throughput:bplustree} shows the results for the B+ tree microbenchmark.
NV-HTM and DudeTM scale poorly compared with \crafty and \HTMonly,
presumably as a result of serializing execution during transaction commit and when persisting writes;
our extended
\iftoggle{extended-version}{results (Appendix~\ref{sec:extended-eval})}{results~\cite{crafty-extended-arxiv}}
do not show significant differences in transaction abort rates.

For both configurations of the benchmark at all thread counts,
\crafty outperforms NV-HTM and DudeTM, and has modest overhead over \HTMonly.

\paragraph{STAMP benchmarks.}

\begin{figure*}
    \vspace*{-1.5em}
    \centering
    \captionsetup[subfloat]{farskip=2pt,captionskip=1pt}

    \subfloat[\bench{kmeans} (high contention)]{
        \includegraphics[width=.5\linewidth]{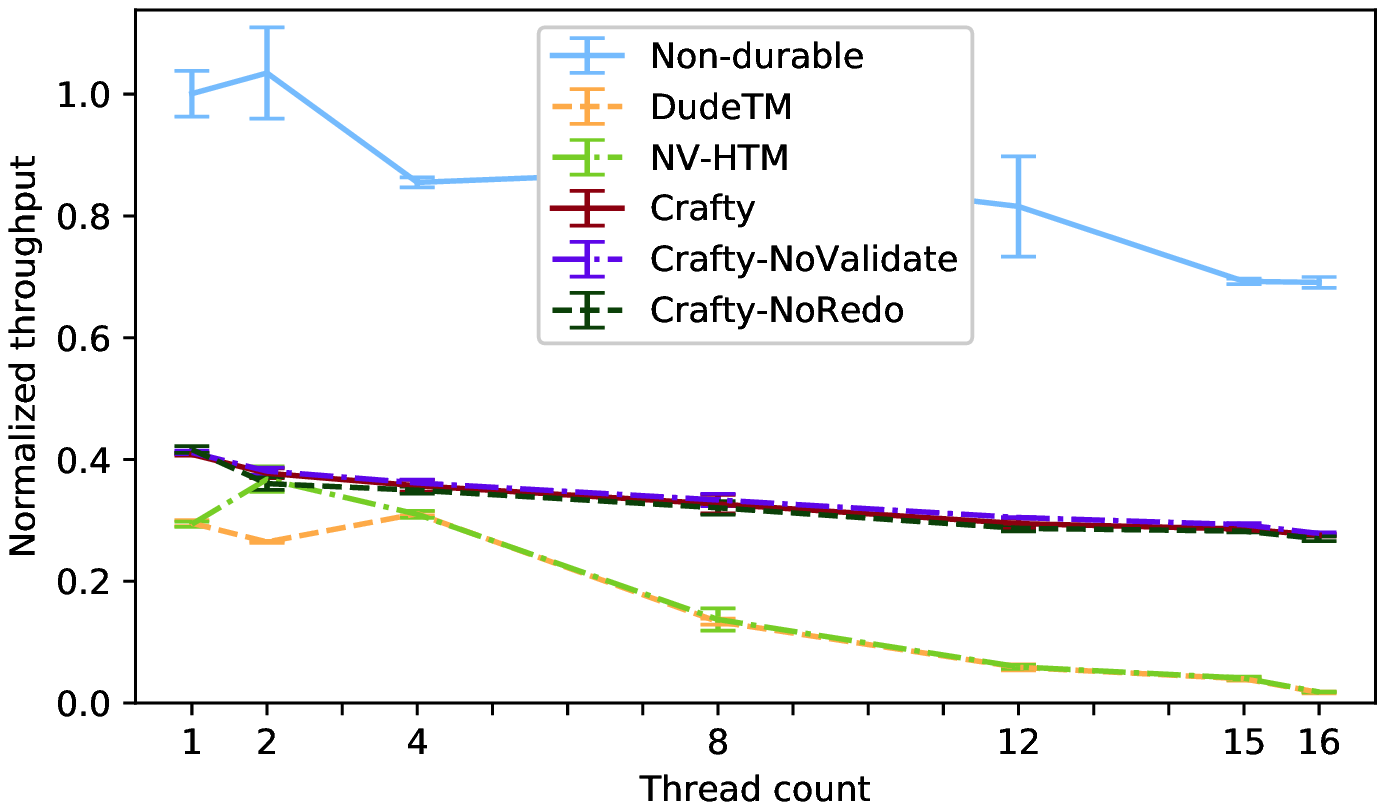}
        \label{fig:throughput:kmeans-high}
    }
    \subfloat[\bench{kmeans} (low contention)]{
        \includegraphics[width=.5\linewidth]{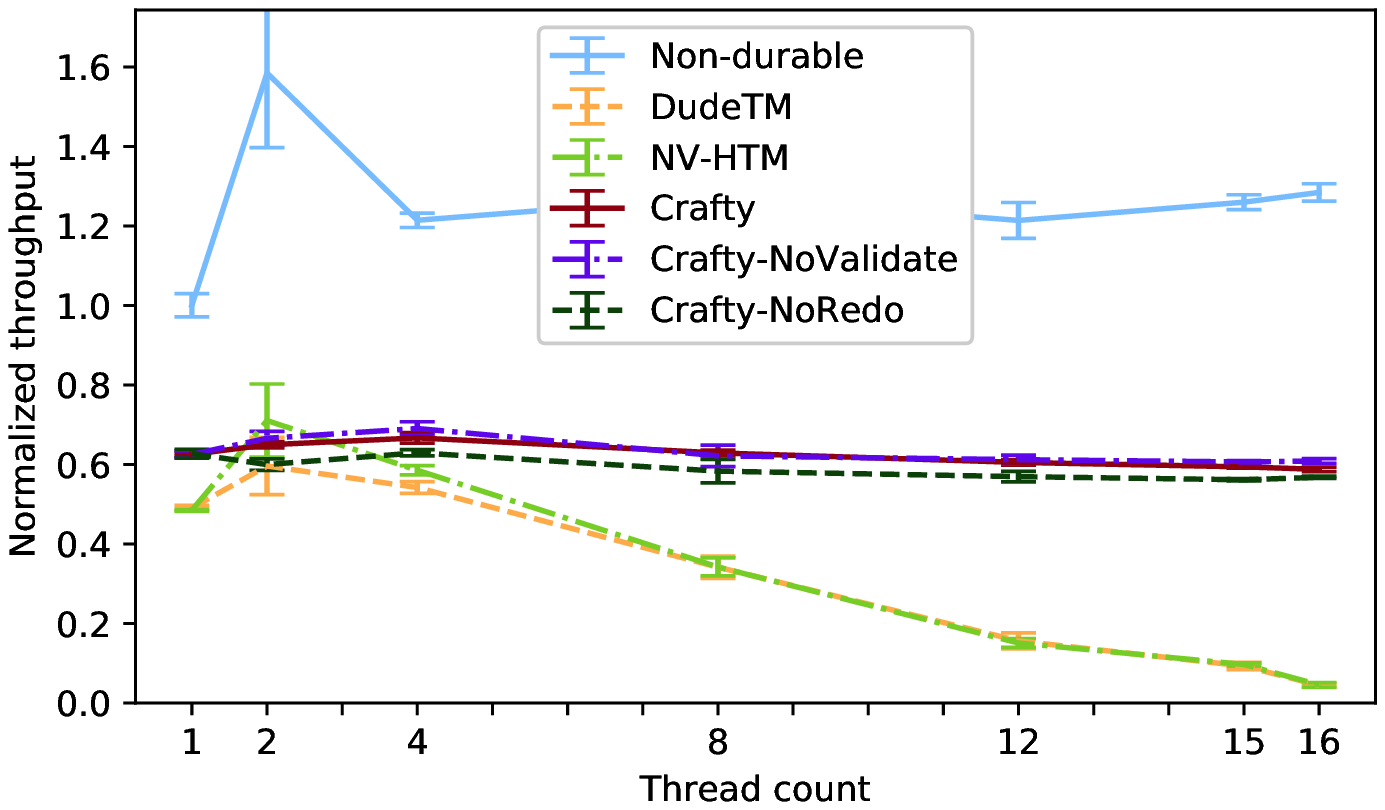}
        \label{fig:throughput:kmeans-low}
    }\\
    \subfloat[\bench{vacation} (high contention)]{
        \includegraphics[width=.5\linewidth]{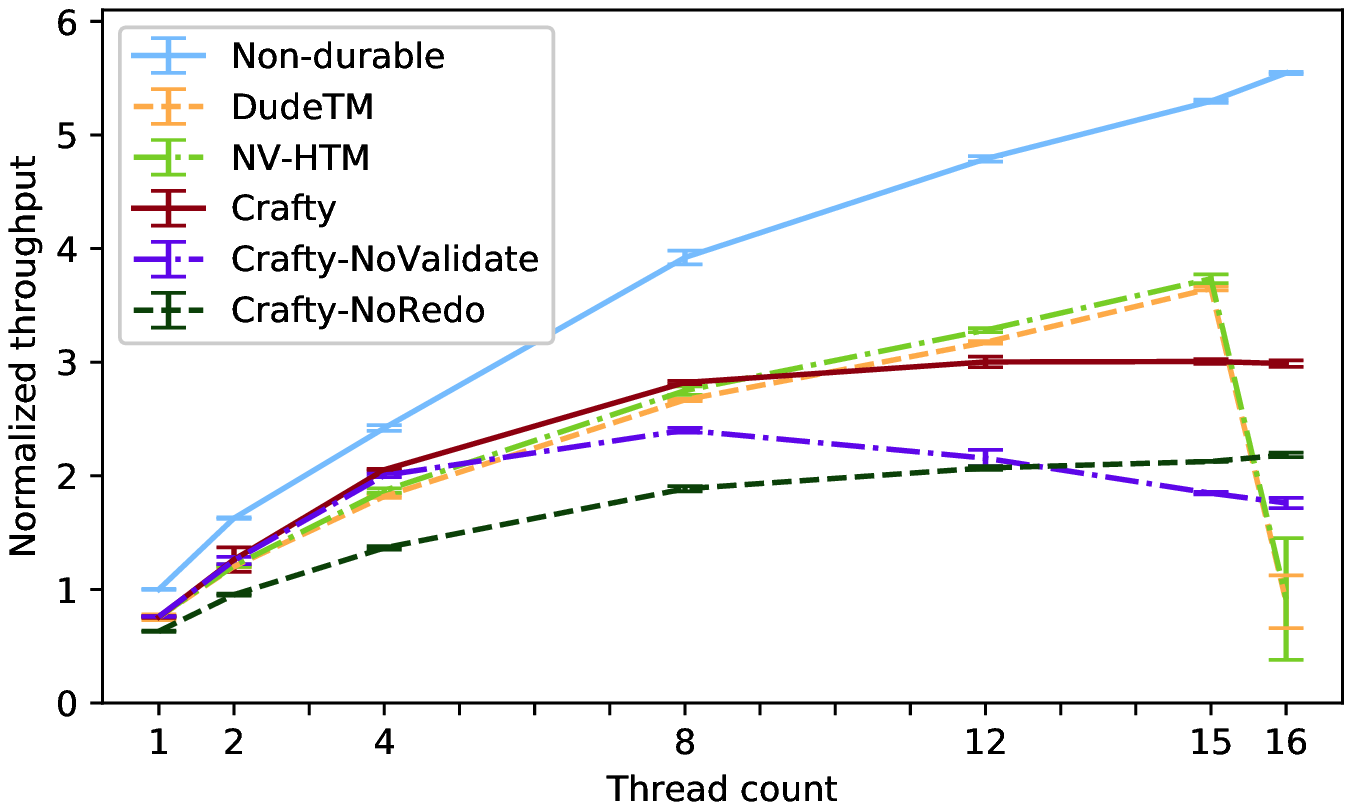}
        \label{fig:throughput:vacation-high}
    }
    \subfloat[\bench{vacation} (low contention)]{
        \includegraphics[width=.5\linewidth]{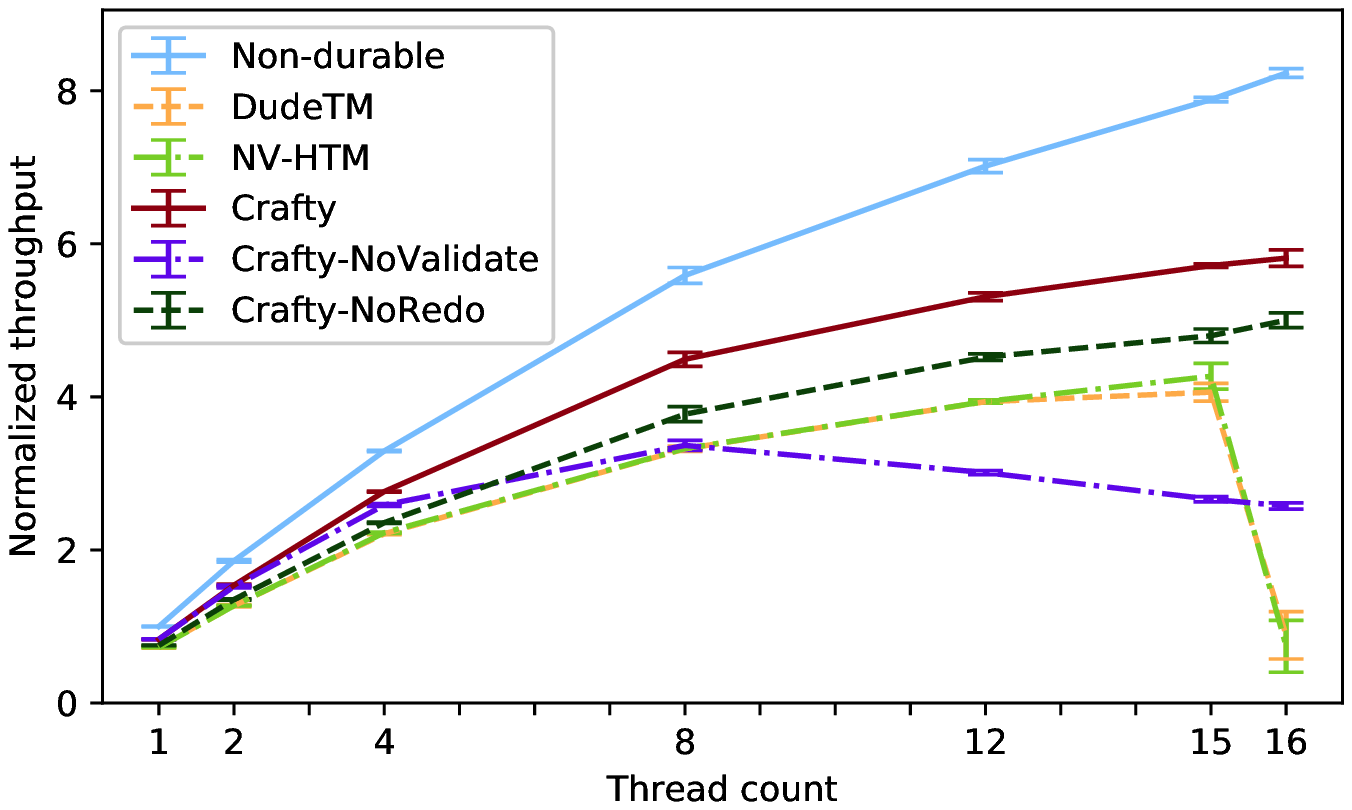}
        \label{fig:throughput:vacation-low}
    }\\
    \subfloat[\bench{labyrinth}]{
        \includegraphics[width=.5\linewidth]{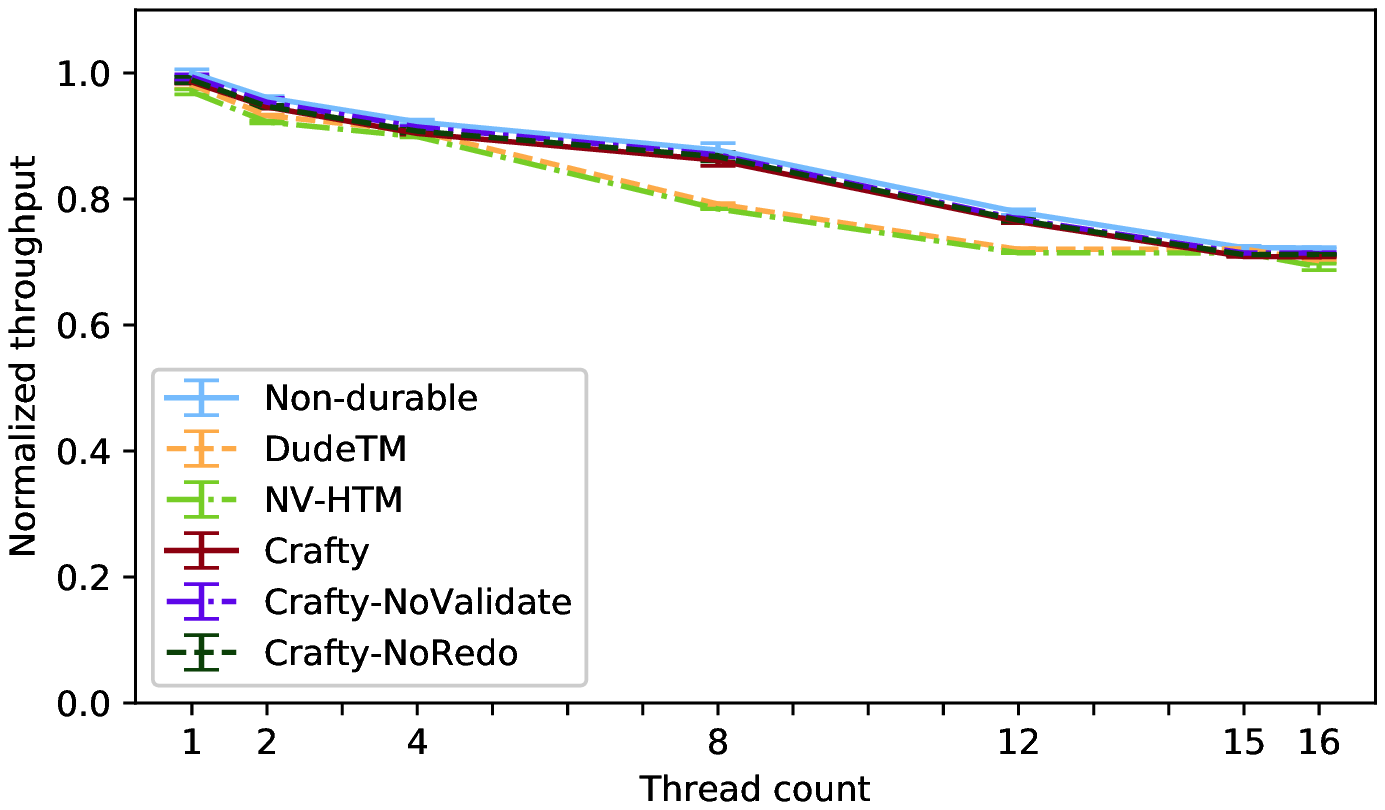}
        \label{fig:throughput:labyrinth}
    }
    \subfloat[\bench{ssca2}]{
        \includegraphics[width=.5\linewidth]{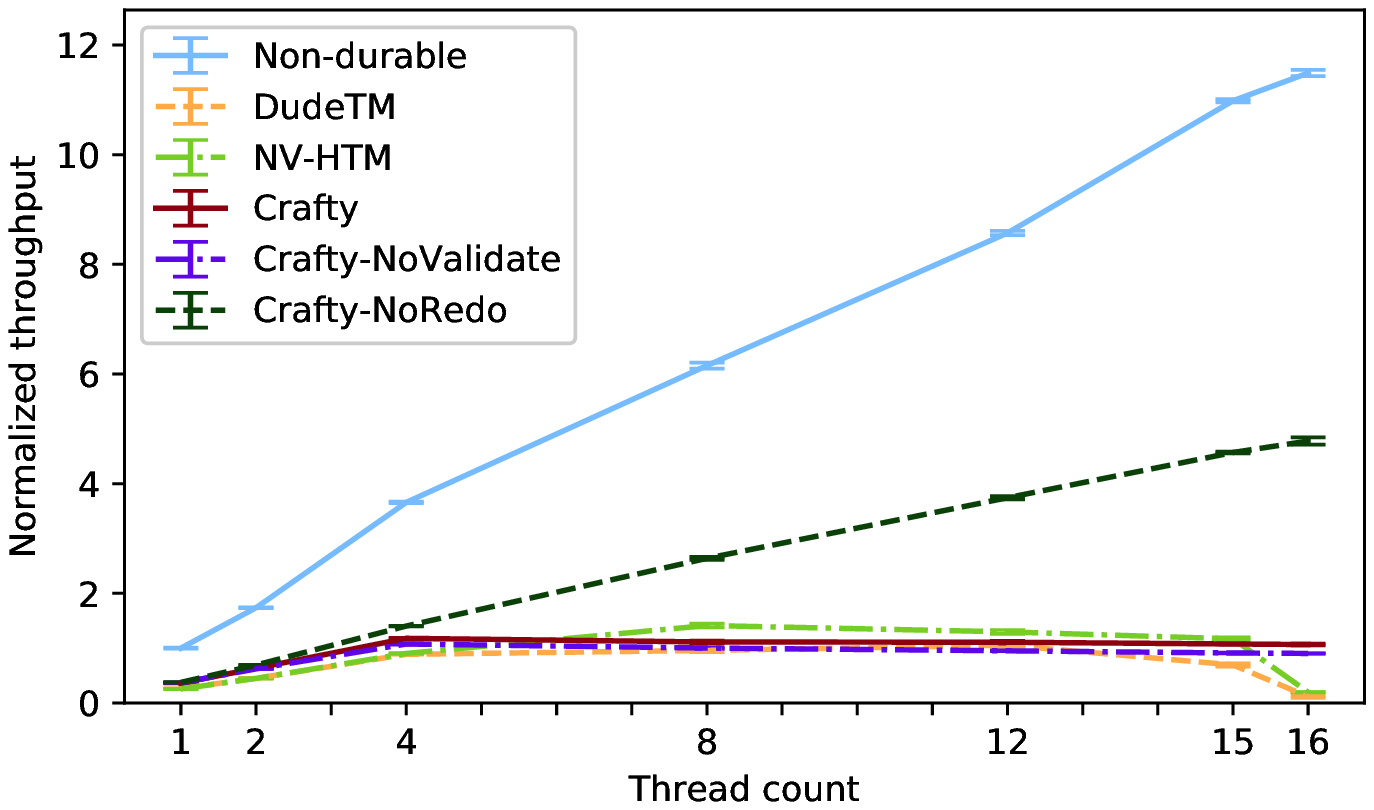}
        \label{fig:throughput:ssca2}
    }\\
    \subfloat[\bench{genome}]{
        \includegraphics[width=.5\linewidth]{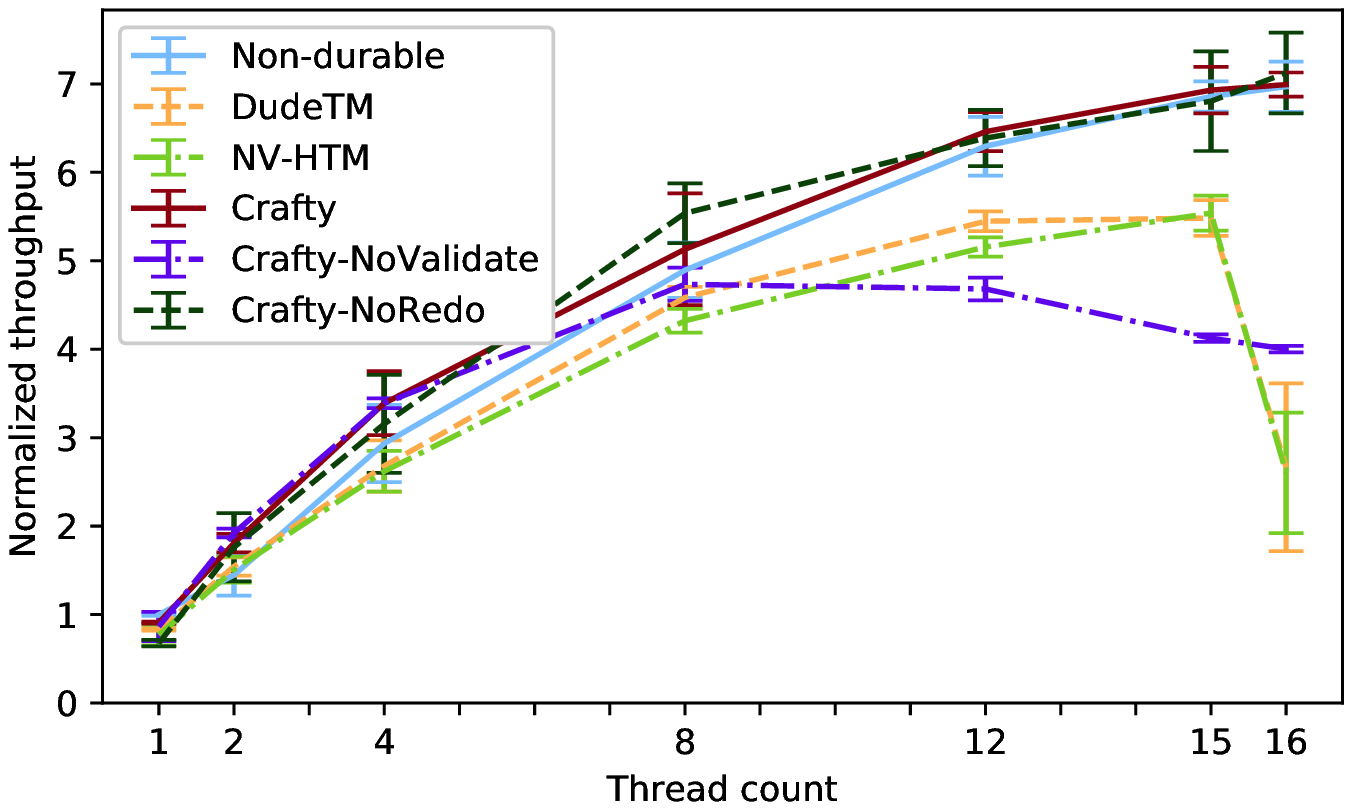}
        \label{fig:throughput:genome}
    }
    \subfloat[\bench{intruder}]{
        \includegraphics[width=.5\linewidth]{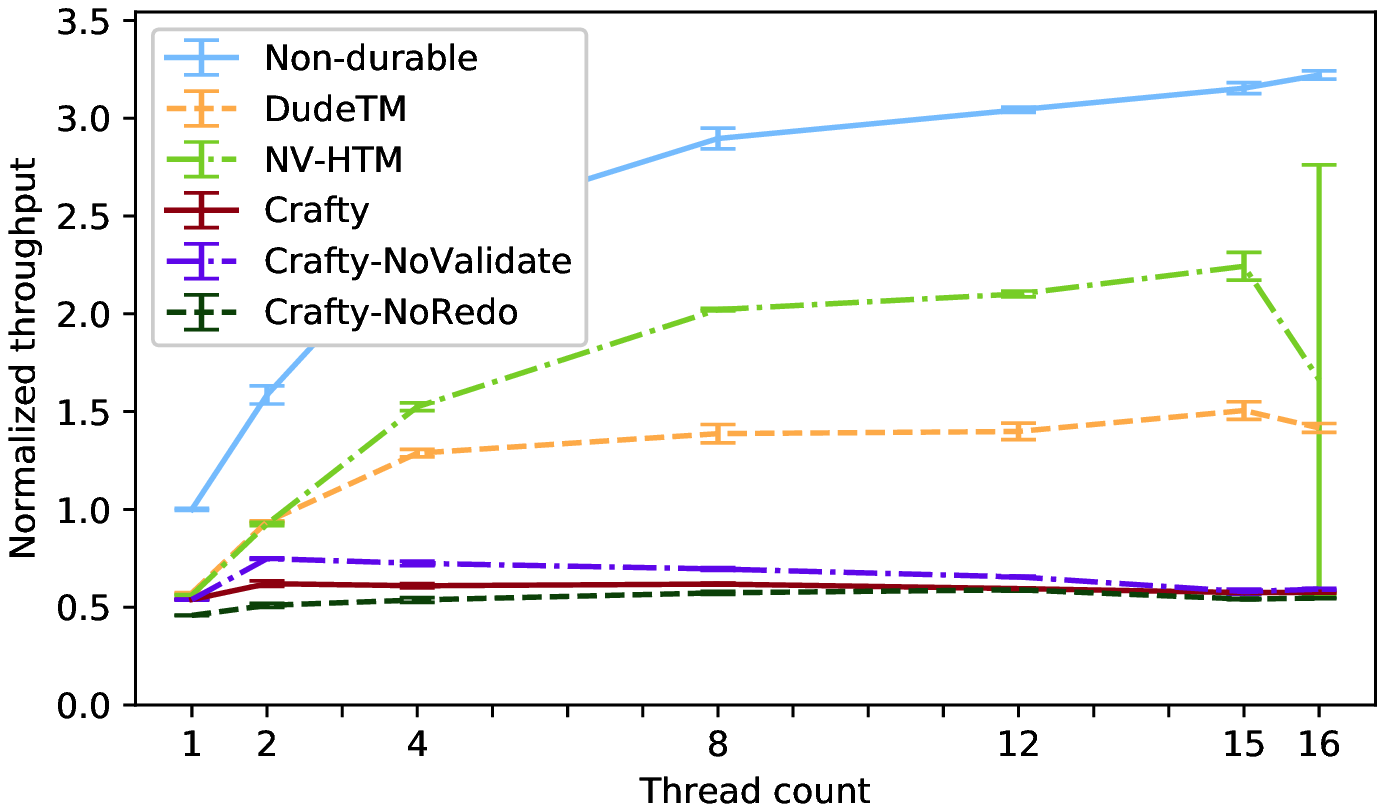}
        \label{fig:throughput:intruder}
    }
    \caption{Throughput of \crafty and competing approaches,
    on the STAMP benchmarks. Crafty has low overhead and scales well at high thread counts.}
    \label{fig:throughput:stamp}
\end{figure*}

Figure~\ref{fig:throughput:stamp} shows results for the STAMP benchmarks.
Across the benchmarks, Crafty generally performs better than NV-HTM and DudeTM
and scales as well as \HTMonly (the exception is \bench{intruder}, discussed below).

Figures~\ref{fig:throughput:kmeans-high} and \ref{fig:throughput:kmeans-low} show
that Crafty outperforms NV-HTM and DudeTM at thread counts above 4 for \bench{kmeans} under both high and low contention.

Figures~\ref{fig:throughput:vacation-high} and \ref{fig:throughput:vacation-low} show that Crafty adds modest overhead over \HTMonly on \bench{vacation}.
Except for the high-contention \bench{vacation} configuration above 8 threads,
for which NV-HTM and DudeTM perform best, Crafty outperforms competing approaches.
(The sudden drop in NV-HTM and DudeTM's throughput at 16 threads is the same issue as described above for the \bankfee microbenchmark.)
Both figures show the benefits of using both the \redo and \validate phases
under higher thread counts but low contention.

Figures~\ref{fig:throughput:labyrinth} and \ref{fig:throughput:ssca2} show similar performance for Crafty, NV-HTM, and DudeTM
on \bench{labyrinth} and \bench{ssca2}.
An exception is that for \bench{ssca2}, which has very low contention,
\Crafty-NoRedo performs significantly better than the other configurations.
\iftoggle{extended-version}{As Figure~\ref{fig:stats:ssca2} in Appendix~\ref{sec:extended-eval} shows,}{As our extended results show~\cite{crafty-extended-arxiv},}
across all thread counts, \Crafty-NoRedo, which uses only \logp and \validate phases,
experiences very few aborts.

\kaan{Should add more here, see the following notes.
\mike{The \textbackslash notes seem inaccurate / out of date.
Looking at the 100 ns results, it looks like \Crafty-NoRedo's overhead over \HTMonly on \bench{ssca2}
mostly \textbf{isn't} coming from write latency.}}
\notes{
a similar but larger effect for \bench{ssca2}:
At low thread counts, \Crafty adds high per-thread overhead.
This overhead tends to parallelize well, narrowing the gap between durable and non-durable transactions
at high thread counts.
While some of \crafty's overhead is due to its instrumentation at writes and its extra phases and hardware transactions,
we note that \bench{ssca2}'s persistent transactions each execute just two writes on average
and executes frequent transactions (supplementary material),
so \crafty cannot amortize write latency over many threads.
That is, per-transaction write latency may be a substantial source of overhead for \bench{ssca2}'s short, frequent transactions.}

Figure~\ref{fig:throughput:genome} shows that \crafty scales well at high thread counts for \bench{genome},
while NV-HTM and DudeTM are unable to scale quite as well.
Crafty-NoValidate scales poorly with more threads, showing the value of the \validate phase
when the \redo phase fails frequently due to numerous simultaneous transactions.

Figure~\ref{fig:throughput:intruder} shows that for \bench{intruder},
\crafty performs worse than NV-HTM and DudeTM.
While detailed statistics in
\iftoggle{extended-version}{Figure~\ref{fig:stats:intruder} in Appendix~\ref{sec:extended-eval}}{our extended results~\cite{crafty-extended-arxiv}}
show that \crafty configurations commit and abort significantly more
hardware transactions than the other configurations,
these results do not explain \crafty's poor performance:
\Crafty inherently commits, and often aborts, more hardware transactions than other approaches
across the other programs, yet generally outperforms NV-HTM and DudeTM.
As of the camera-ready deadline, we have not been able to understand this issue better
(we fixed an issue just days before the deadline that allowed our implementation to run \bench{intruder} without error).
\mike{TODO: Say something better about \bench{intruder} if possible.
\kaan{I tried configuring the benchmark to have fewer conflicts (which is one thing that grows much more quickly for Crafty than other solutions), but that hasn't improved Crafty's scalability on intruder (might be because the configuration doesn't really allow you to adjust the conflict rate, I think).
I also tried disabling the code for supporting recovery (committing into other threads logs), including disabling the checks for it. Also doesn't help.
Disabling NVM latency for Crafty doesn't help.
It's very likely not due to malloc/free logging, there is at most 3 allocations logged in a transaction and on average less than 1 malloc per transaction.
I'm open to suggestions on what to investigate that might help.}
\mike{It appears to be a scalability issue: \crafty's single-thread performance
is close to NV-HTM's, but their scalability is completely different,
so it makes sense that latency and malloc/free logging wouldn't explain it.
The extended results don't explain it.
\medskip\\
I'd guess it's a bug somewhere.
Trying to find a bug in the \crafty implementation at this point seems unproductive.
I'd suggest spot-checking the raw output of \bench{intruder} for \crafty and NV-HTM at moderate thread counts
and double-checking relevant parts of your execution and graph processing scripts
to make sure there's not a bug (or a manual mistake, if you did any manual processing
of results) that's specific to \bench{intruder}, if you haven't already done those things.}
\kaan{I have double-checked the results manually. The graphs for both the throughput and the stats are correct.}}

\notes{Figure~\ref{fig:throughput:bayes} shows that all configurations perform similarly on \bench{bayes}.
This is because around half of the persistent transactions for all methods
fall back to the SGL, as can be seen in the results in \iftoggle{extended-version}{Figure~\ref{fig:stats-ptx:bayes} in Section~\ref{sec:extended-eval}}{our extended arXiv version~\cite{crafty-extened-arxiv}}.
This behavior is caused by many hardware transactions executing the
HTM-incompatible \code{vzeroupper} x86 instruction.}

\later{
\mike{DudeTM has an inherent scalability bottleneck: The Reproduce stage is basically serialized in its write-back of state,
based on timestamp order. That should be an issue for highly parallel workloads (\ie, many threads but not a lot of contention).
Note that even though the Reproduce stage (like the Persist stage) is decoupled from transaction execution (the Perform stage),
the Perform stage can only run as fast as the Reproduce stage because logs will be circular / will be reused for practicality/efficiency reasons.
The experimental section of the DudeTM paper~\cite{dudetm} isn't all that clear,
but it looks like most experiments use just 4 threads. The scalability graph's normalization choices are unnecessarily confusing:
It's unimpressive to scale well when adding run-time overhead, and it's unclear what the ``conflict mitigated'' configuraton is.}
\mike{DudeTM reports pretty low run-time overheads over STM, but that's not really all that impressive
considering that STM adds high run-time overhead to begin with.}
\mike{I investigated why NV-HTM is often slower than \crafty for single-threaded execution.
It seems to be two factors:
(1) NV-HTM's periodic persisting of logs and the persistent copies of program state and
(2) the fact that (in place(s) that I'm having trouble tracking down precisely)
NV-HTM performs waiting proportional to a number of persisted writes
(by calling \code{SPIN\_PER\_WRITE(nb\_writes)} with a value of \code{nb\_writes > 1}).
I had trouble understanding exactly which code paths and behavior NV-HTM was using,
but I came to these conclusions by testing the performance of
(1) turning off the checkpoint manager (by setting \code{DO\_CHECKPOINT = 1}) and/or
(2) forcing \code{nb\_writes = 1} in \code{SPIN\_PER\_WRITE(nb\_writes)}.
When \code{SPIN\_PER\_WRITE(nb\_writes)} gets called with \code{nb\_writes > 1},
I'm not sure whether that's overkill or actually needed by NV-HTM.}
}


\section{Related Work}

\Crafty leverages hardware transactional memory (HTM) to control persist ordering,
while also supporting the use of commodity HTM for concurrency control in persistent transactions.
To our knowledge, no prior work has used HTM to control persist ordering.
Prior work supports commodity HTM for concurrency control in
persistent transactions~\cite{dudetm,nv-htm,persistent-htm-giles-2017}.
DudeTM and NV-HTM use shadow-paging-based copy-on-write mechanisms
and incur scalability bottlenecks~\cite{dudetm,nv-htm};
we compared with them qualitatively and quantitatively in this paper.
Giles \etal\ introduced an approach for HTM-based persistent transactions
that requires instrumenting program reads~\cite{persistent-htm-giles-2017},
arguably forgoing a key benefit of using HTM instead of STM.
In contrast with the prior work, which works around
the challenges of combining persistence and HTM,
\crafty \emph{leverages} HTM to control persist ordering,
as realized in the new \ndundologging mechanism.

\paragraph{Modifying HTM}

Several research efforts propose nontrivial modifications to commodity HTM to
support persistent transactions~\cite{htpm,phtm,ptm,phytm,dhtm}.
In contrast, \crafty shows how to leverage and work with contemporary systems.

\notes{
Another proposed hardware modification is using \emph{nonvolatile caches} to eliminate the costs
of persist ordering~\cite{kiln}.
In contrast, the proposed work aims to reduce persistence costs to similar levels
without incurring the costs of nonvolatile caches.
}

\paragraph{Software persistent transactions.}

Many existing systems provide persistent
transactions~\cite{mnemosyne,NV-heaps,kamino-tx,pmthreads,persistent-transactions-kolli-2016,persistent-memory-programming,pmfs,romulus,pisces,softwrap,transactions-for-pm,persistent-memory-transactions,pser,onefile}.
These approaches use undo, redo, or copy-on-write mechanisms to provide failure atomicity.
The approaches either assume that programs provide thread atomicity through locks or another concurrency control mechanism,
or they apply STM to provide thread atomicity together with failure atomicity.

\notes{
\mike{Already covered in Background:}
Marathe \etal\ evaluated the costs of undo logging, redo logging,
and (non-shadow-paging-based) copy-on-write mechanisms~\cite{persistent-memory-transactions}.
They found that these crash-consistency mechanisms had different advantages in different settings.
\Crafty introduces a new crash-consistency mechanism (\ndundologging)
that offers advantages including compatibility with commodity HTM.
}

\notes{
Frameworks for persistent memory programming and for emulating persistent memory
also provide persistent
transactions~\cite{pmfs,persistent-memory-programming}.\footnote{\url{https://pmem.io/pmdk/}, \url{https://github.com/linux-pmfs/pmfs}}
}


\paragraph{Failure atomicity of critical sections.}

Several approaches including \emph{Atlas} provide failure atomicity for
lock-based critical sections~\cite{atlas,nvthreads,atlas-follow-up,justdo-logging,ido} or
synchronization-free regions~\cite{persistency-sfr}.
\Crafty (in its thread-unsafe mode) could likewise provide failure atomicity for lock-based regions.


\notes{
\paragraph{Logging.}

Prior work improves logging to reduce the numbers of writes and persists to
NVM~\cite{logging-algorithms-oopsla17,proteus,hardware-assisted-logging,picl,atom-logging}.
\Crafty's logging employs techniques introduced by
Cohen \etal~\cite{logging-algorithms-oopsla17}.
Hu \etal\ show how to modify data structures so that the
data structure itself is stored in the log~\cite{log-structured-nvmm}.
}

\notes{
\paragraph{Miscellaneous.}

\emph{ThyNVM} enables crash consistency for unmodified programs~\cite{thynvm}.

The \emph{WHISPER} benchmark suite enables evaluating persistent memory program characteristics~\cite{pm-whisper}.
}

\paragraph{Failure ordering.}

This paper focuses on providing failure atomicity.
Providing failure atomicity relies on 
the lower-level property of \emph{failure ordering},
which refers to the order of persisted writes that the recovery observer sees.
This paper's \ndundologging leverages HTM to control persist ordering.
Prior work introduces \emph{memory persistency models}, 
which extend memory consistency models to incorporate the recovery
observer~\cite{memory-persistency,language-level-persistency,efficient-persist-barriers,delegated-persist-ordering}.

\notes{
\paragraph{Non-durable transactional memory.}

Prior work also uses STM or HTM or hybrid TM for \emph{non-durable} transactions\dots

Prior work uses transactional memory~\cite{tm-herlihy-moss} to support
\emph{non-durable} transactions that provide only thread atomicity.
Commodity HTM extends the caches and cache coherence protocol
to detect and resolve conflicts and roll back aborted transactions,
and falls back to software transactional memory (STM) or a global lock
for transactions that cannot complete using HTM~\cite{yoo-rtm-2013,hybrid-tm-openjdk,invyswell,reduced-hardware-norec}.
STM detects and resolves conflicts in
software and uses undo or redo logs (\cf~\cite{lightweight-transact,norec-stm,mcrt-stm,stm-only-research-toy,stm-more-than-research-toy}).
\notes{
For example, \emph{NOrec} buffers stores in a redo log during transaction execution,
and before committing a transaction validates its reads and commits its writes~\cite{norec-stm}.
}%
In contrast, \Crafty uses hardware transactional isolation to control persist ordering,
an it supports using HTM to support rollback and to detect and resolve conflicts.

\mike{Mentioned this elsewhere:}
\Crafty's thread-unsafe mode adaptively executes hardware transactions of different sizes (different numbers of persistent writes),
to balance the latency of persist operations with the risk of aborting.
Prior work in other contexts splits costs to balance per-transaction costs with time lost to aborts~\cite{legato-cgo-2017,block-chop}.
}

\section{Conclusion}

\Ndundologging is a new crash-consistency mechanism
that leverages commodity HTM to persist a transaction's undo log entries before
its persistent writes.
\Crafty is a new design that uses \ndundologging to provide persistent transactions.
An evaluation shows that \crafty
performs well compared with non-durable transactions and
has better performance than
state-of-the-art persistent transaction designs.
These results show the potential for efficient persistent transactions
using today's computing systems.


\begin{acks}

Many thanks to Daniel Castro for making the NV-HTM implementation publicly available
and providing help using it.
Thanks to Steve Blackburn, Jake Roemer, and Tomoharu Ugawa for helpful discussions and feedback.
We thank the anonymous reviewers and our shepherd, Erez Petrank,
for feedback and suggestions that improved the final paper.

This material is based upon work supported by the National Science
Foundation under Grants CAREER-1253703, XPS-1629126, CNS-1613023,
CNS-1703598, and CNS-1763172, and by the Office of Naval Research
under Grants N00014-16-1-2913 and N00014-18-1-2037.

\end{acks}

\newcommand{\showDOI}[1]{\unskip}
\bibliographystyle{ACM-Reference-Format}
\bibliography{bib/conf-abbrv,bib/plass,crafty}

\iftoggle{extended-version}{
\appendix

\later{
\section{Correctness}

Suppose we wanted to prove the correctness of \crafty.
What would that look like? Seems like there are two different properties:
\begin{itemize}
\item Transactions are thread-safe / consistent, \ie, the ACI properties of ACID are preserved.
\item The recovery observer returns to a consistent persistent state,
\ie, the durability property of ACID is preserved.
\end{itemize}

The first property seems doable given the current algorithms.
I think it's pretty straightforward to see that the \logp and \validate phases
provide ACI. The \redo phase provides ACI because of the local and global timestamp variables,
which probably isn't too hard to prove.

The second property seems less obviously true and thus more important to prove.
The proof would seem to rely on having a more concrete recovery algorithm than we have now.
We could at least start with respect to the simpler recovery algorithm
that doesn't worry about log entry reuse, persist granularity, and whatever else.
}

\section{Additional Results}
\label{sec:extended-eval}

This section contains additional results
that supplement the main paper's results.

\paragraph{Persistent writes per transaction.}

Table~\ref{tab:write-per-ptx}
shows the average numbers of writes
executed by each persistent transaction. Because \Crafty amortizes persist
latency across all writes in a transaction, it reduces
latency compared with approaches that incur per-write overhead if each
transaction executes multiple writes.  On the other hand, long
transactions are more likely to abort due to capacity constraints and
conflicts with other threads.


    \begin{table}
    \footnotesize
    \begin{tabular}{@{}l|r@{\quad}r@{\quad}r@{\quad}r@{\quad}r@{\quad}r@{\quad}r@{}}
    & 1 & 2 & 4 & 8 & 12 & 15 & 16 \\
    \hline
    
Bank (medium) & 10.0  & 10.0  & 10.0  & 10.0  & 10.0  & 10.0  & 10.0 \\
Bank (high) & 10.0  & 10.0  & 10.0  & 10.0  & 10.0  & 10.0  & 10.0 \\
Bank (none) & 10.0  & 10.0  & 10.0  & 10.0  & 10.0  & 10.0  & 10.0 \\\hline
B+ tree (mixed) & 13.3  & 13.3  & 13.3  & 13.3  & 13.2  & 13.2  & 13.2 \\
B+ tree (insert only) & 14.0  & 14.0  & 14.0  & 14.0  & 14.0  & 14.0  & 14.0 \\\hline
\bench{kmeans} (high) & 25.0  & 25.0  & 25.0  & 25.0  & 25.0  & 25.0  & 25.0 \\
\bench{kmeans} (low) & 25.0  & 25.0  & 25.0  & 25.0  & 25.0  & 25.0  & 25.0 \\
\bench{vacation} (high) & 8.0  & 8.0  & 8.0  & 8.0  & 8.0  & 8.0  & 8.0 \\
\bench{vacation} (low) & 5.5  & 5.5  & 5.5  & 5.5  & 5.5  & 5.5  & 5.5 \\
\bench{labyrinth} & 177.6  & 177.4  & 177.1  & 176.3  & 175.4  & 175.1  & 174.9 \\
\bench{ssca2} & 2.0  & 2.0  & 2.0  & 2.0  & 2.0  & 2.0  & 2.0 \\
\bench{genome} & 2.1  & 2.1  & 2.0  & 2.0  & 2.1  & 2.1  & 2.0 \\
\bench{intruder} & 1.8  & 1.8  & 1.8  & 1.8  & 1.8  & 1.8  & 1.8 \\
    \end{tabular}
    \caption{Number of writes per executed persistent transaction on average,
    for each evaluated thread count.
    \label{tab:write-per-ptx}}
    \end{table}

\paragraph{Transaction breakdowns.}

The following pages contain figures that
present the breakdowns for persistent transactions
and hardware transactions executed.
For each benchmark, the figure contains two bar graphs: one for the breakdown of persistent transactions
and the other for the breakdown of hardware transactions.

The persistent transaction breakdown shows how each persistent transaction was \emph{completed}.
For \HTMonly, Dude\-TM, and NV-HTM, persistent transactions
can be completed using a hardware transaction (labeled \emph{Non-\crafty})
or the SGL fallback.
For \crafty, persistent transactions can be completed
using the \redo or \validate phase or the SGL fallback.
An exception is for \emph{Read Only} transactions,
for which \crafty skips the \redo and \validate phases.
(\HTMonly, DudeTM, and NV-HTM also perform read-only transactions, but
the graphs categorize them as \emph{Non-\crafty}.)

The hardware transaction breakdown shows the outcome of each hardware transaction:
either a commit or a conflict, capacity, explicit, or ``zero'' abort.
Conflict aborts occur if multiple concurrent transactions performing conflicting accesses to
the same cache line.
Capacity aborts occur if the transaction accesses more cache lines than HTM can handle.
Explicit aborts occur if the program explicitly requests an abort
as part of its programming, or if a \redo or \validate transaction aborts due to failed checks (\ie, line~\ref{line:redo-abort} in Algorithm~\ref{alg:redo-phase} or line~\ref{line:validate-abort} in Algorithm~\ref{alg:validate-phase}).
\later{
\mike{Is that a thing, or are these validation failures or something else?
\kaan{I previously checked and confirmed that some benchmarks do use explicit aborts in their programming.}
\mike{Ah okay. I guess validation failures are rare in practice,
and \redo aborts are counted in a different way for now.}}
}%
A ``zero'' abort is an abort that does not fit into any
of these categories.
For example, a transaction that triggers a page fault, executes a system call,
or receives an interrupt will cause a zero abort.
The figures count every executed hardware transaction;
for \crafty, these counts include transactions performed 
\crafty's \logp, \redo, and \validate phases---including for the \logp phase in an SGL section.

\paragraph{Sensitivity to NVM latency.}

The last several figures present the same performance results as the main paper,
but emulate an NVM persist latency of 100 ns (instead of 300 ns as in the main paper).
These results help to show how much performance cost is due to NVM latency,
and they represent the expected performance if the NVM controller
includes a buffer as part of the persistence domain~\cite{persistent-memory-programming}
(Section~\ref{subsec:background-mechanisms}).

\begin{figure*}
    \centering
    \subfloat[Persistent transaction breakdowns.]{
        \includegraphics[width=\linewidth]{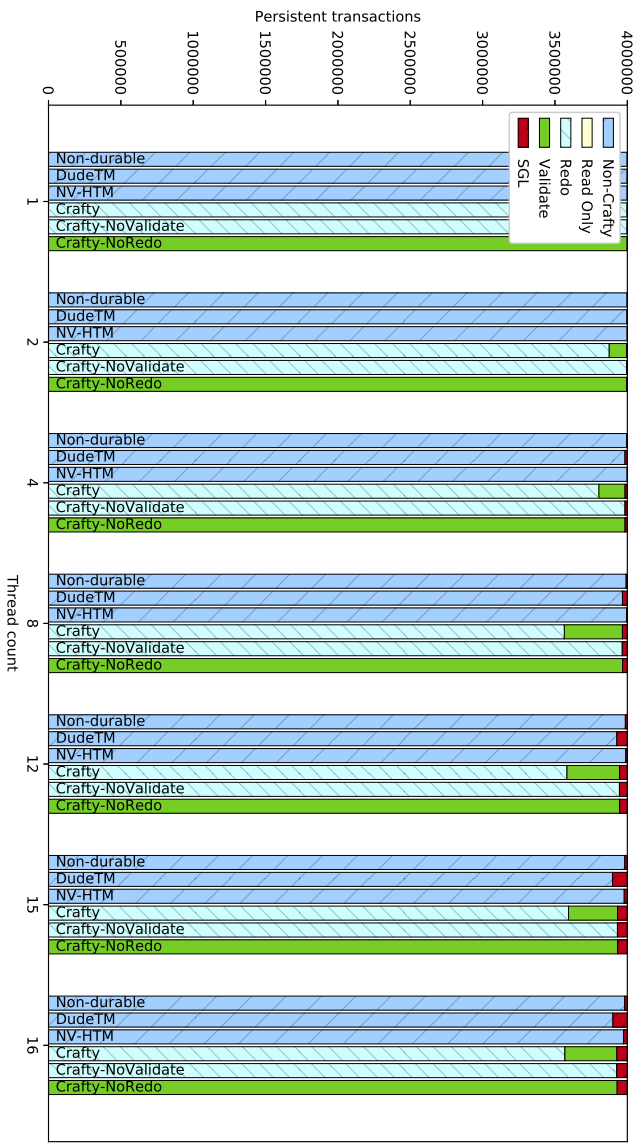}
    }

    \subfloat[Hardware transaction breakdowns.]{
        \includegraphics[width=\linewidth]{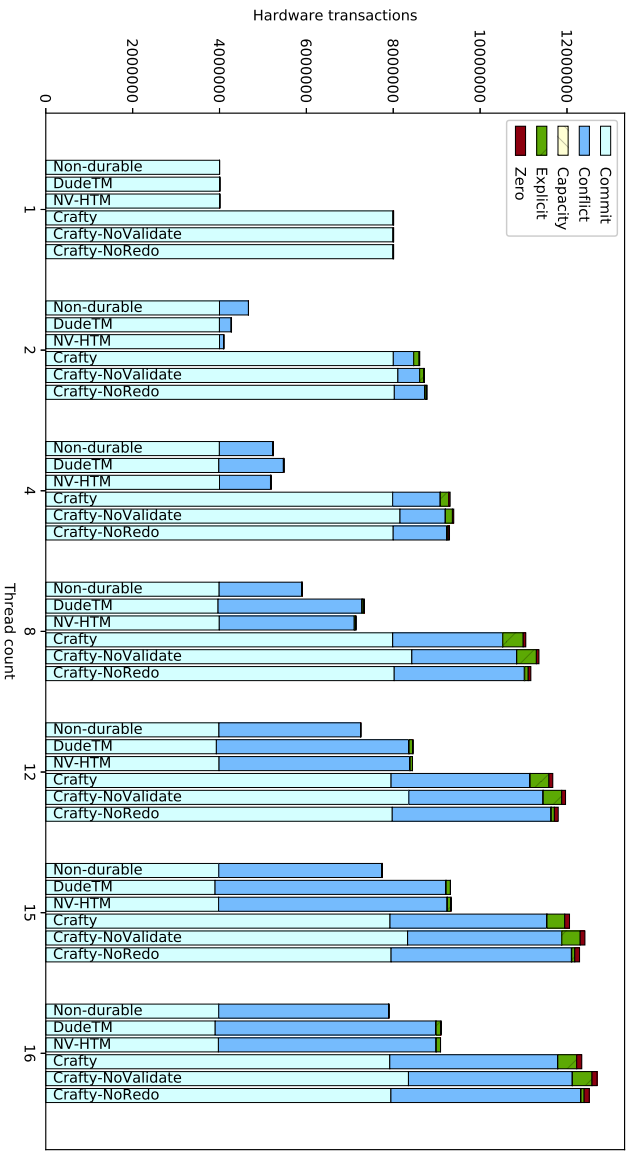}
    }
    \caption{Persistent and hardware transaction breakdowns for the \bankfee microbenchmark (high contention).}
\end{figure*}

\begin{figure*}
    \centering
    \subfloat[Persistent transaction breakdowns.]{
        \includegraphics[width=\linewidth]{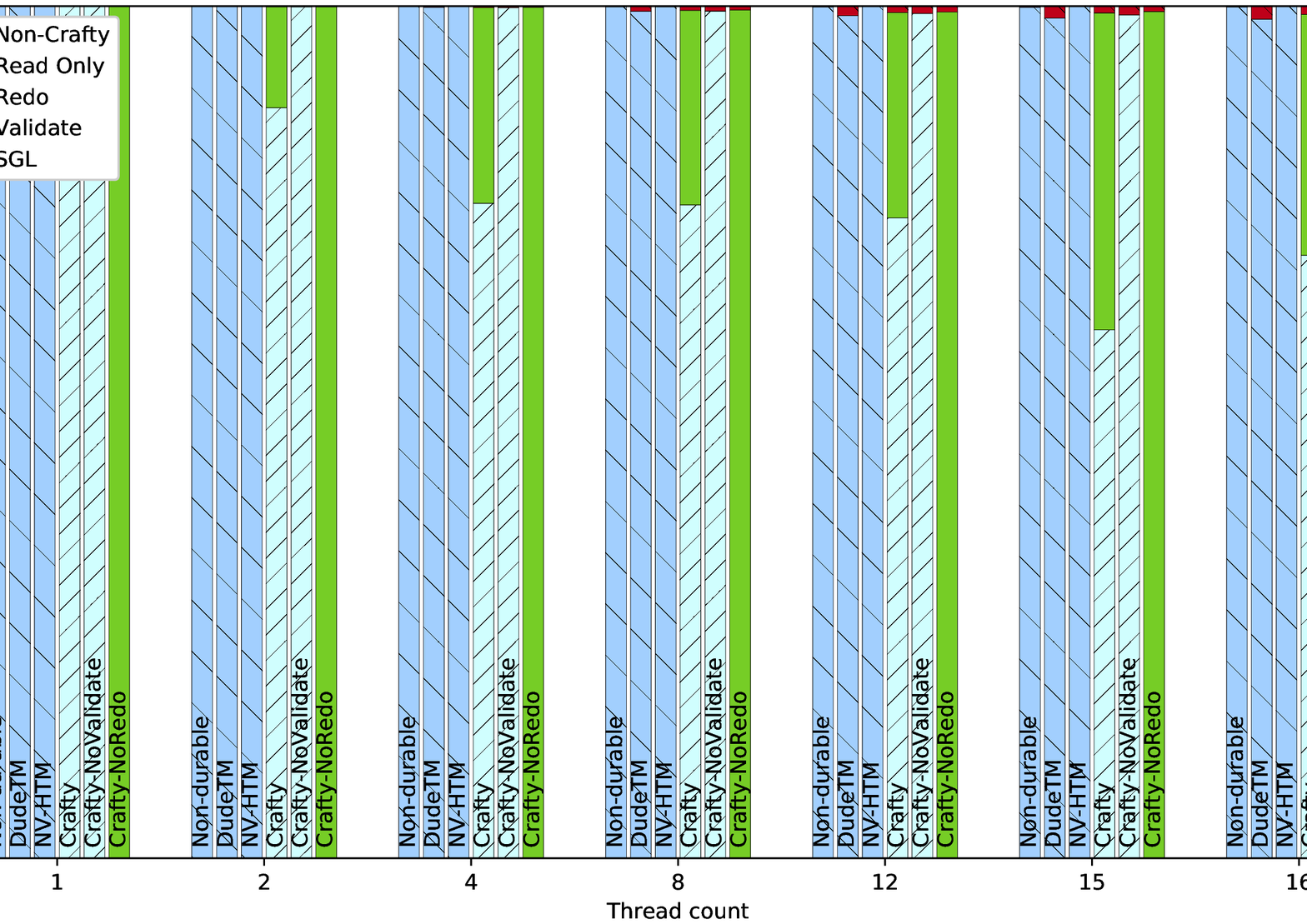}
    }

    \subfloat[Hardware transaction breakdowns.]{
        \includegraphics[width=\linewidth]{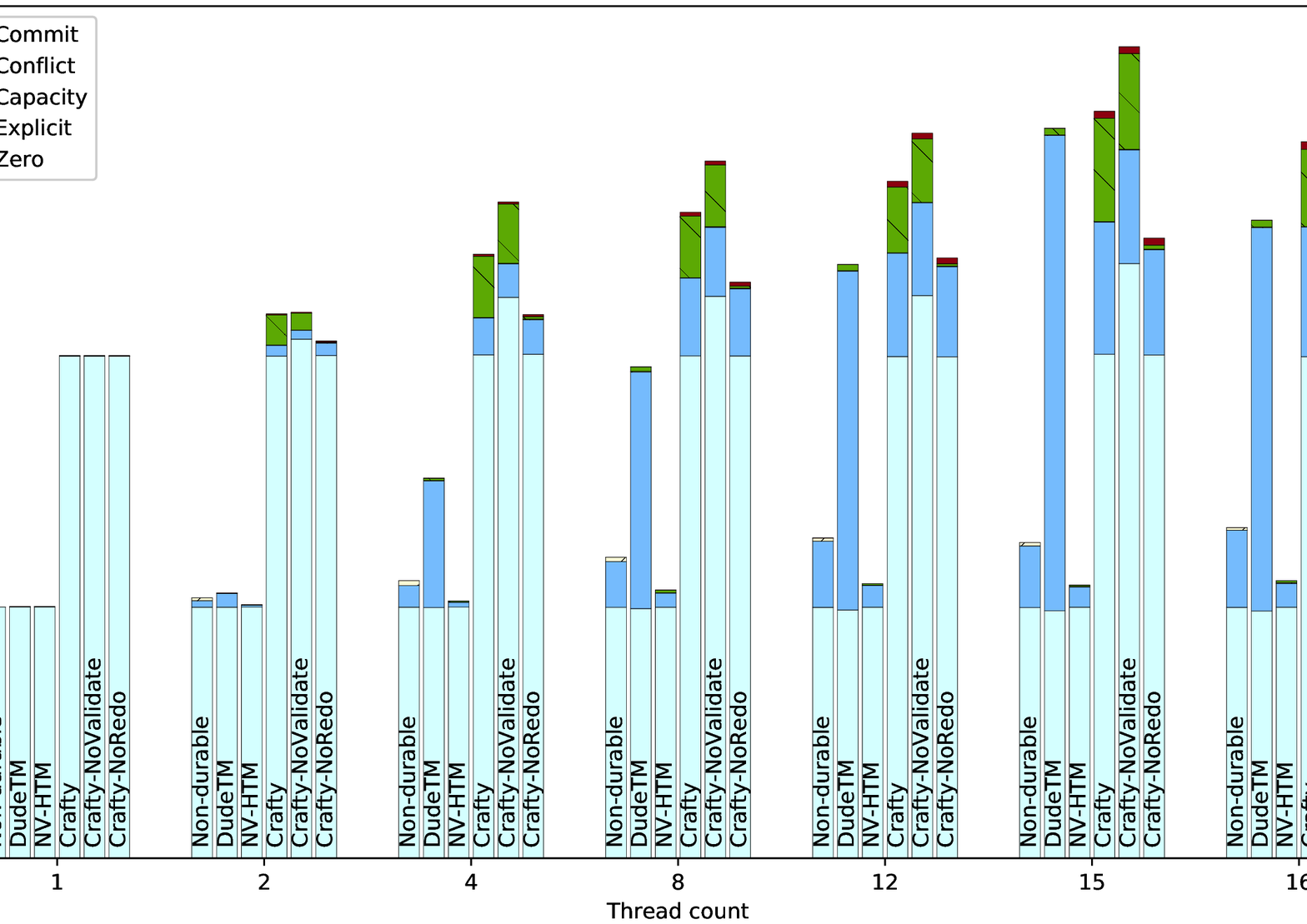}
    }
    \caption{Persistent and hardware transaction breakdowns for the \bankfee microbenchmark (medium contention).}
\end{figure*}

\begin{figure*}
    \centering
    \subfloat[Persistent transaction breakdowns.]{
        \includegraphics[width=\linewidth]{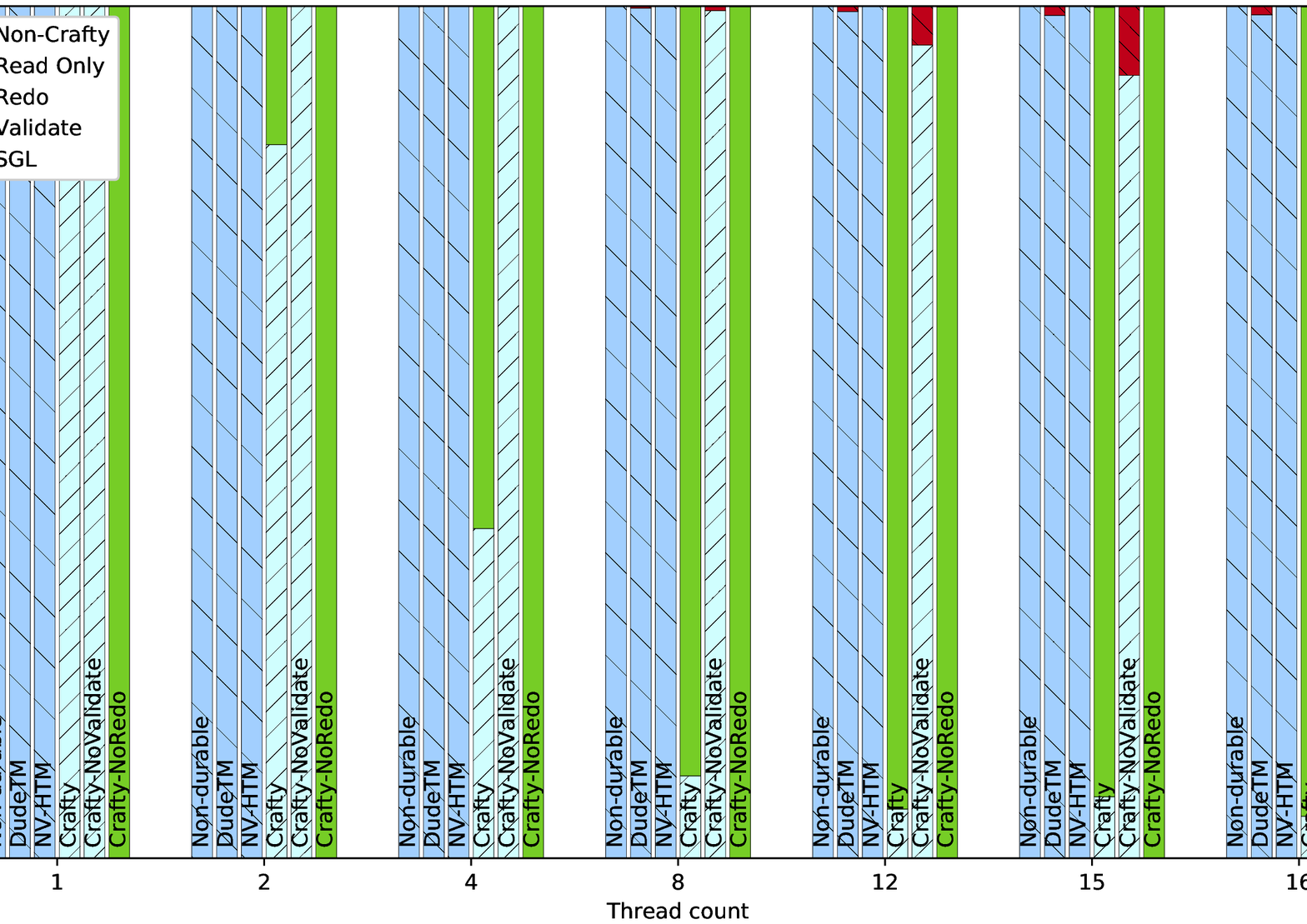}
    }

    \subfloat[Hardware transaction breakdowns.]{
        \includegraphics[width=\linewidth]{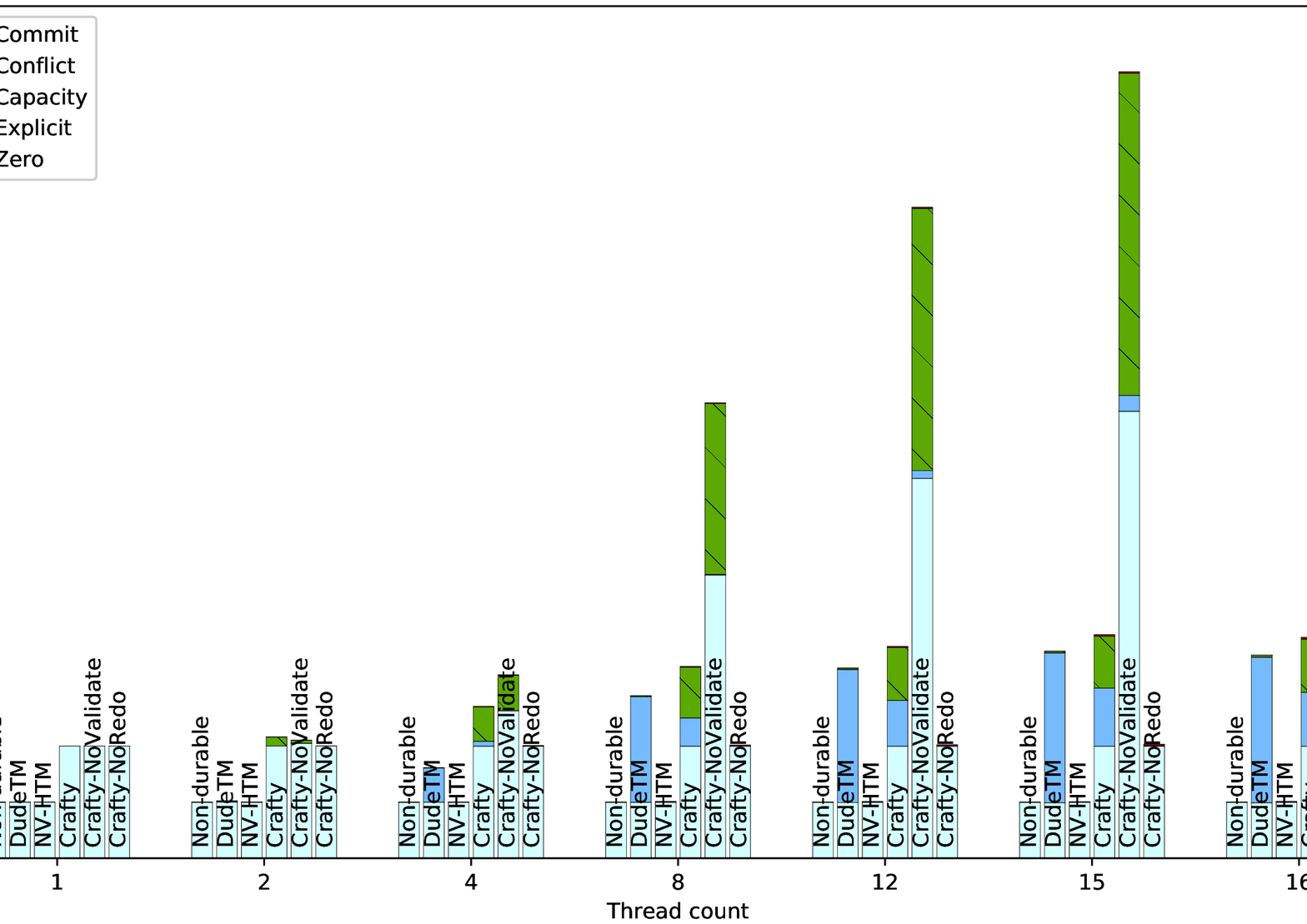}
    }
    \caption{Persistent and hardware transaction breakdowns for the \bankfee microbenchmark (no contention).}
    \label{fig:stats:bank-fee-nc}
\end{figure*}

\begin{figure*}
    \centering
    \subfloat[Persistent transaction breakdowns.]{
        \includegraphics[width=\linewidth]{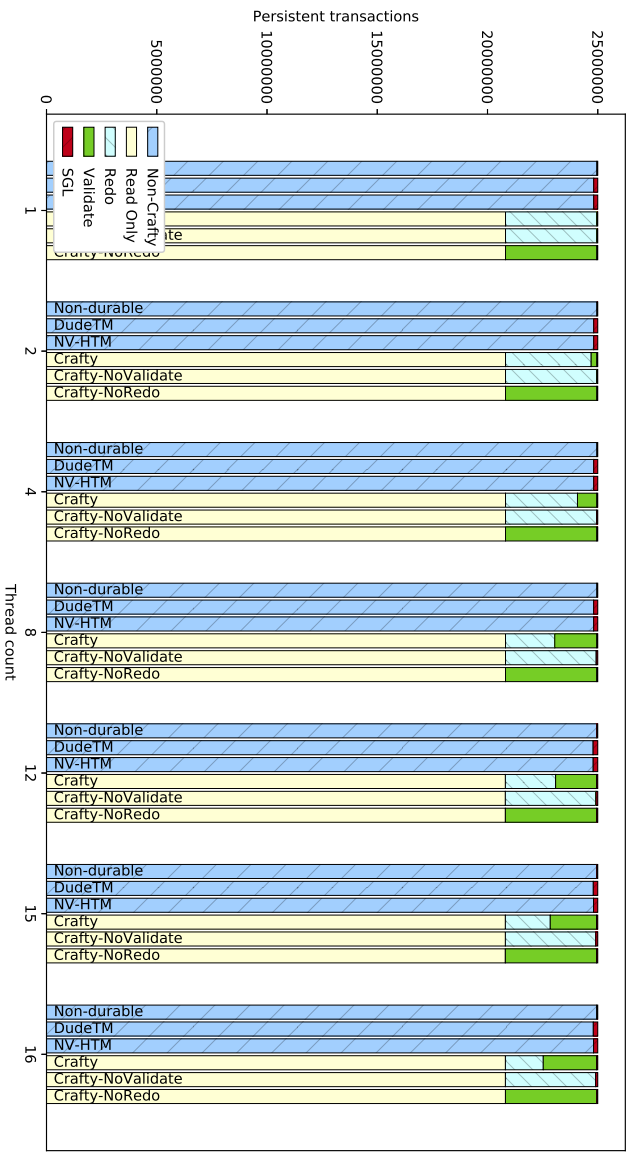}
    }

    \subfloat[Hardware transaction breakdowns.]{
        \includegraphics[width=\linewidth]{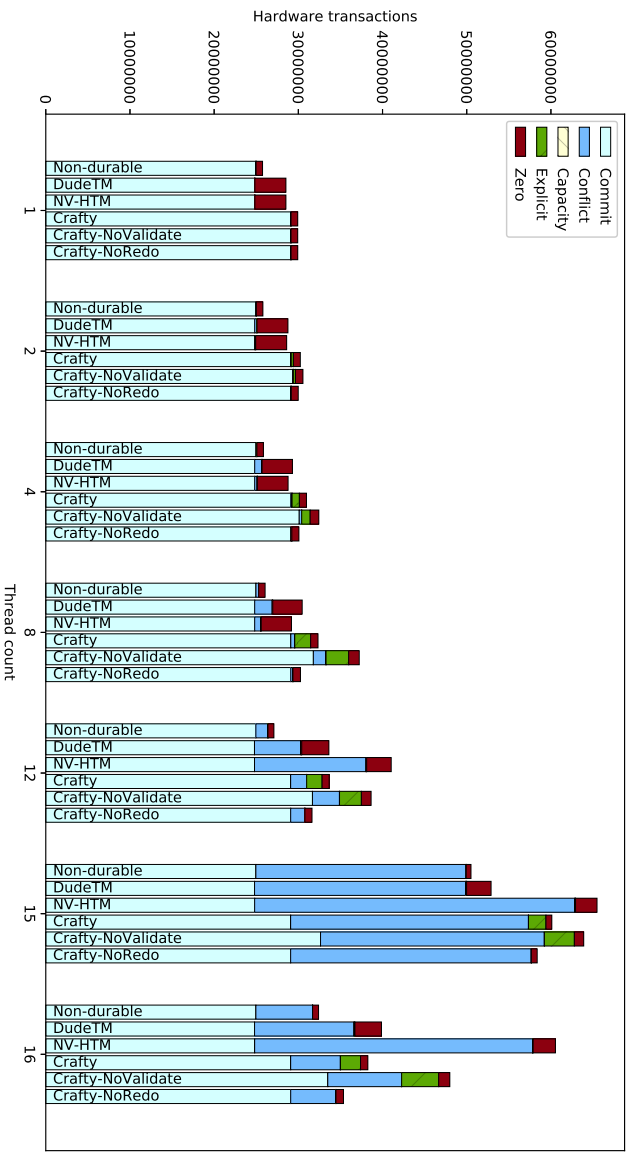}
    }
    \caption{Persistent and hardware transaction breakdowns for the B+ tree microbenchmark with insert operations only.}
\end{figure*}

\begin{figure*}
    \centering
    \subfloat[Persistent transaction breakdowns.]{
        \includegraphics[width=\linewidth]{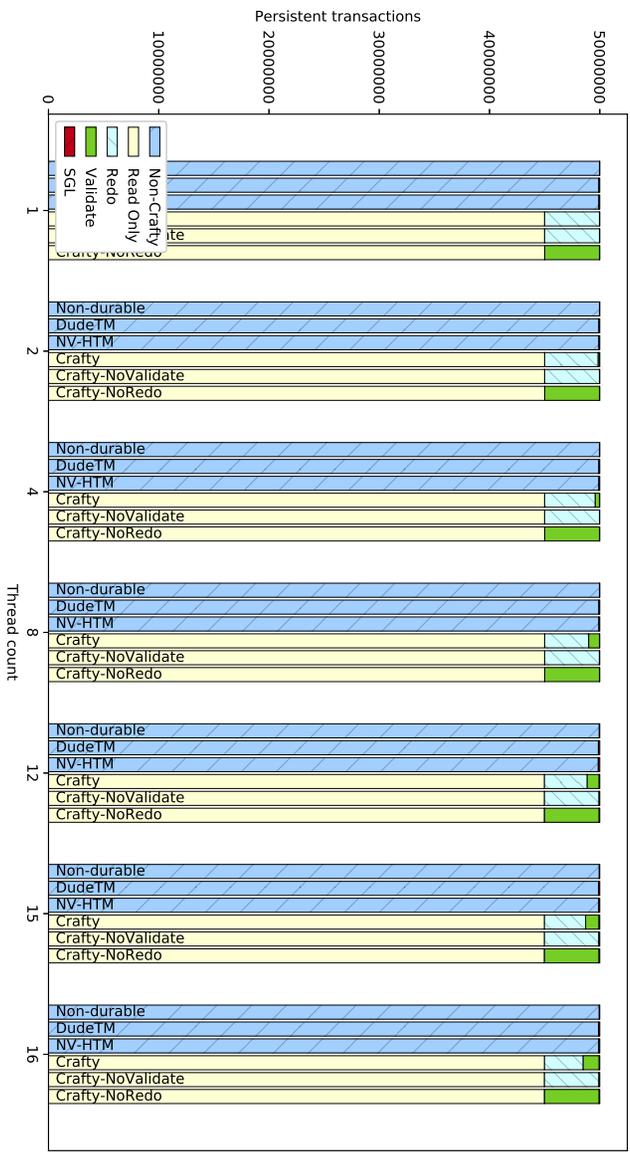}
    }

    \subfloat[Hardware transaction breakdowns.]{
        \includegraphics[width=\linewidth]{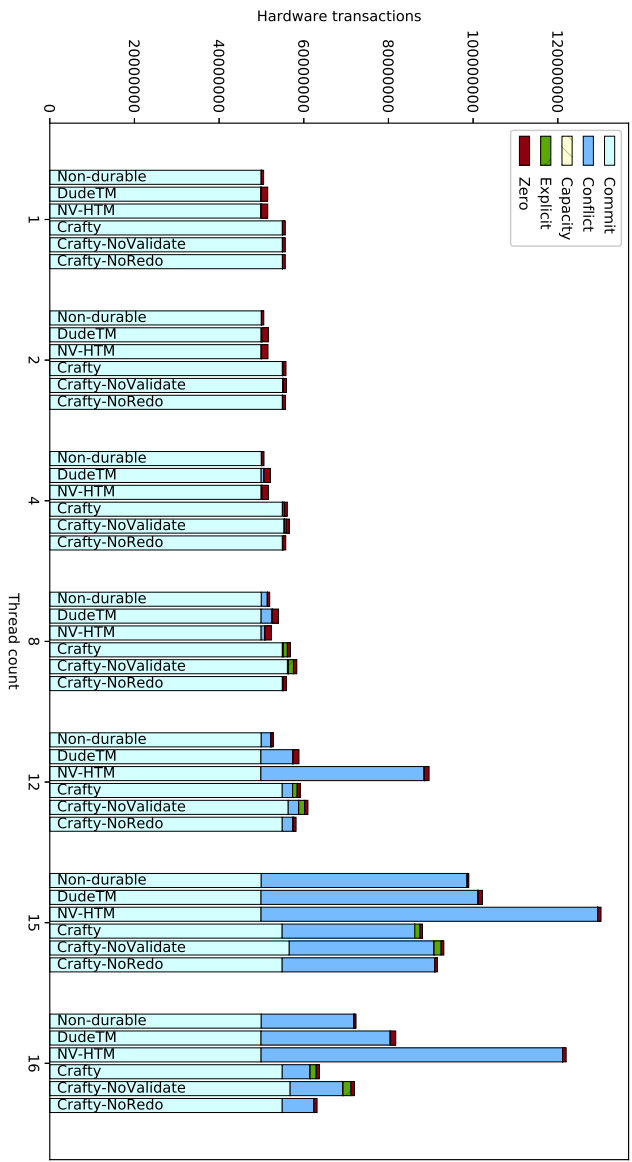}
    }
    \caption{Persistent and hardware transaction breakdowns for the B+ tree microbenchmark with mixed operations.}
\end{figure*}

\notes{\begin{figure*}
    \centering
    \subfloat[Persistent transaction breakdowns.]{
        \includegraphics[width=\linewidth]{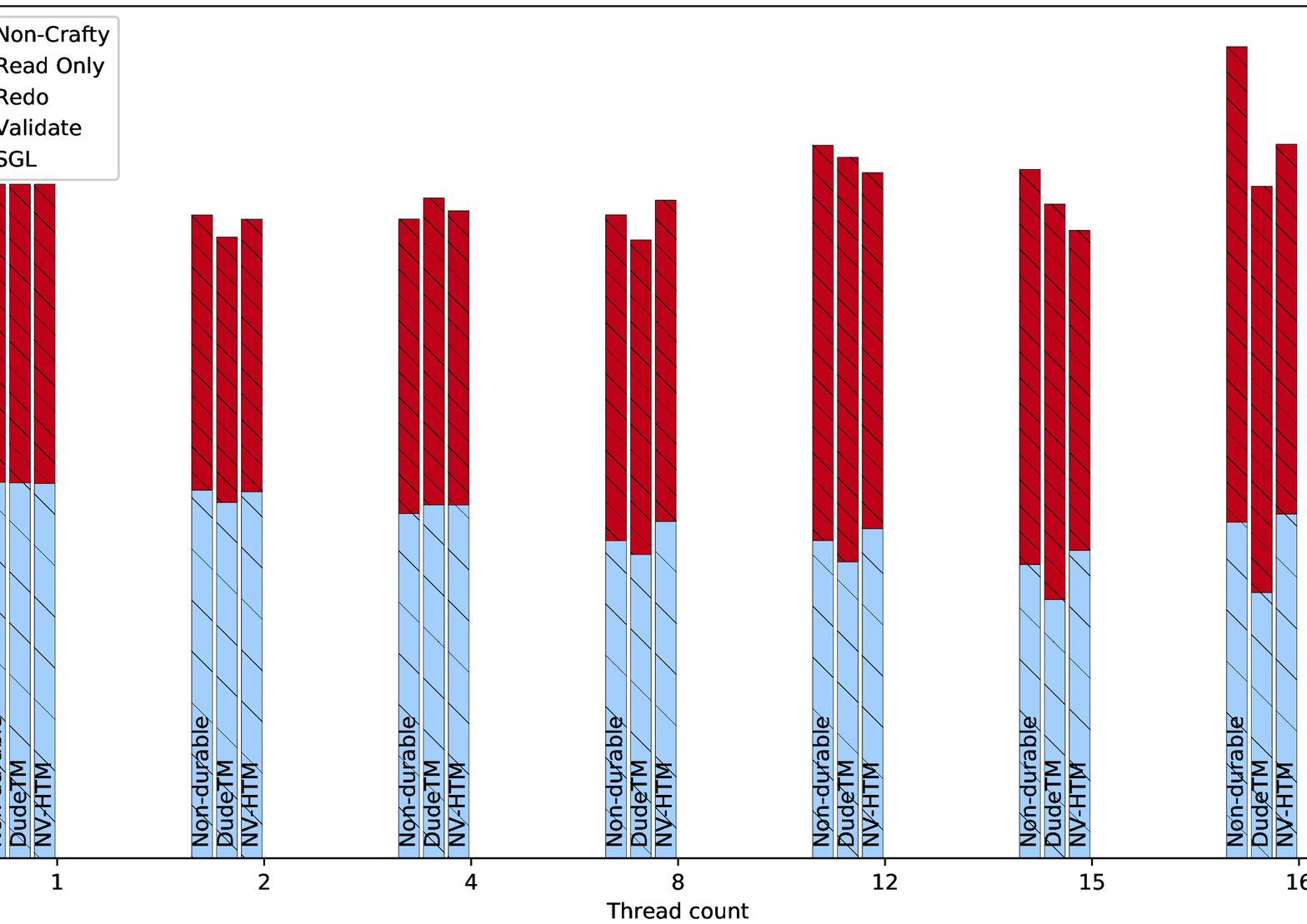}
        \label{fig:stats-ptx:bayes}
    }

    \subfloat[Hardware transaction breakdowns.]{
        \includegraphics[width=\linewidth]{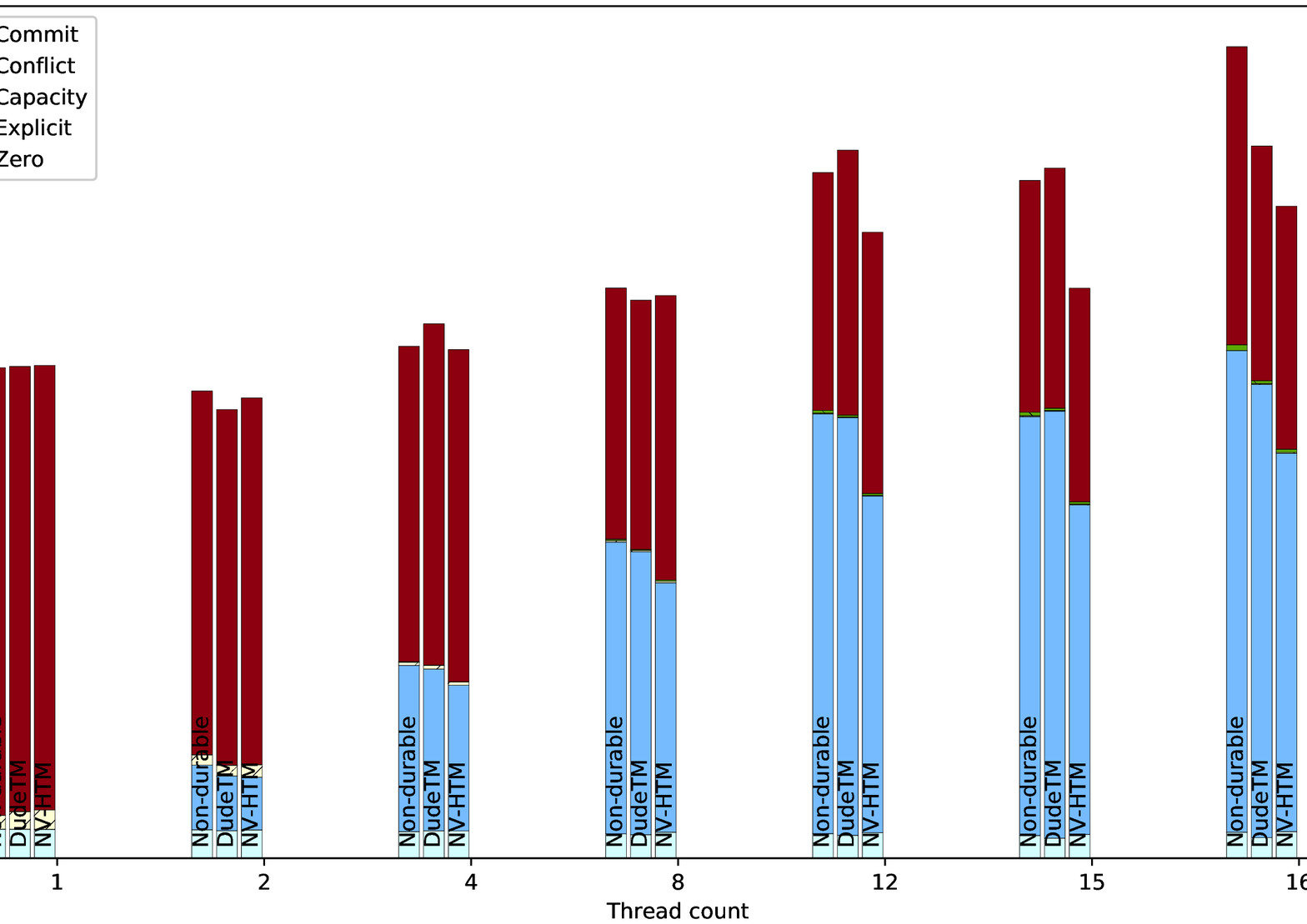}
    }

    \caption{Persistent and hardware transaction breakdowns for \bench{bayes}.}
\end{figure*}}

\begin{figure*}
    \centering
    \subfloat[Persistent transaction breakdowns.]{
        \includegraphics[width=\linewidth]{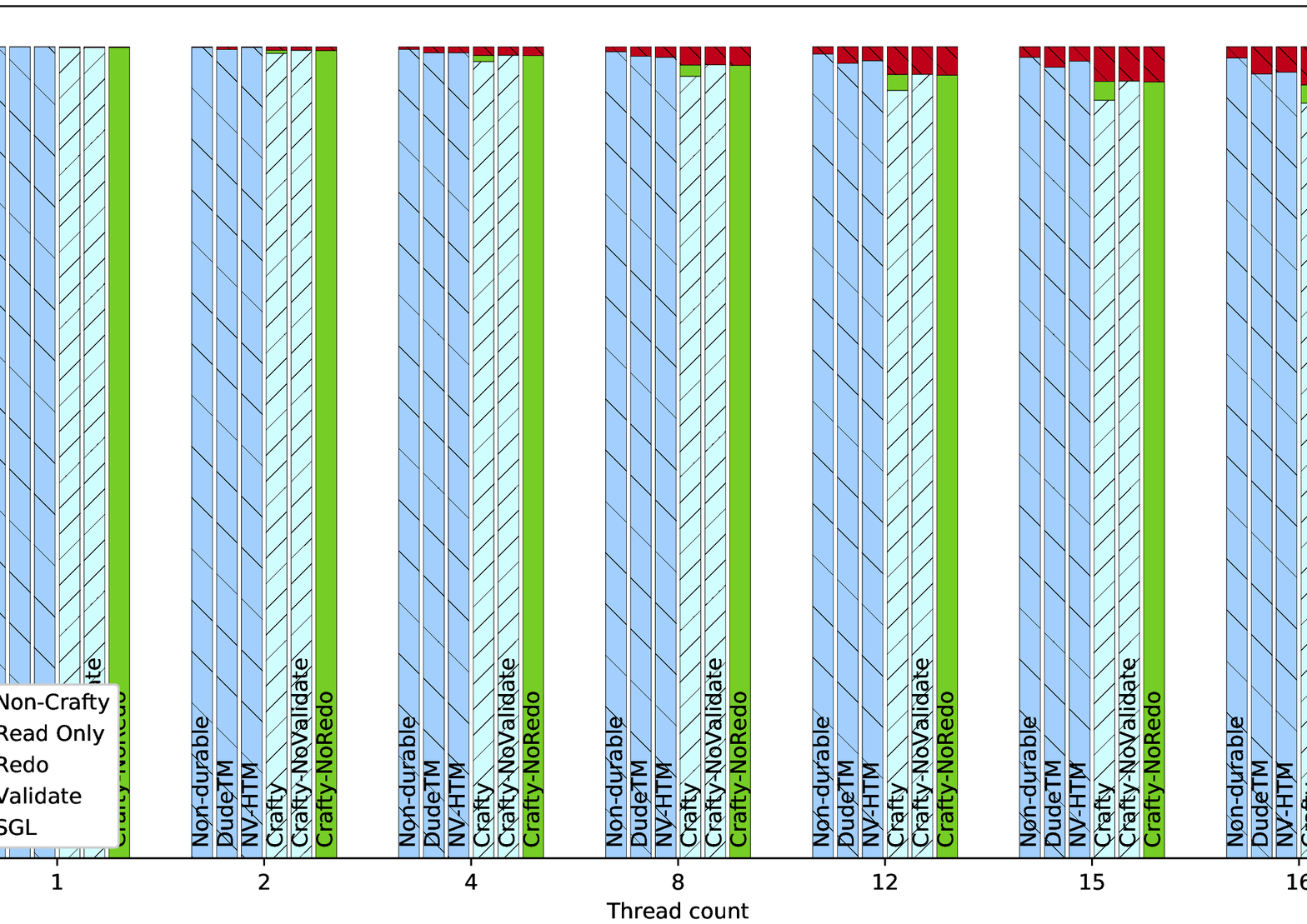}
    }

    \subfloat[Hardware transaction breakdowns.]{
        \includegraphics[width=\linewidth]{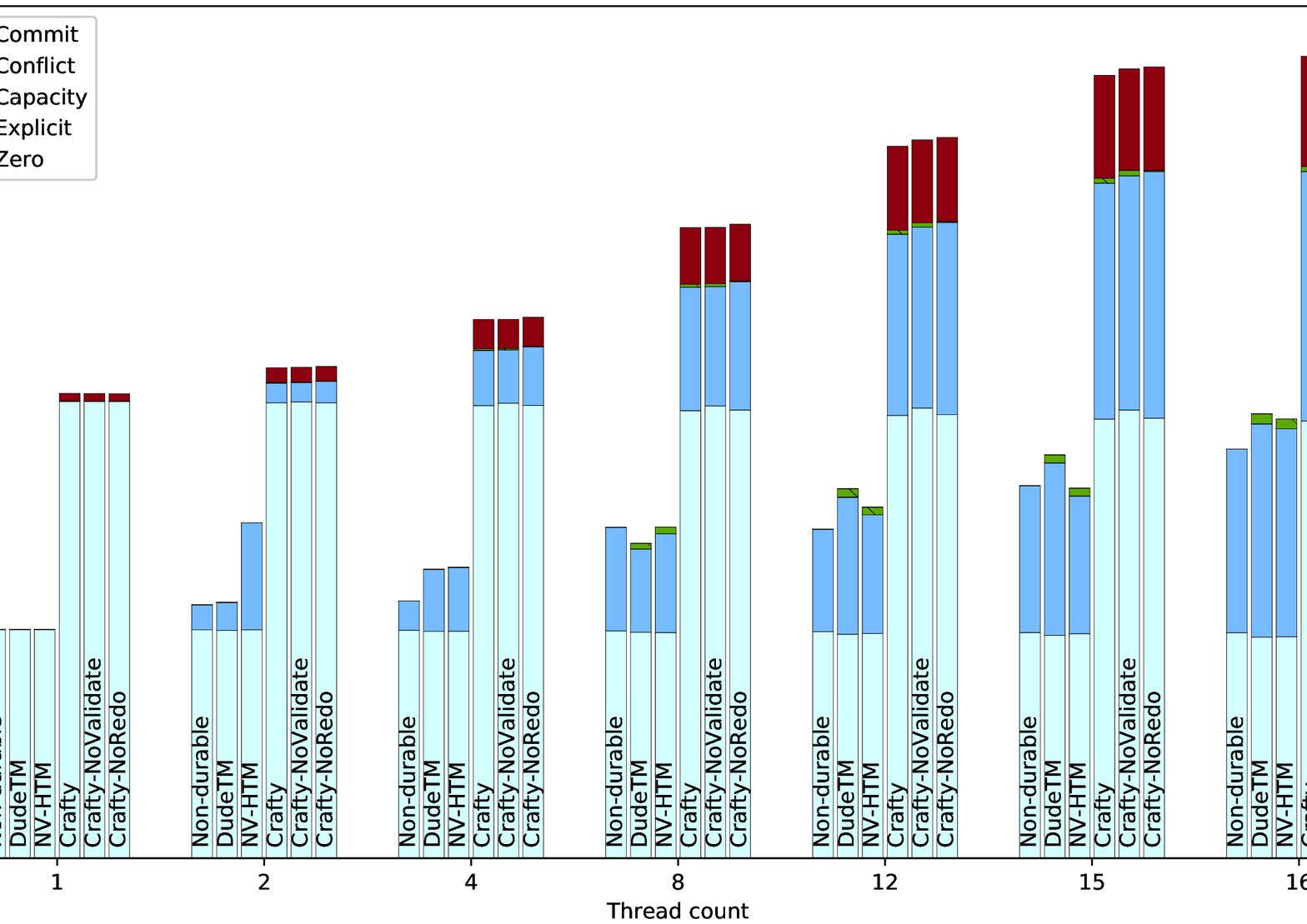}
    }
    \caption{Persistent and hardware transaction breakdowns for \bench{kmeans} (high contention).}
\end{figure*}

\begin{figure*}
    \centering
    \subfloat[Persistent transaction breakdowns.]{
        \includegraphics[width=\linewidth]{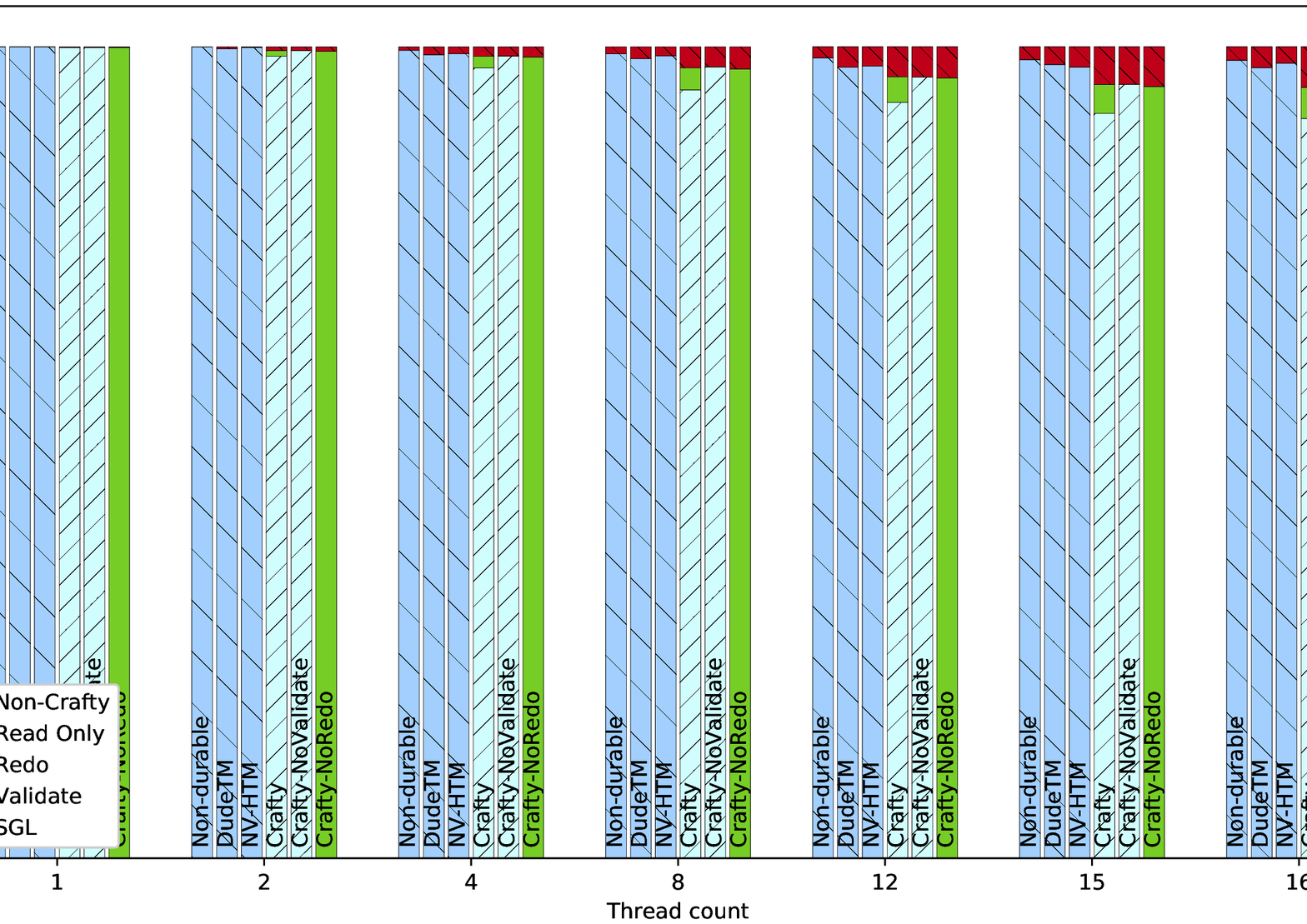}
    }

    \subfloat[Hardware transaction breakdowns.]{
        \includegraphics[width=\linewidth]{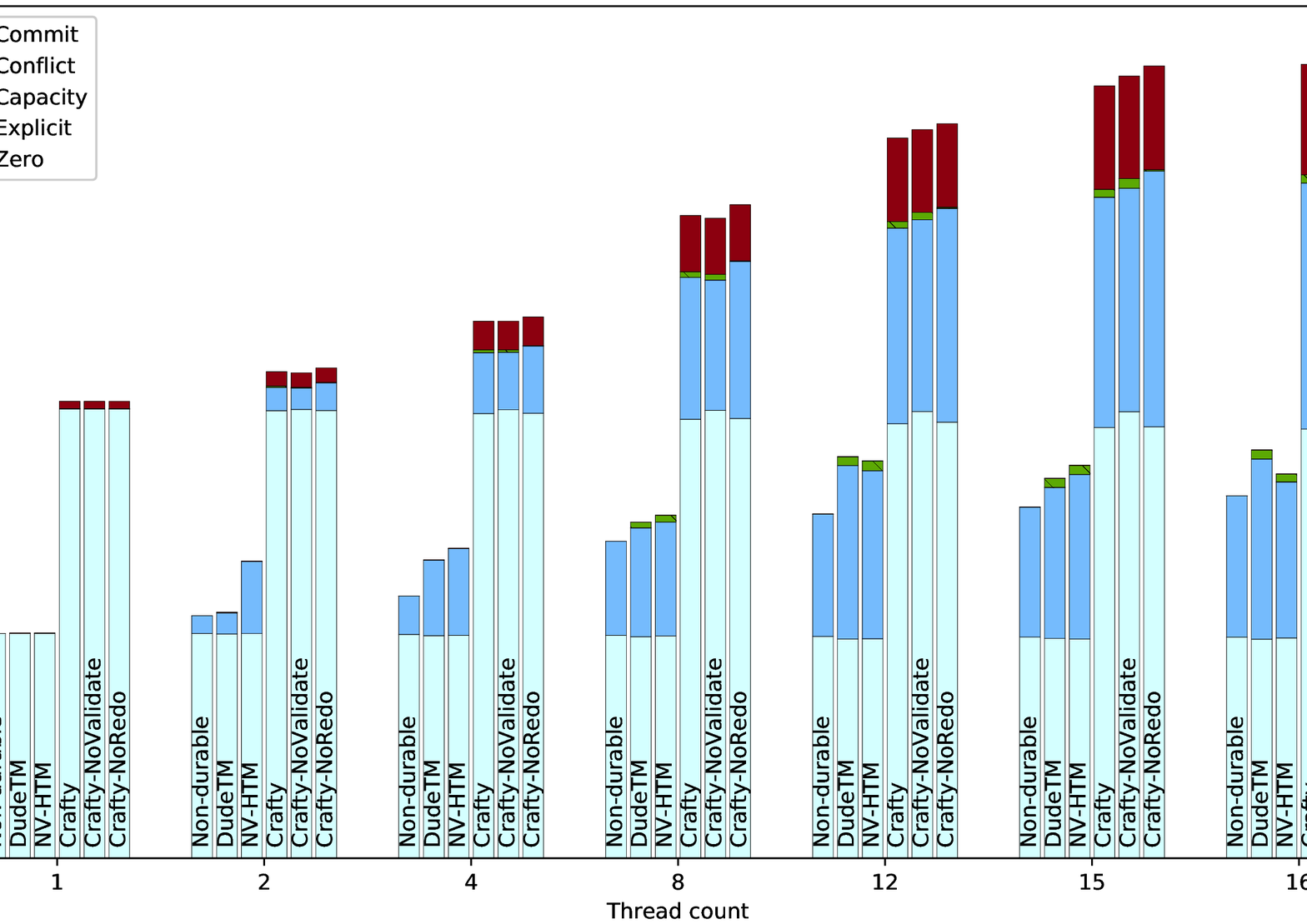}
    }
    \caption{Persistent and hardware transaction breakdowns for \bench{kmeans} (low contention).}
\end{figure*}

\begin{figure*}
    \centering
    \subfloat[Persistent transaction breakdowns.]{
        \includegraphics[width=\linewidth]{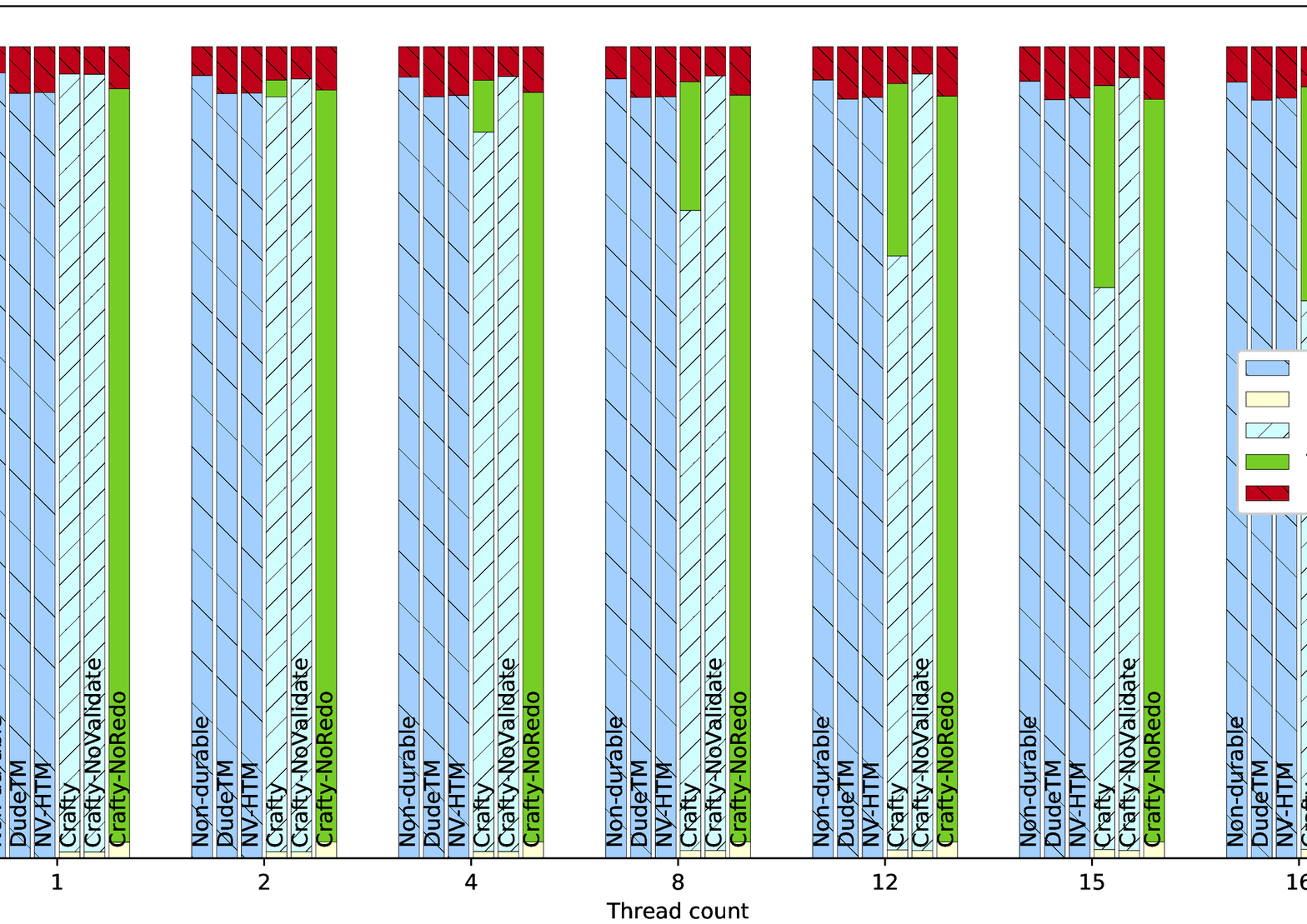}
    }

    \subfloat[Hardware transaction breakdowns.]{
        \includegraphics[width=\linewidth]{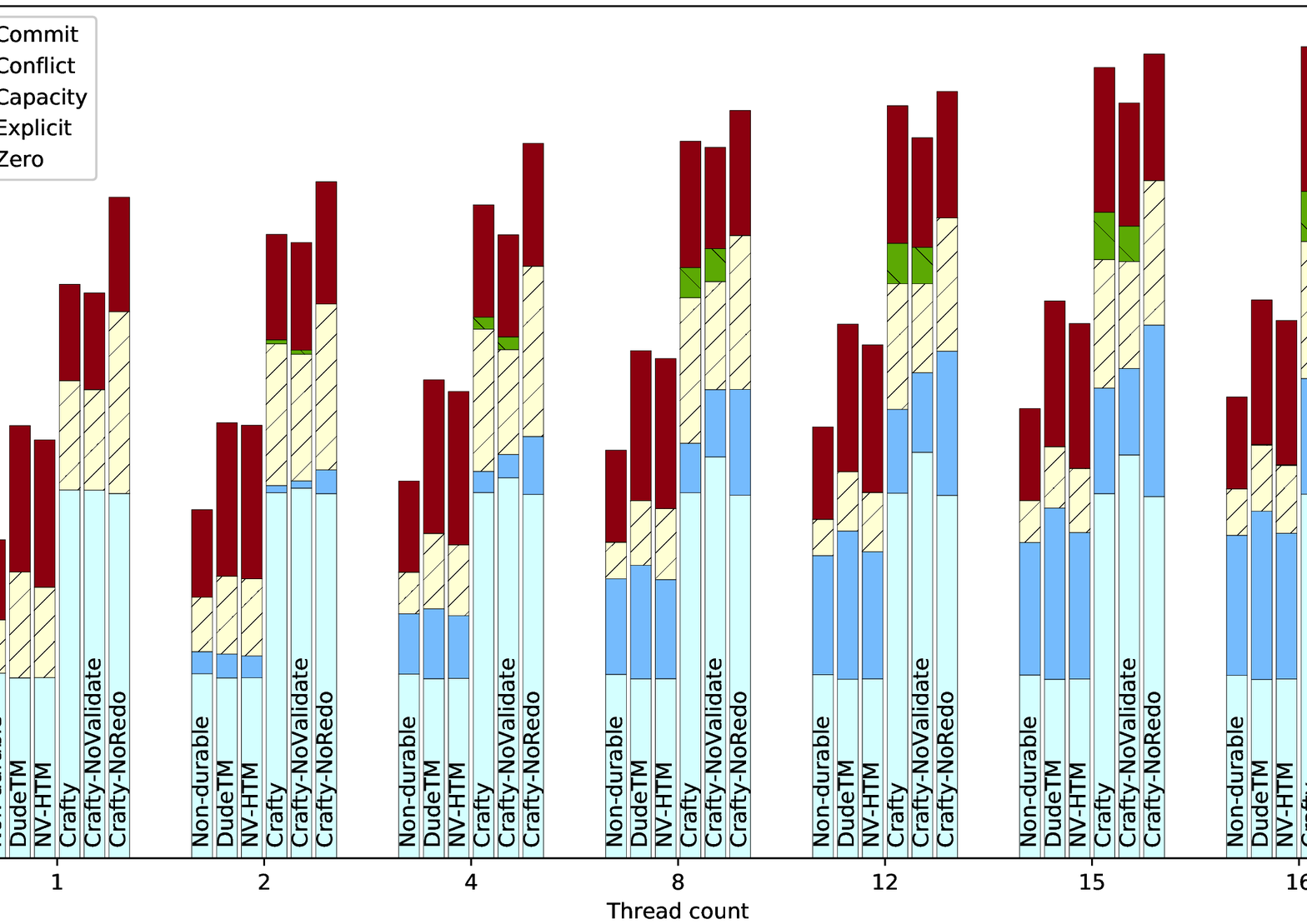}
    }
    \caption{Persistent and hardware transaction breakdowns for \bench{vacation} (high contention).}
\end{figure*}

\begin{figure*}
    \centering
    \subfloat[Persistent transaction breakdowns.]{
        \includegraphics[width=\linewidth]{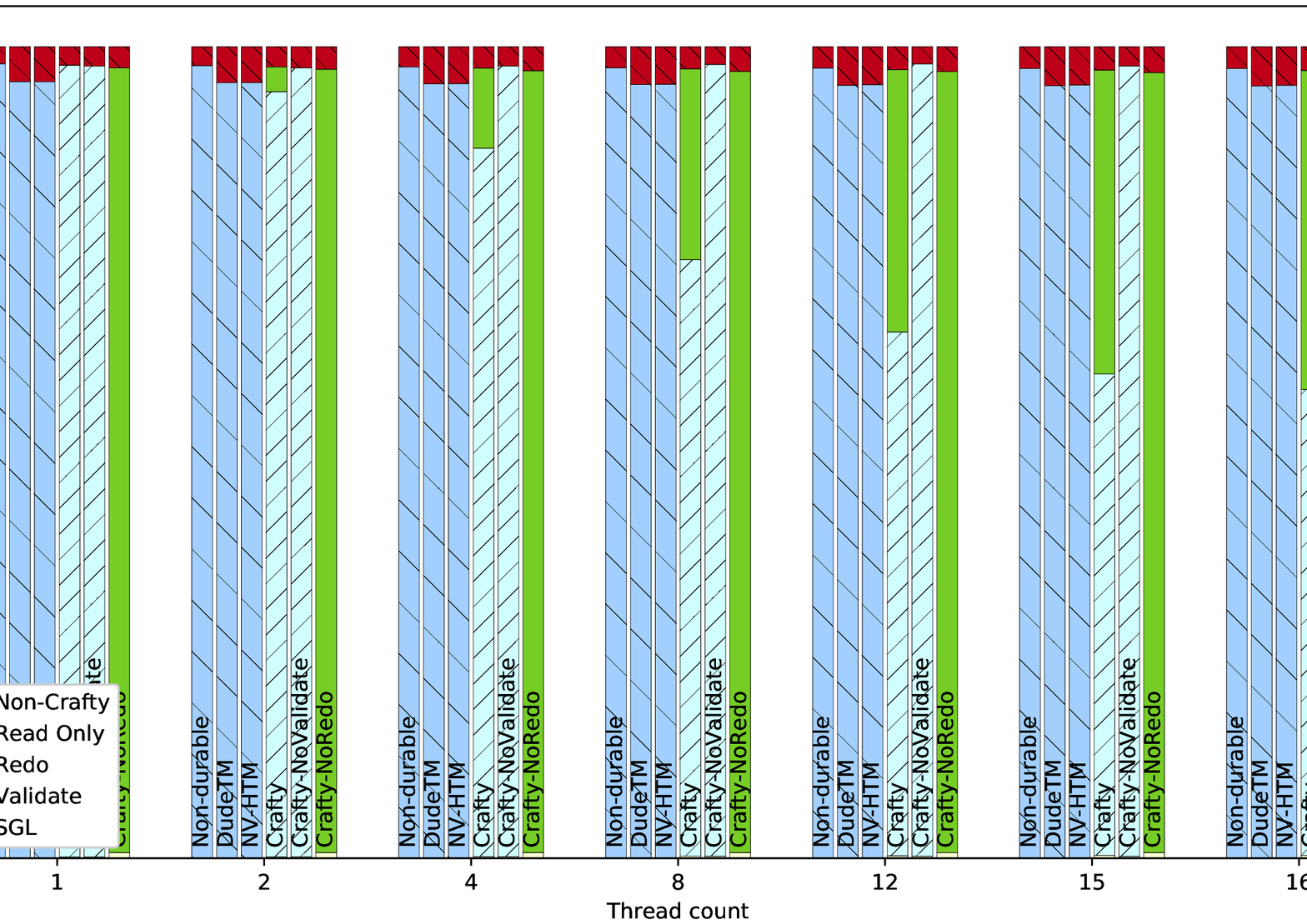}
    }

    \subfloat[Hardware transaction breakdowns.]{
        \includegraphics[width=\linewidth]{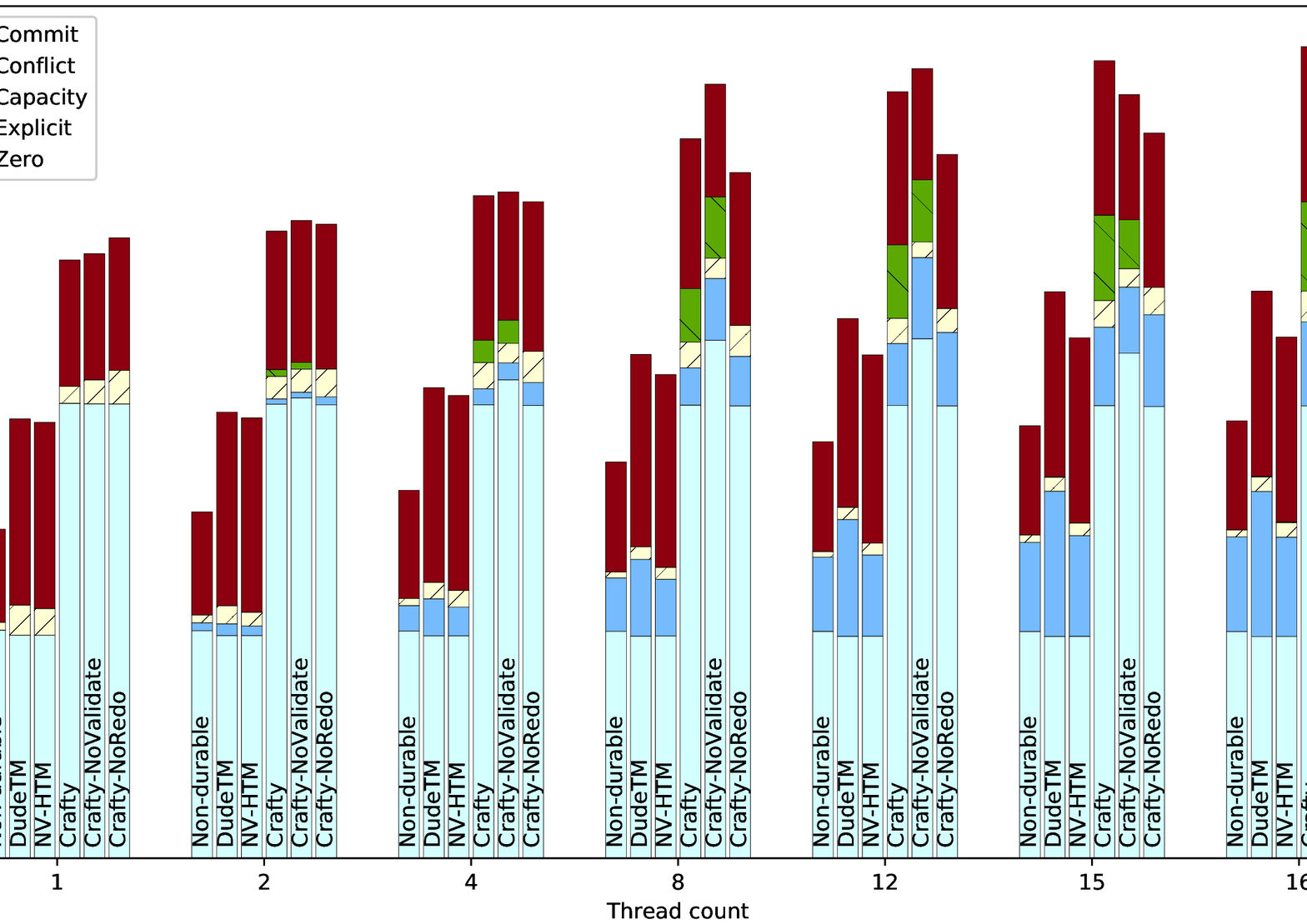}
    }
    \caption{Persistent and hardware transaction breakdowns for \bench{vacation} (low contention).}
\end{figure*}

\begin{figure*}
    \centering
    \subfloat[Persistent transaction breakdowns.]{
        \includegraphics[width=\linewidth]{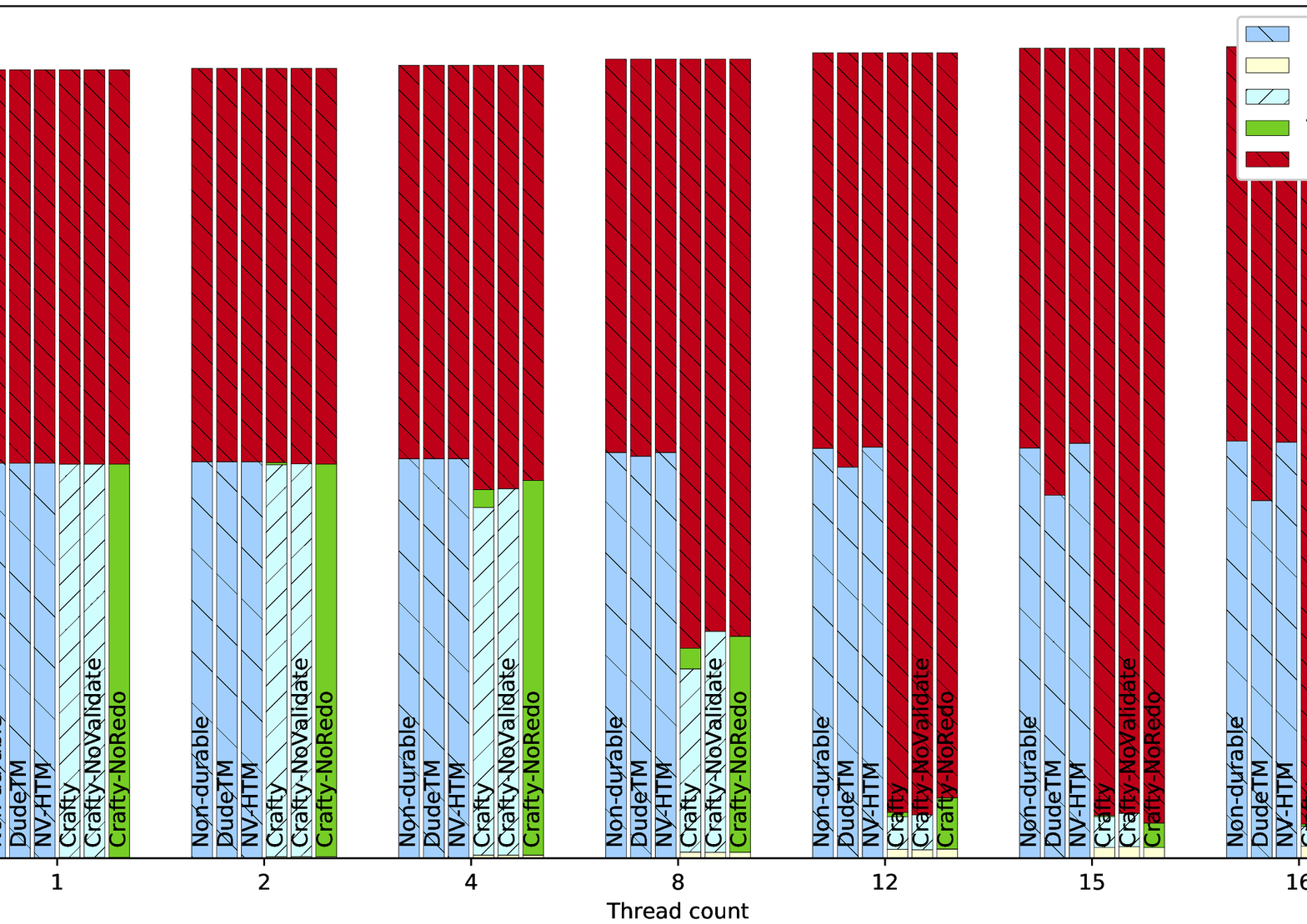}
    }

    \subfloat[Hardware transaction breakdowns.]{
        \includegraphics[width=\linewidth]{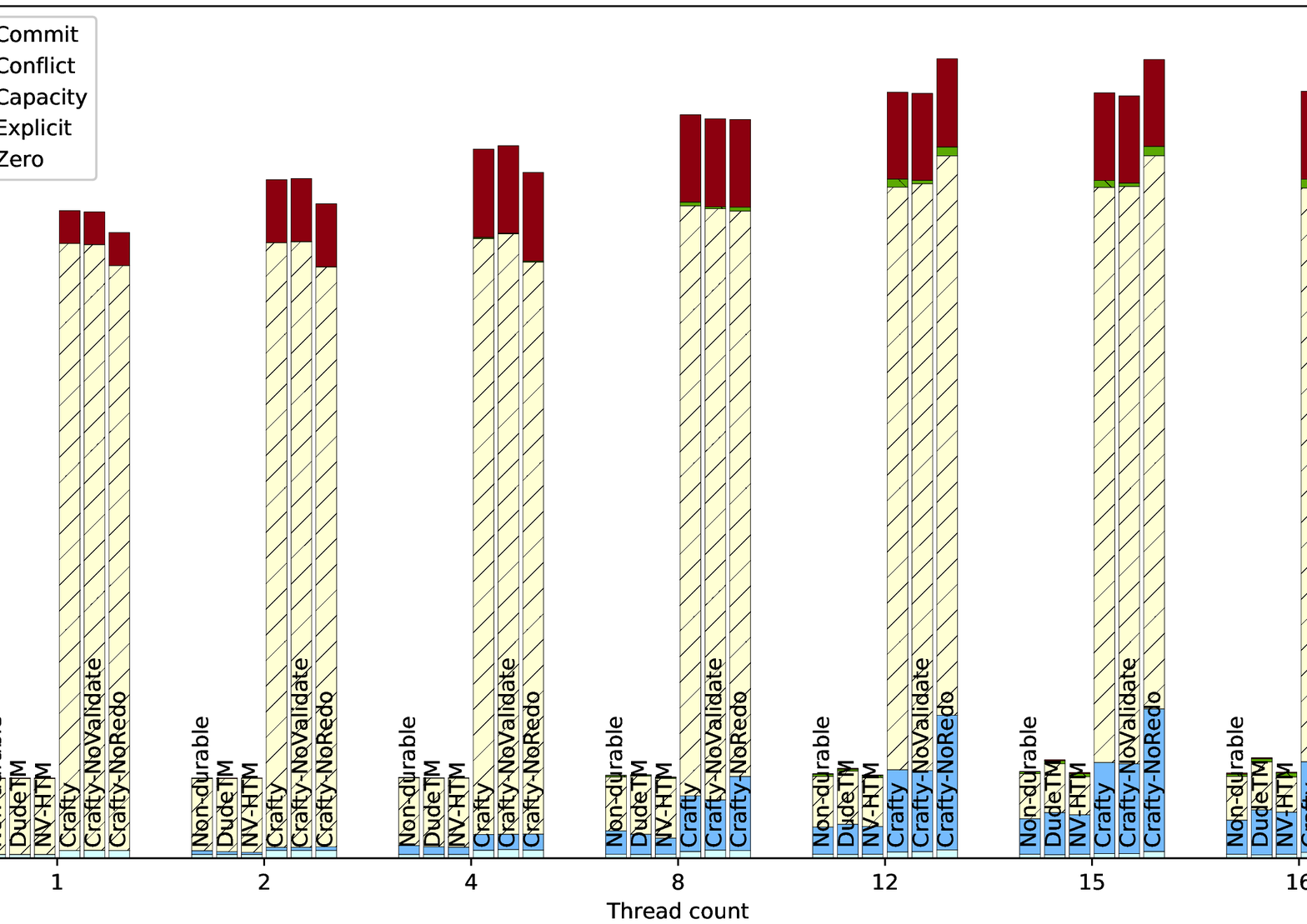}
    }
    \caption{Persistent and hardware transaction breakdowns for \bench{labyrinth}.}
\end{figure*}

\begin{figure*}
    \centering
    \subfloat[Persistent transaction breakdowns.]{
        \includegraphics[width=\linewidth]{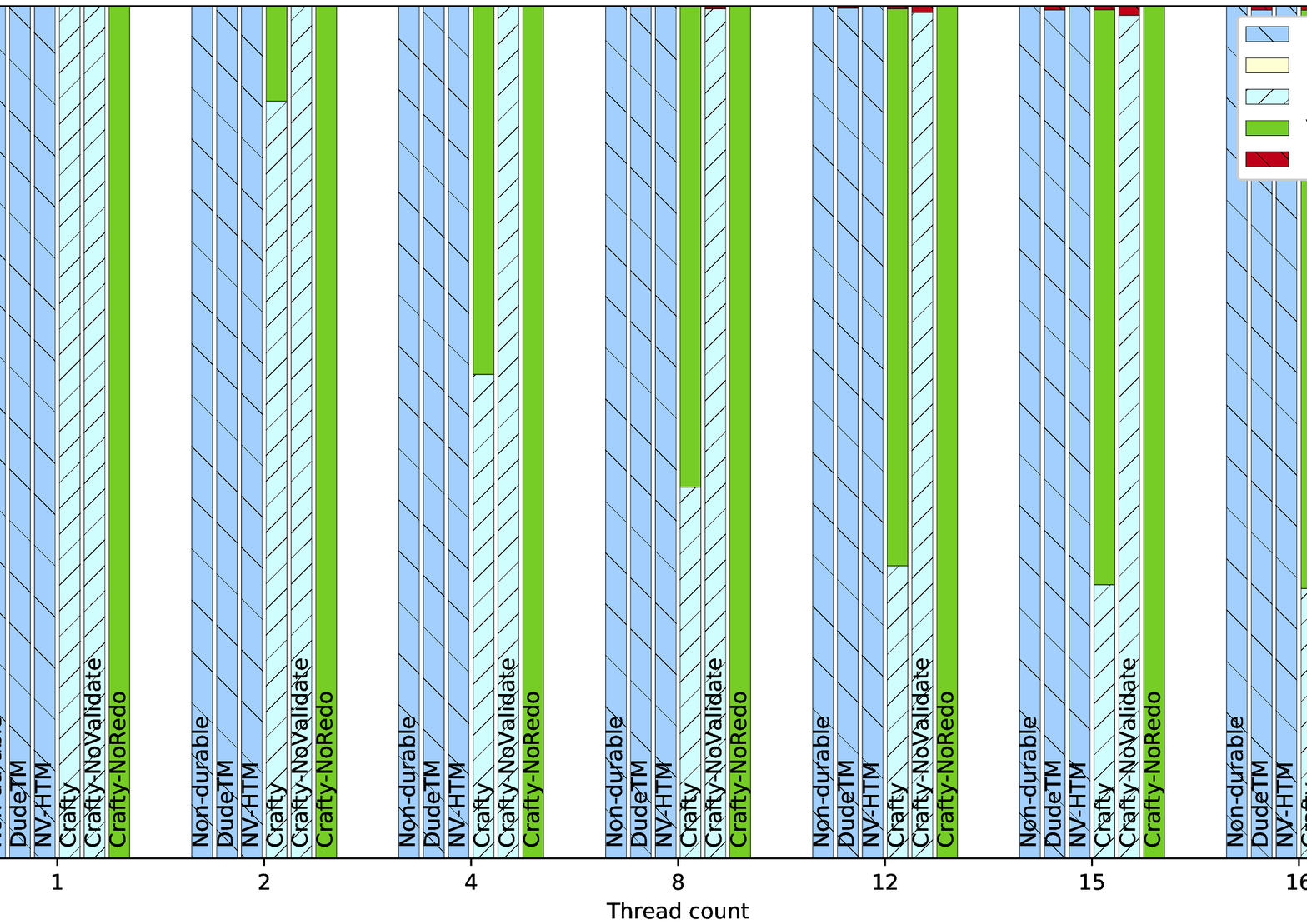}
    }

    \subfloat[Hardware transaction breakdowns.]{
        \includegraphics[width=\linewidth]{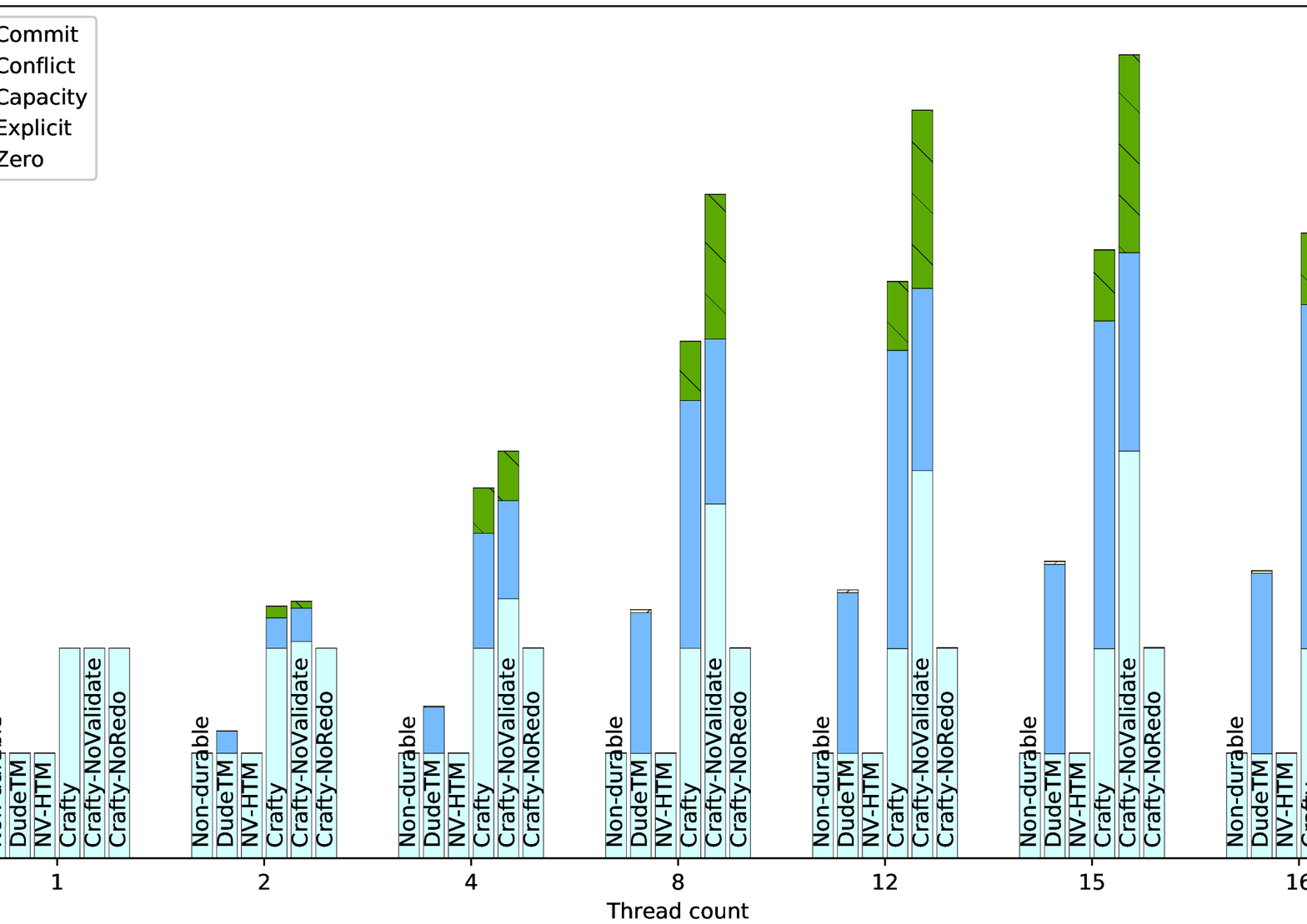}
    }
    \caption{Persistent and hardware transaction breakdowns for \bench{ssca2}.}
    \label{fig:stats:ssca2}
\end{figure*}

\begin{figure*}
    \centering
    \subfloat[Persistent transaction breakdowns.]{
        \includegraphics[width=\linewidth]{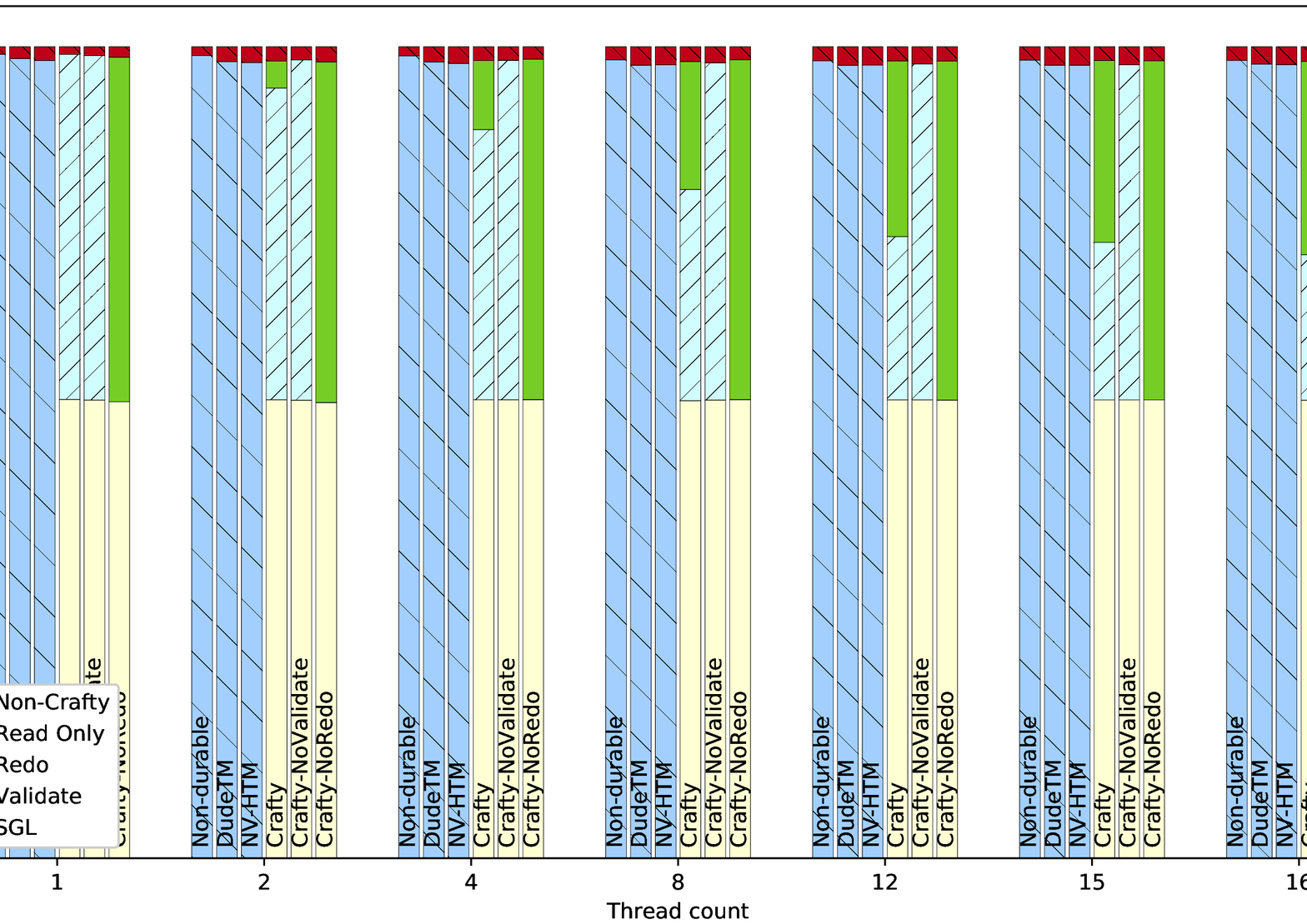}
    }

    \subfloat[Hardware transaction breakdowns.]{
        \includegraphics[width=\linewidth]{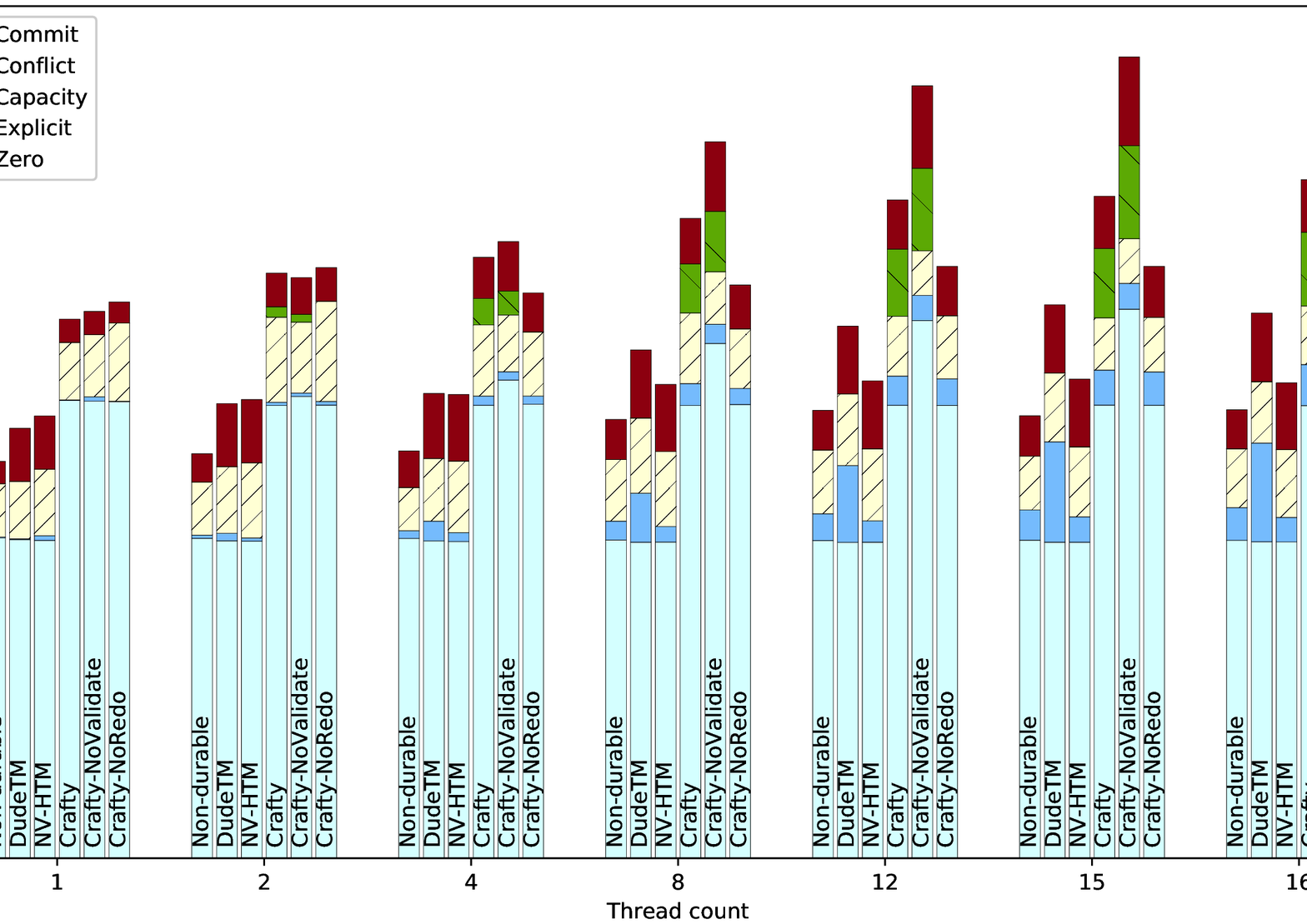}
    }
    \caption{Persistent and hardware transaction breakdowns for \bench{genome}.}
\end{figure*}

\begin{figure*}
    \centering
    \subfloat[Persistent transaction breakdowns.]{
        \includegraphics[width=\linewidth]{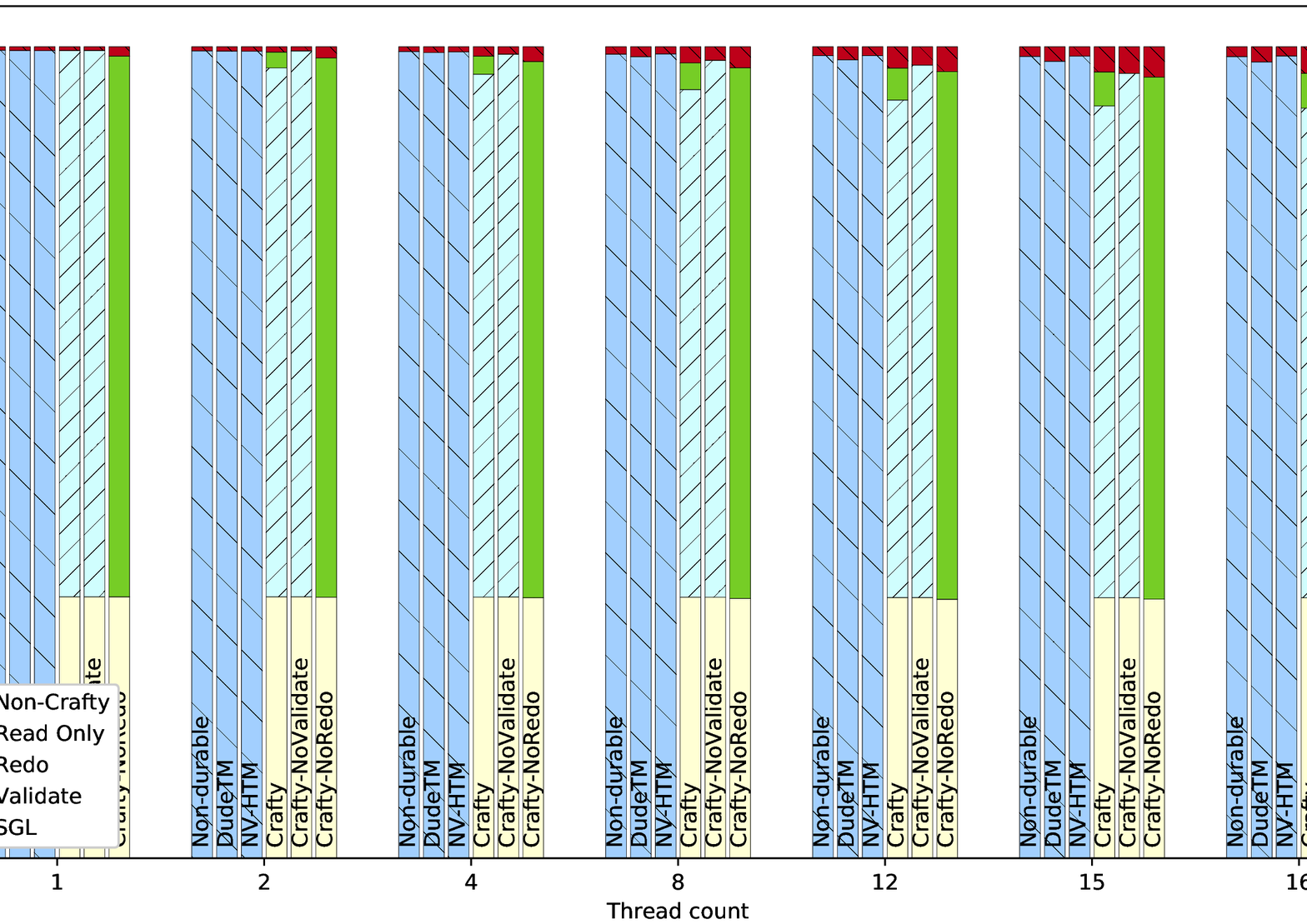}
    }

    \subfloat[Hardware transaction breakdowns.]{
        \includegraphics[width=\linewidth]{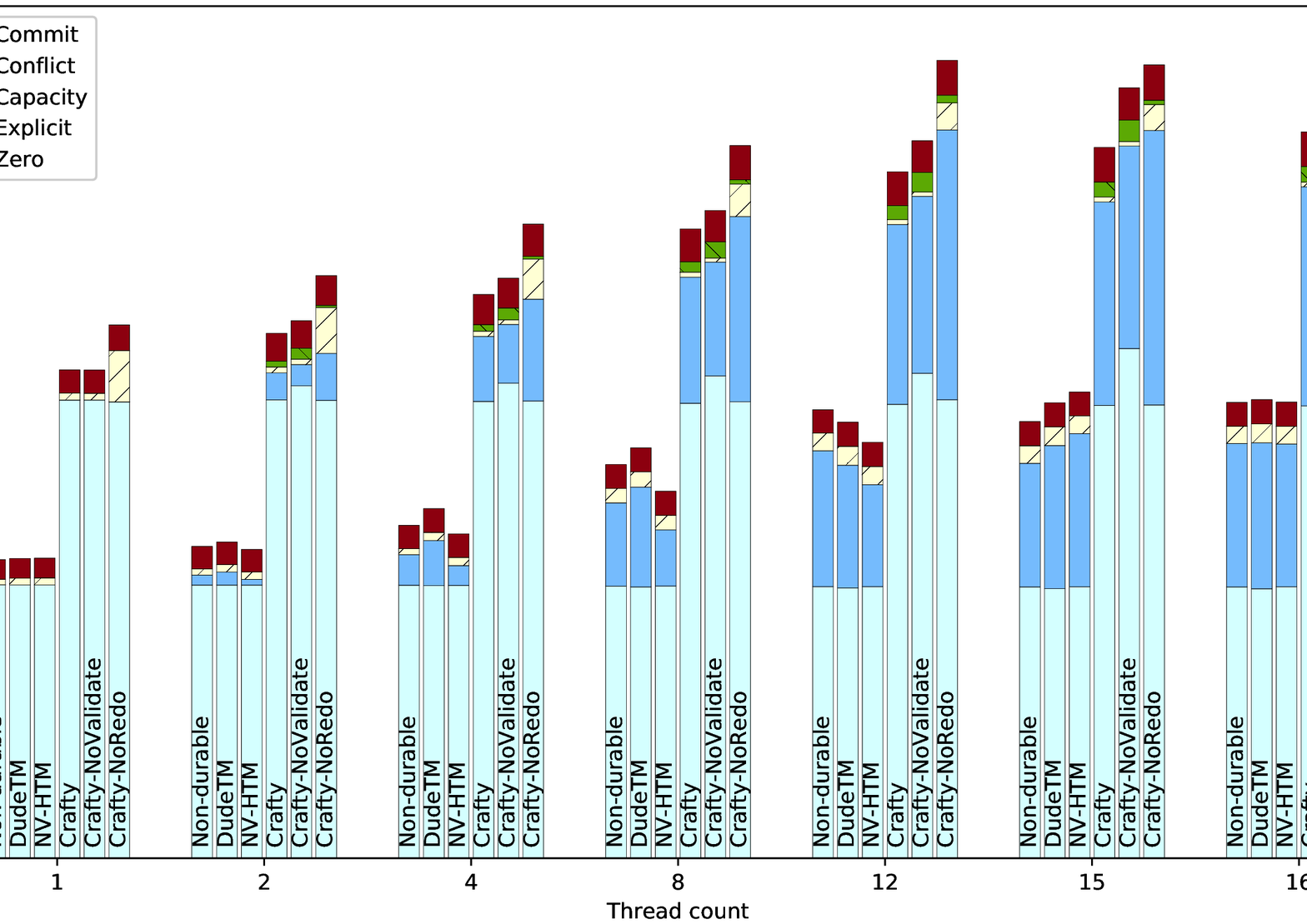}
    }

    \caption{Persistent and hardware transaction breakdowns for \bench{intruder}.}
    \label{fig:stats:intruder}
\end{figure*}

\begin{figure*}
    \vspace*{-1.5em}
    \centering
    \captionsetup[subfloat]{farskip=2pt,captionskip=1pt}
    \subfloat[High contention]{
        \includegraphics[width=.5\linewidth]{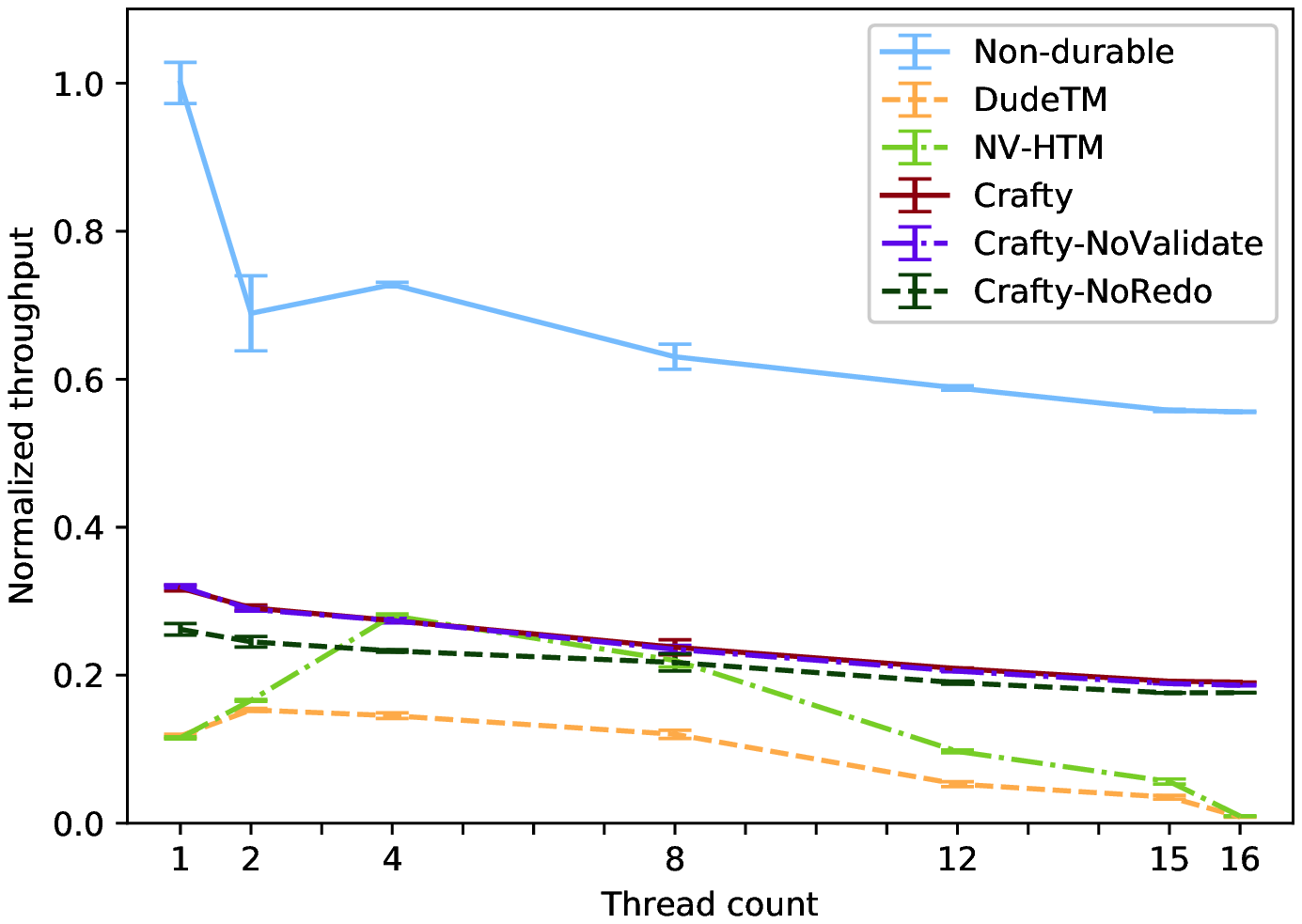}
        \label{fig:throughput-100ns:bank-fee-hc}
    }
    \\
    \subfloat[Medium contention]{
        \includegraphics[width=.5\linewidth]{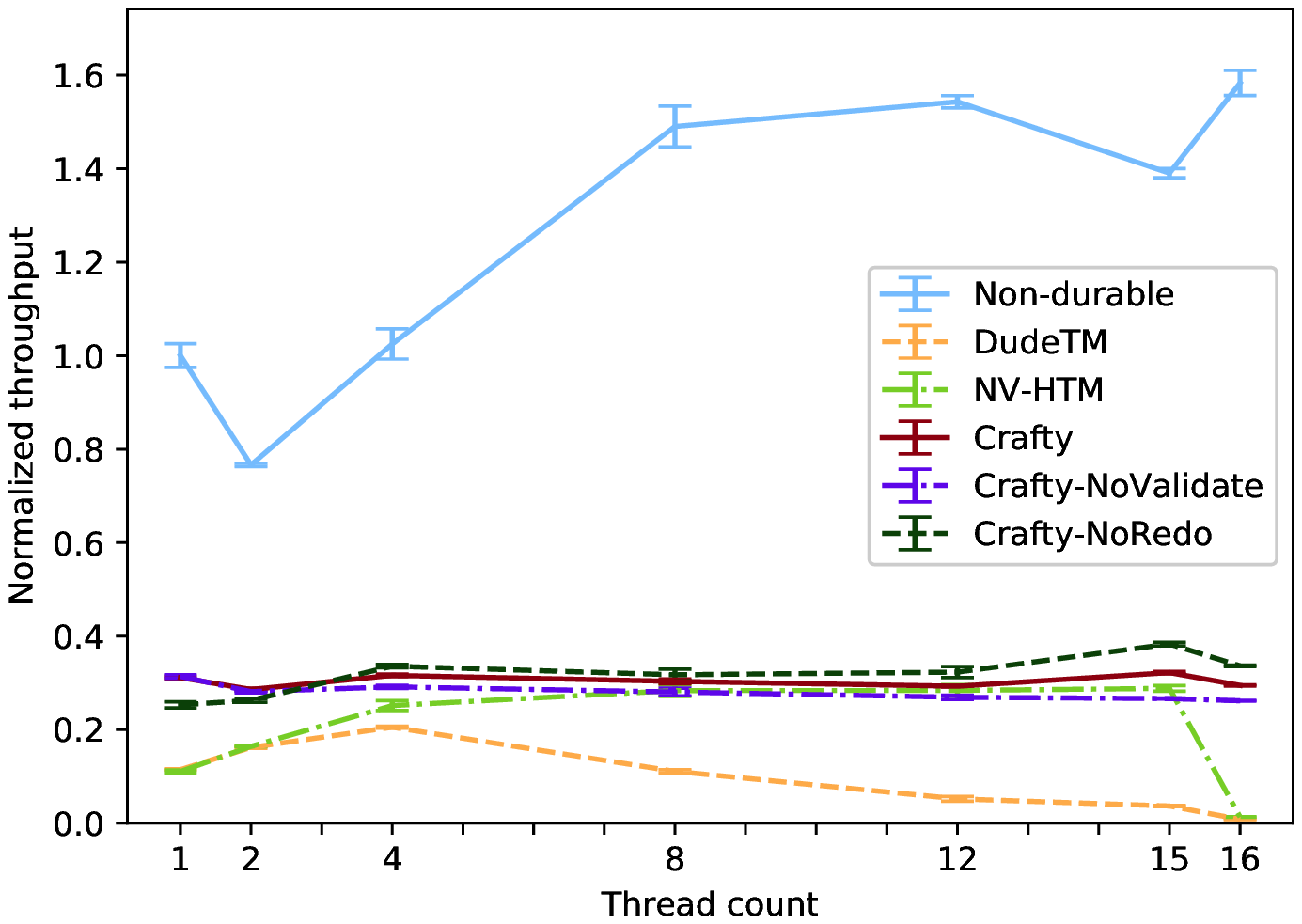}
        \label{fig:throughput-100ns:bank-fee}
    }
    \\
    \subfloat[No contention]{
        \includegraphics[width=.5\linewidth]{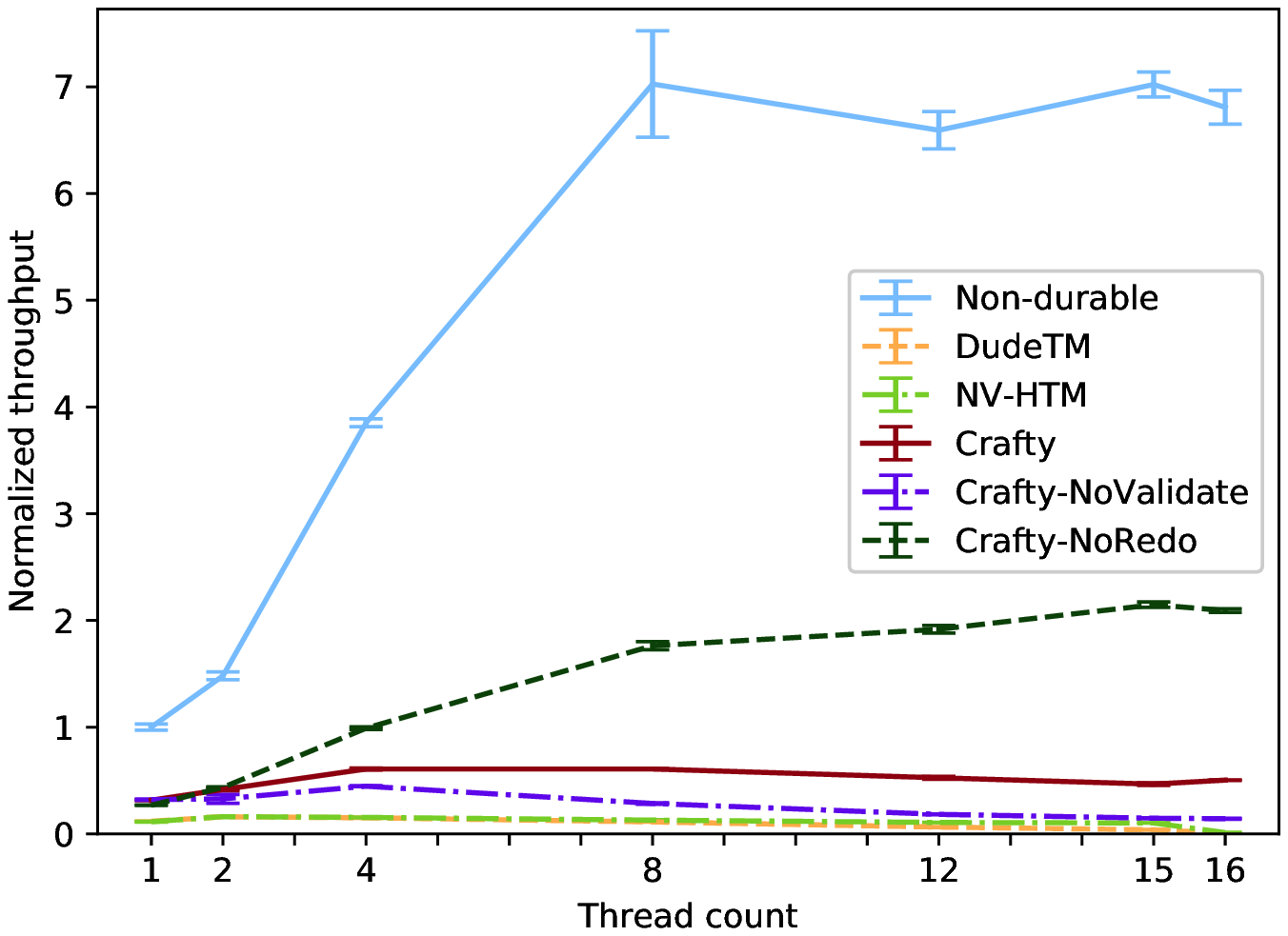}
        \label{fig:throughput-100ns:bank-fee-nc}
    }
    \caption{Throughput of Crafty and competing approaches,
    using the \bankfee microbenchmark at three contention levels,
    emulating an NVM latency of 100 ns (instead of 300 ns as in Figure~\ref{fig:throughput:bank}).}
    \label{fig:throughput-100ns:bank}
\end{figure*}

\begin{figure*}
    \centering
    \captionsetup[subfloat]{farskip=2pt,captionskip=1pt}
    \subfloat[Insert operations only]{
        \includegraphics[width=.5\linewidth]{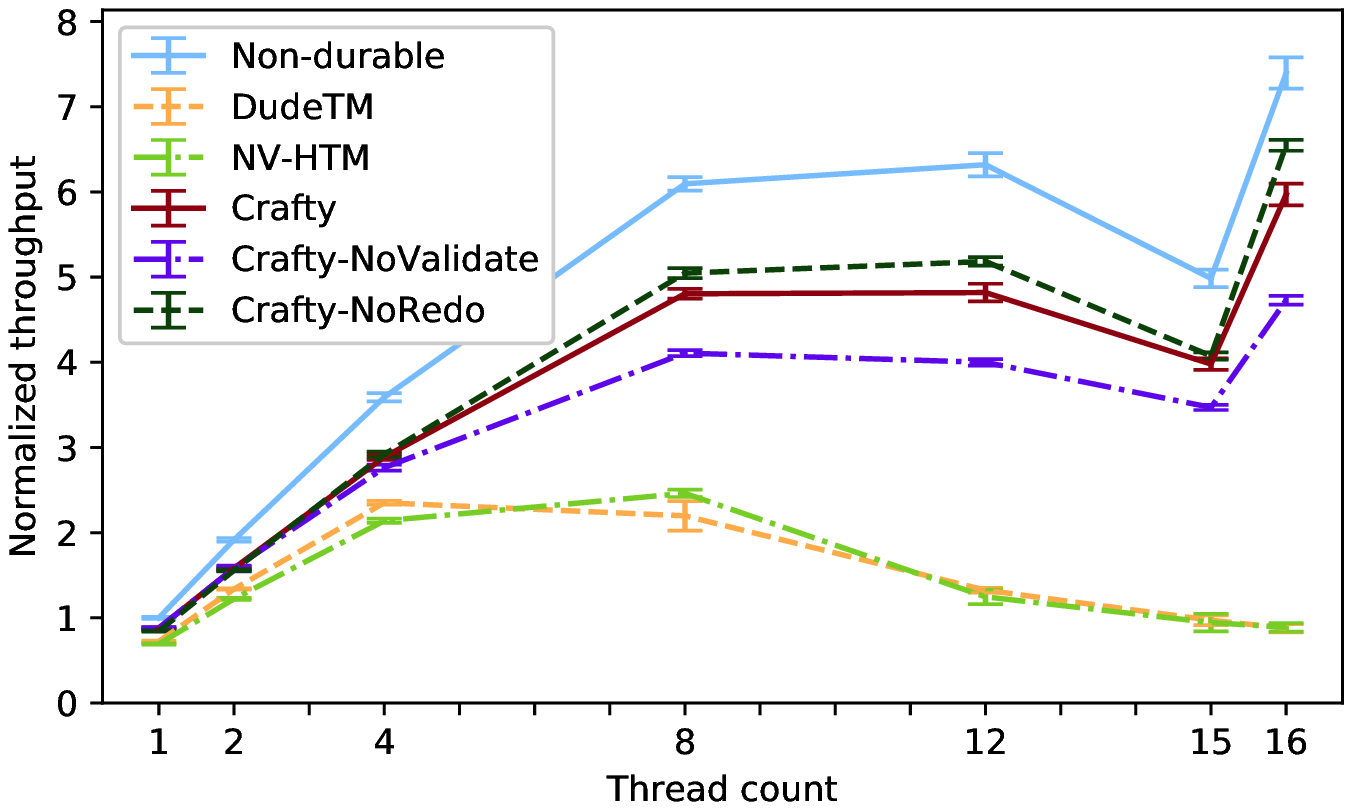}
    }
    \subfloat[Lookup, insert, and remove operations]{
        \includegraphics[width=.5\linewidth]{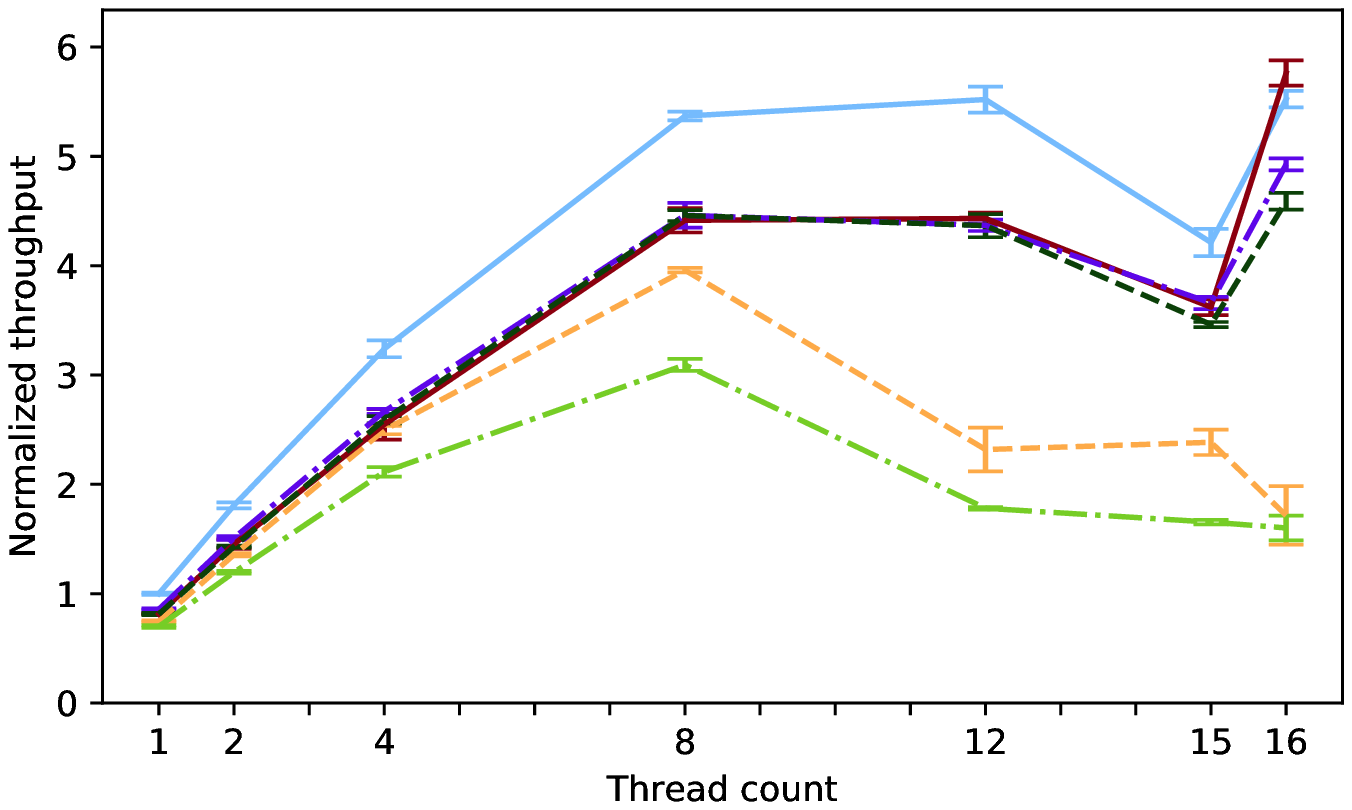}
    }
    \caption{Throughput of \crafty and competing approaches,
    on the B+ tree microbenchmark, for mixed operations and insert only,
    emulating an NVM latency of 100 ns (instead of 300 ns as in Figure~\ref{fig:throughput:bplustree}).}
\end{figure*}

\begin{figure*}
    \vspace*{-1.5em}
    \centering
    \captionsetup[subfloat]{farskip=2pt,captionskip=1pt}

    \subfloat[\bench{kmeans} (high contention)]{
        \includegraphics[width=.5\linewidth]{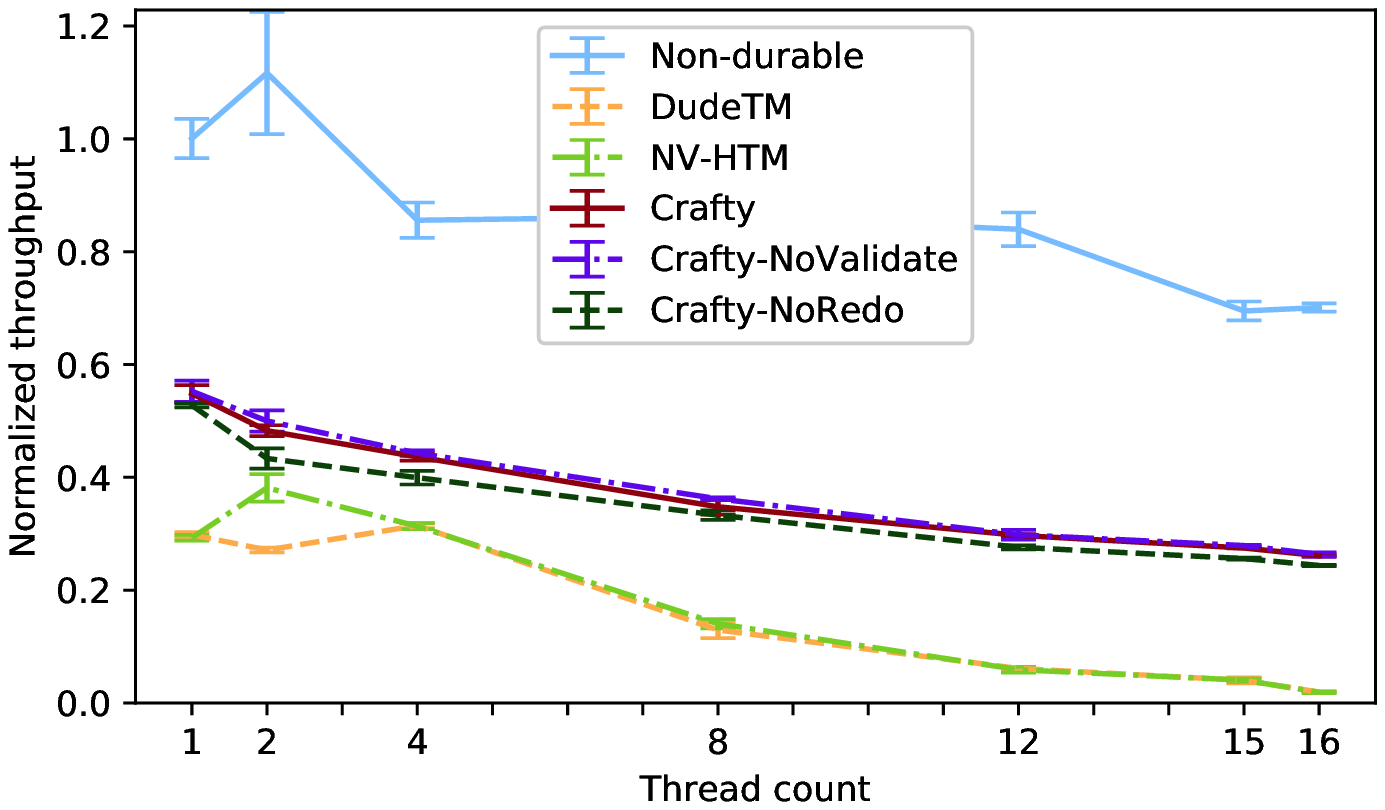}
        \label{fig:throughput-100ns:kmeans-high}
    }
    \subfloat[\bench{kmeans} (low contention)]{
        \includegraphics[width=.5\linewidth]{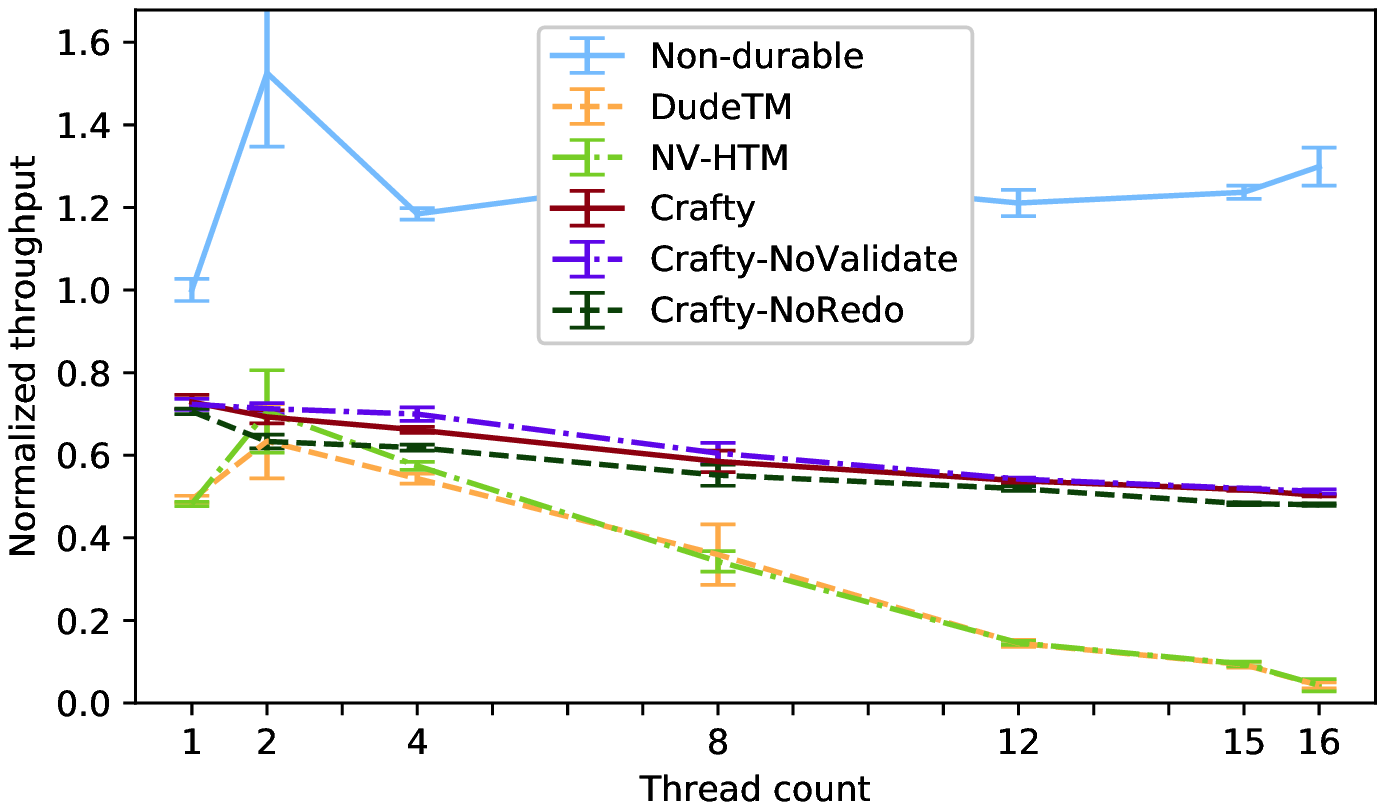}
        \label{fig:throughput-100ns:kmeans-low}
    }\\
    \subfloat[\bench{vacation} (high contention)]{
        \includegraphics[width=.5\linewidth]{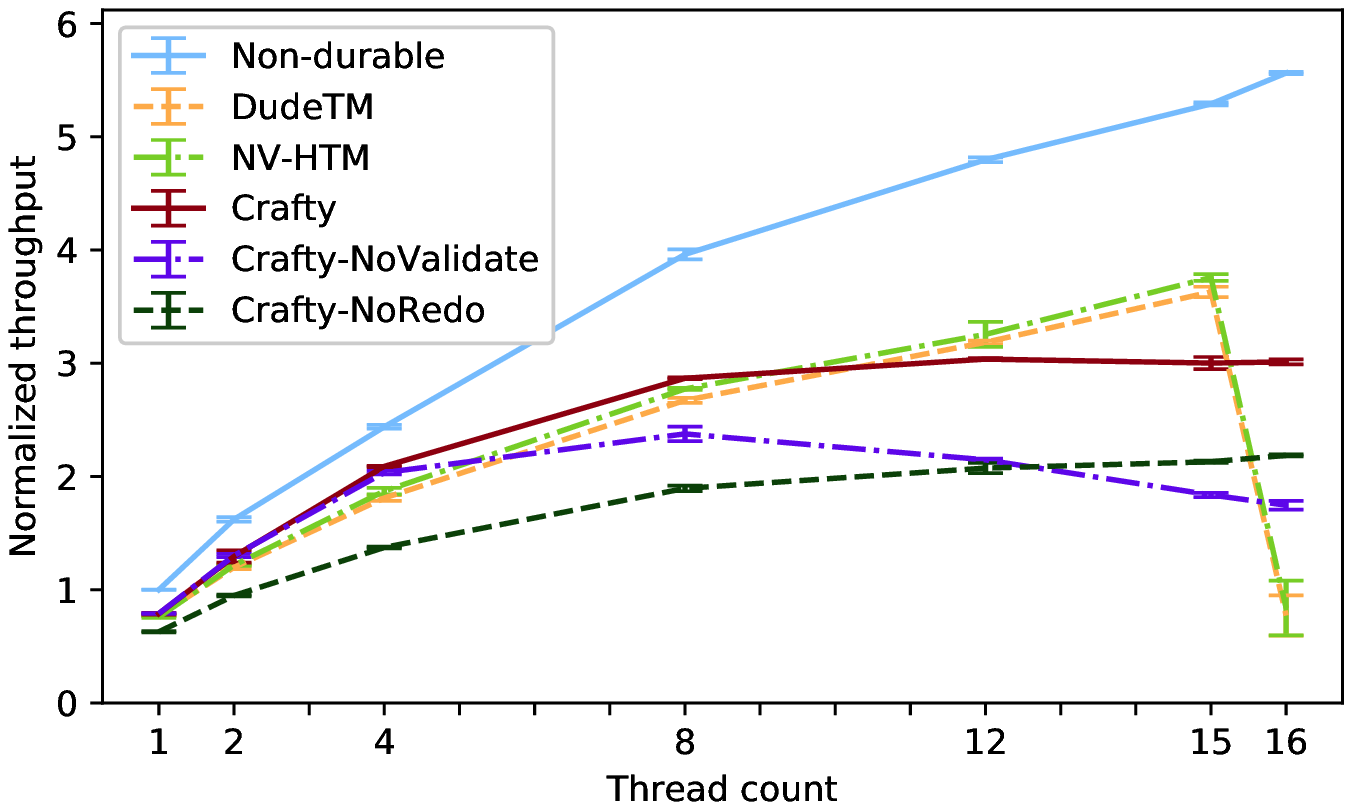}
        \label{fig:throughput-100ns:vacation-high}
    }
    \subfloat[\bench{vacation} (low contention)]{
        \includegraphics[width=.5\linewidth]{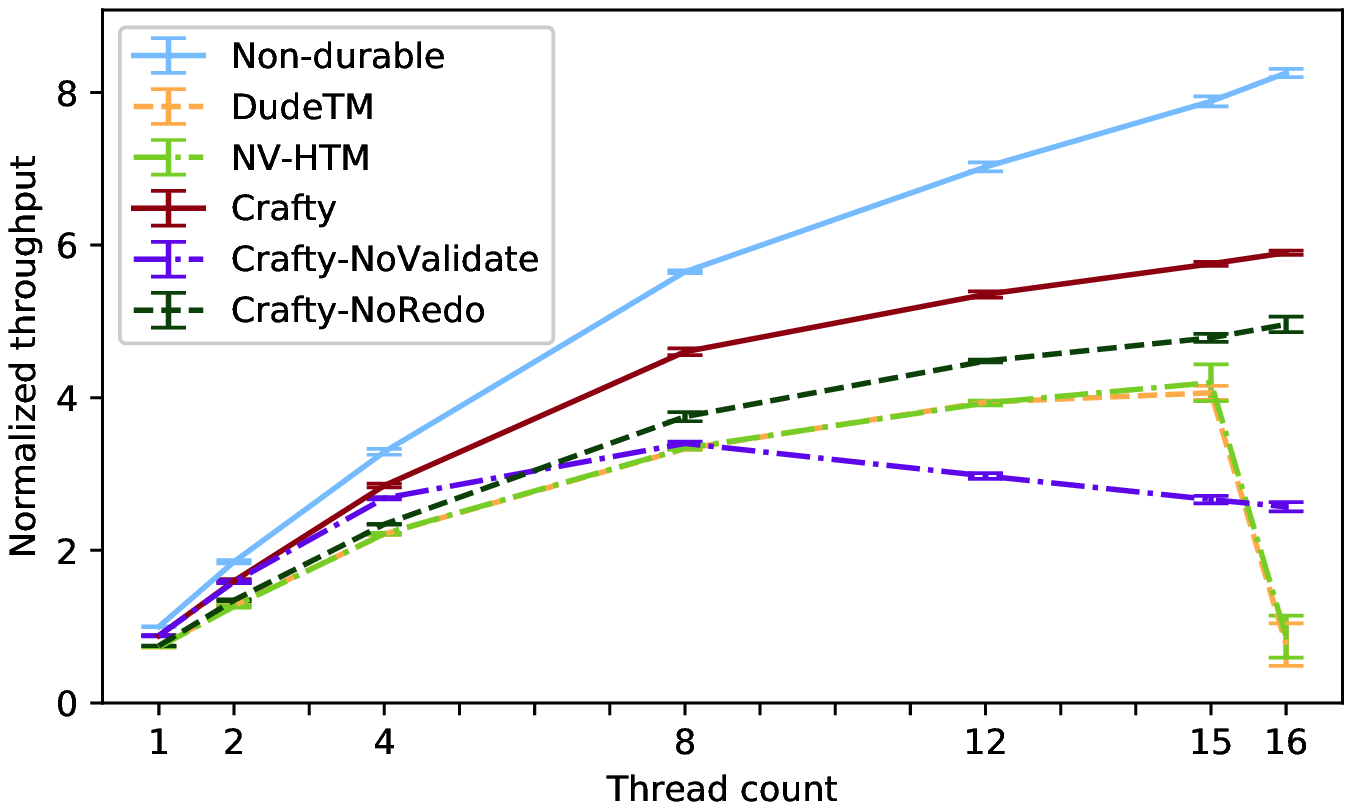}
        \label{fig:throughput-100ns:vacation-low}
    }\\
    \subfloat[\bench{labyrinth}]{
        \includegraphics[width=.5\linewidth]{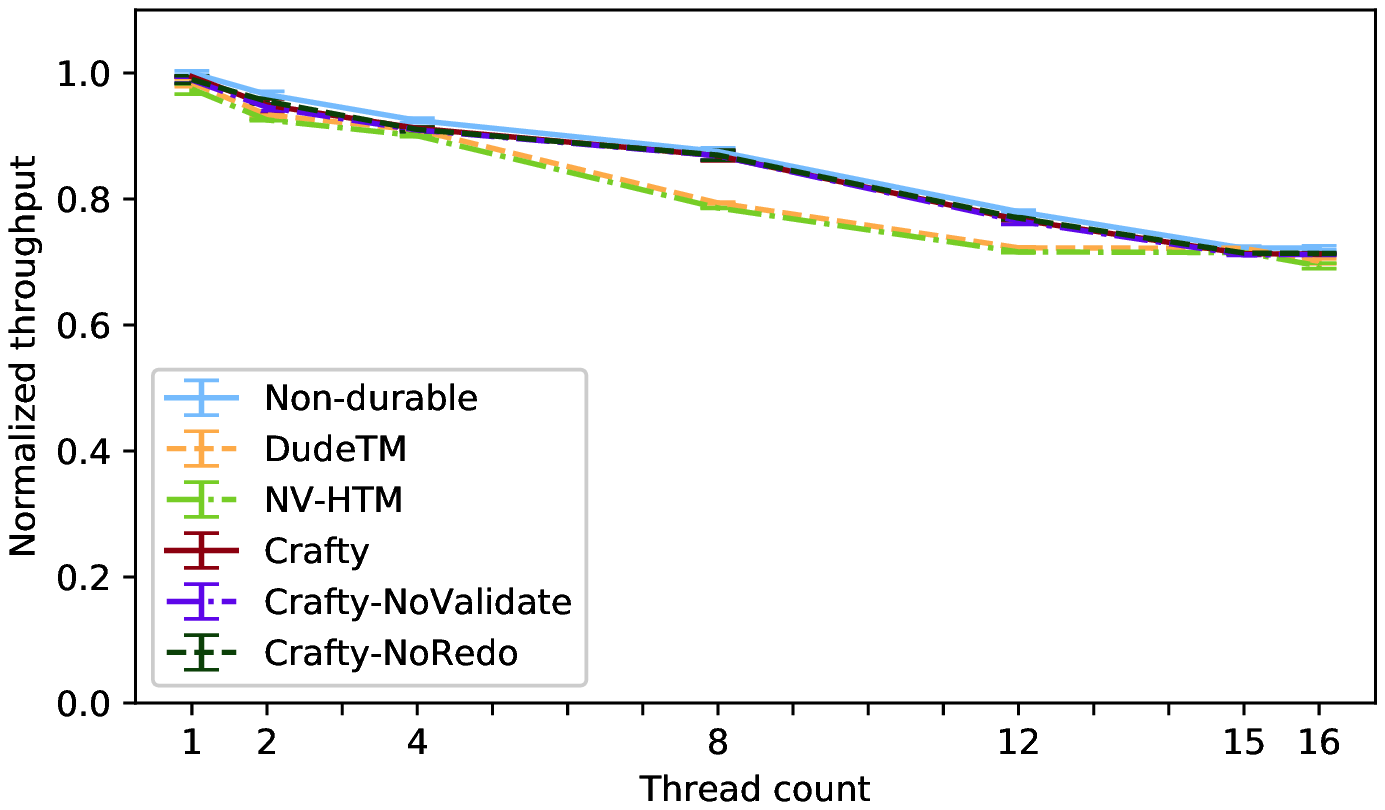}
        \label{fig:throughput-100ns:labyrinth}
    }
    \subfloat[\bench{ssca2}]{
        \includegraphics[width=.5\linewidth]{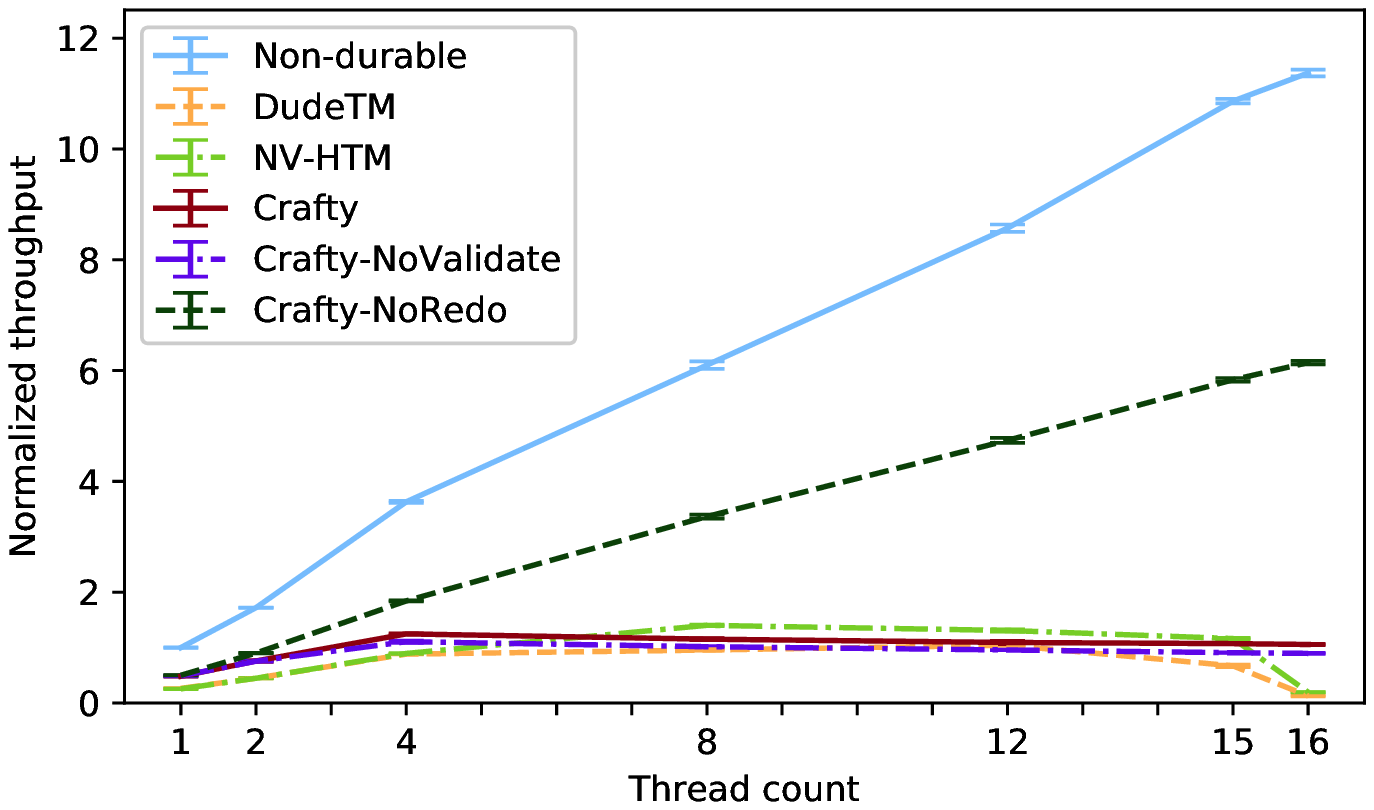}
        \label{fig:throughput-100ns:ssca2}
    }\\
    \subfloat[\bench{genome}]{
        \includegraphics[width=.5\linewidth]{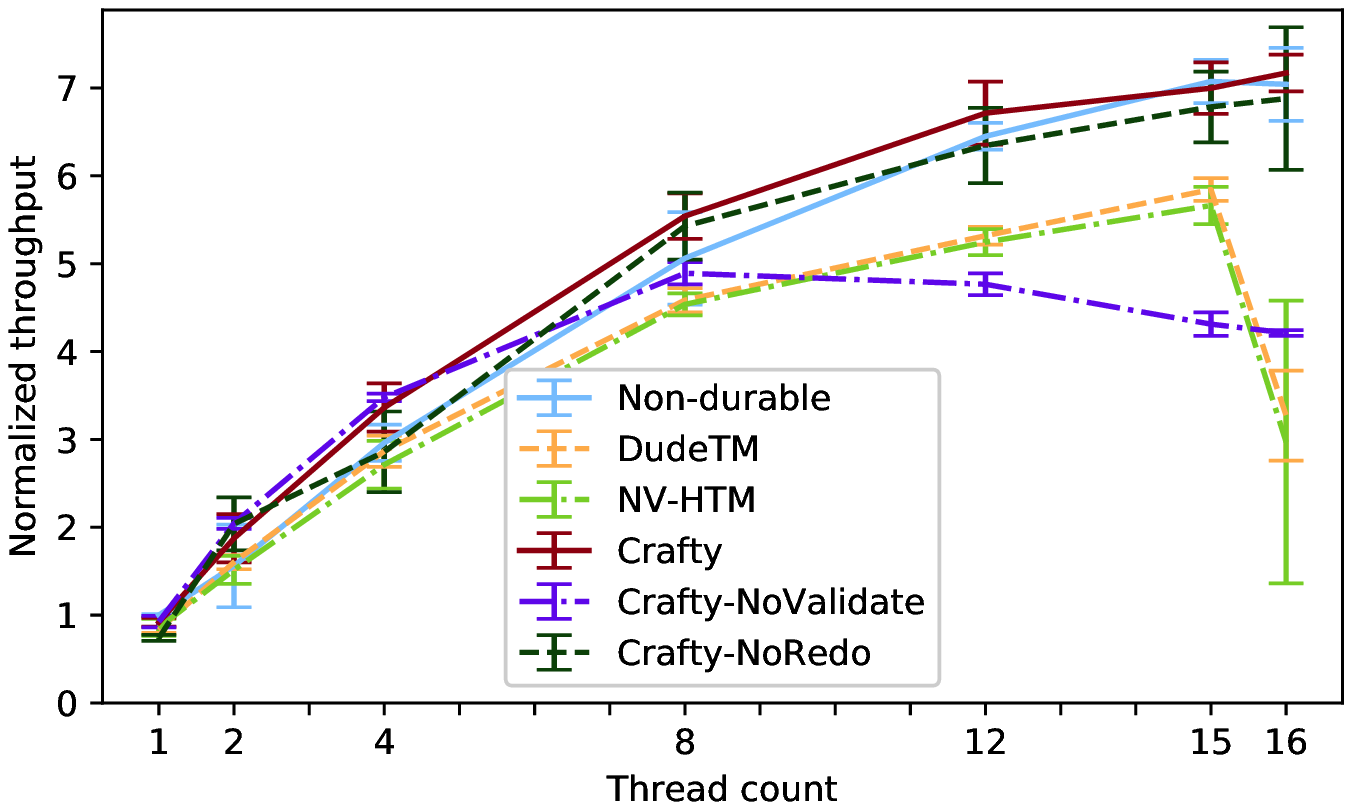}
        \label{fig:throughput-100ns:genome}
    }
    \subfloat[\bench{intruder}]{
        \includegraphics[width=.5\linewidth]{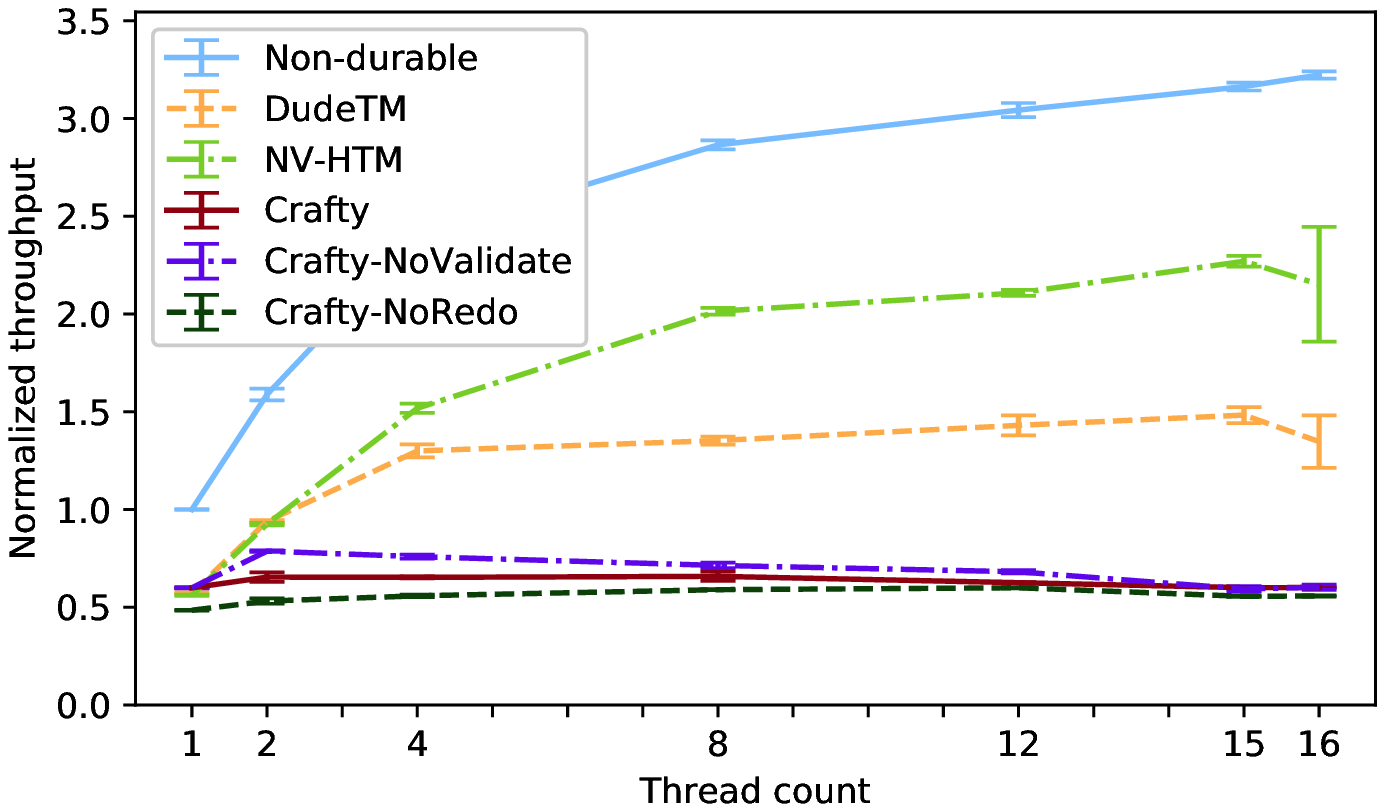}
        \label{fig:throughput-100ns:intruder}
    }
    \caption{Throughput of \crafty and competing approaches,
    on the STAMP benchmarks,
    emulating an NVM latency of 100 ns (instead of 300 ns as in Figure~\ref{fig:throughput:stamp}).}
    \label{fig:throughput-100ns:stamp}
\end{figure*}

\later{
\begin{figure*}
    \vspace*{-1.5em}
    \centering
    \captionsetup[subfloat]{farskip=2pt,captionskip=1pt}
    \subfloat[\bench{TPC-C}]{
        \includegraphics[width=.5\linewidth]{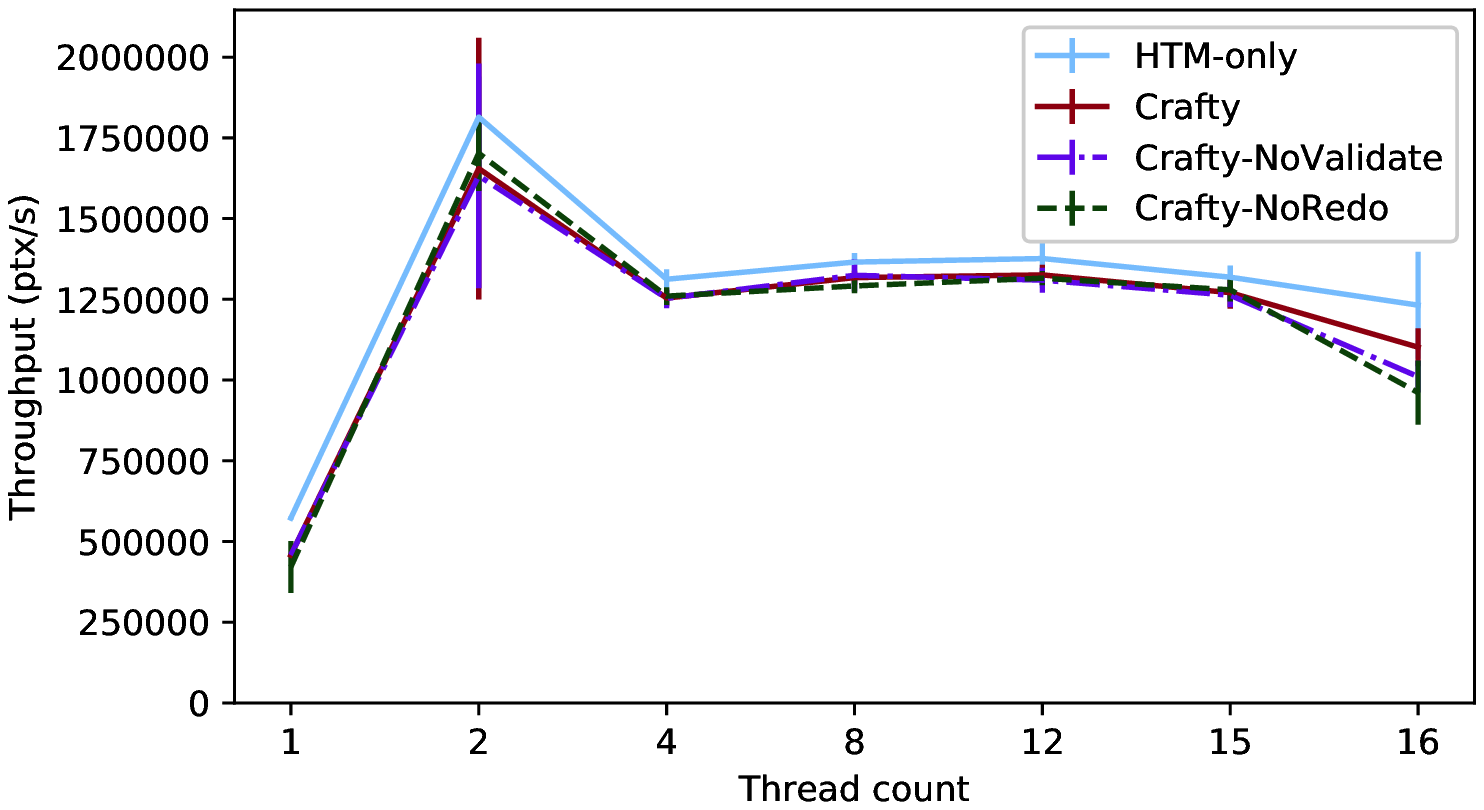}
        \label{fig:time:bayes}
    }
    \caption{Throughput of \crafty and other approaches,
    on the TPC-C benchmark.}
    \label{fig:time:stamp}
    \kaan{Not up to date.}
\end{figure*}

\begin{figure*}
    \centering
    \captionsetup[subfloat]{farskip=2pt,captionskip=1pt}
    \includegraphics[width=\linewidth]{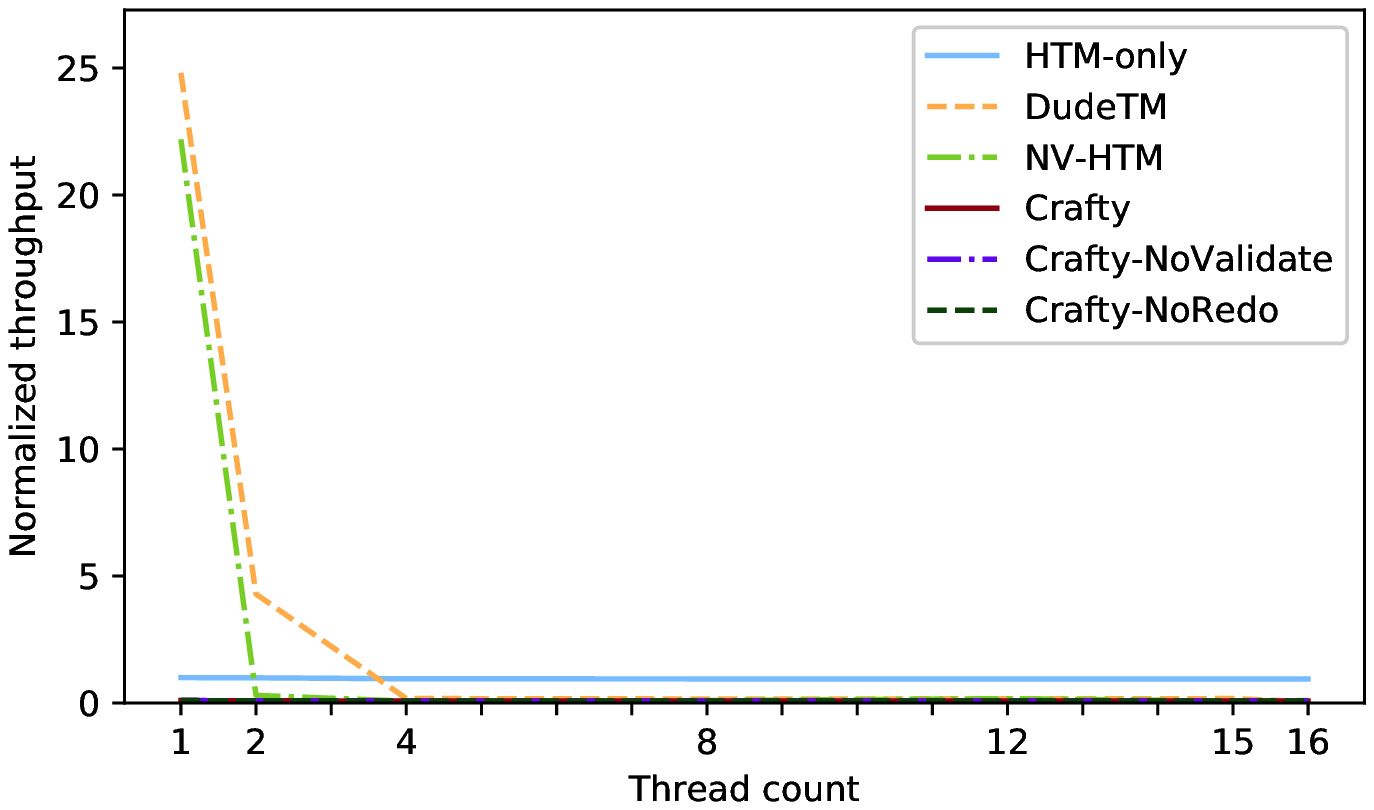}
    \caption{Throughput of \crafty and other approaches on memcached.}
    \kaan{I know that the graph is correct here, so there's some issue in the benchmark itself.}
\end{figure*}

\begin{figure*}
    \centering
    \subfloat[Persistent transaction breakdowns.]{
        \includegraphics[width=\linewidth]{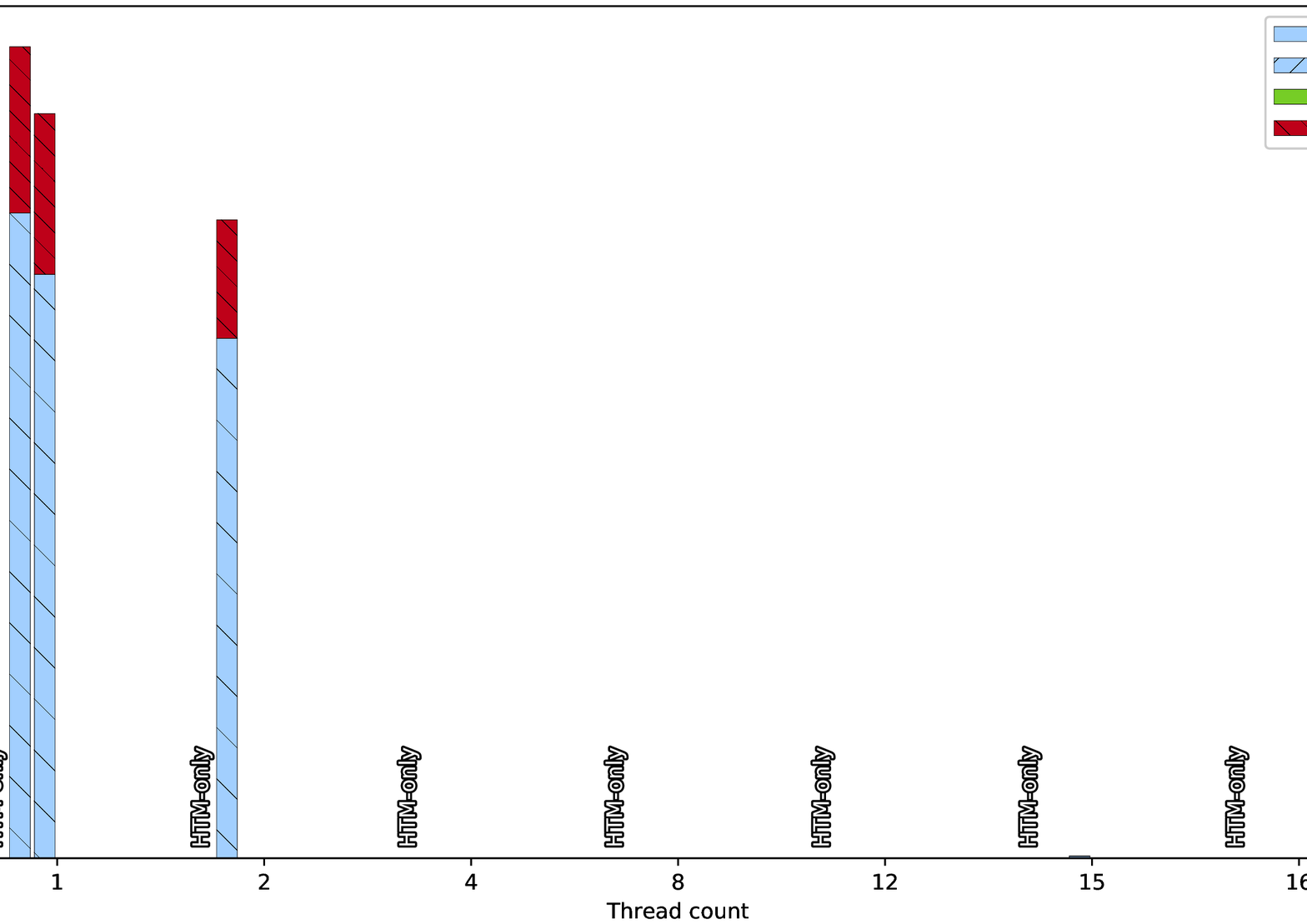}
    }

    \subfloat[Hardware transaction breakdowns.]{
        \includegraphics[width=\linewidth]{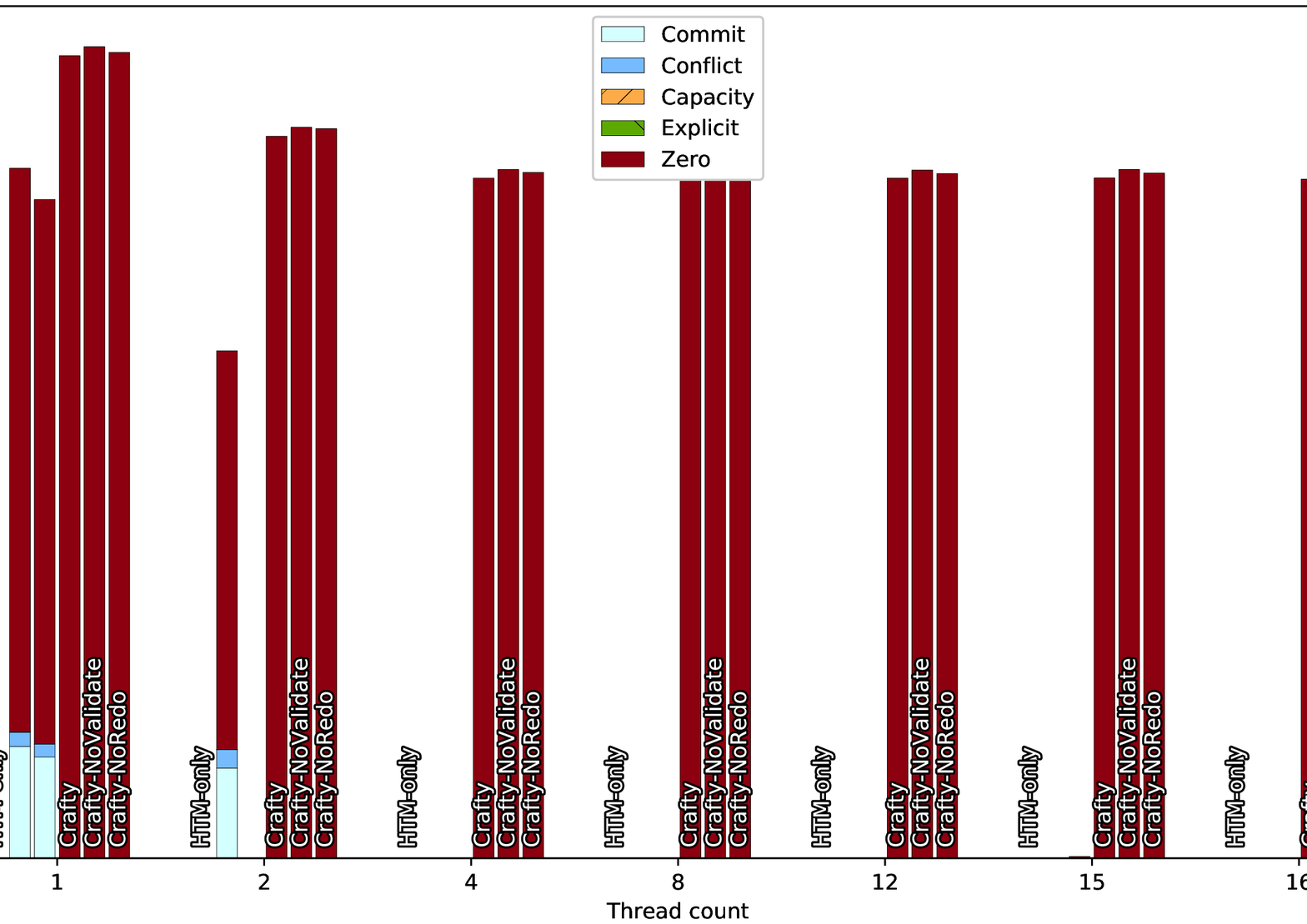}
    }
    \caption{Persistent and hardware transaction breakdowns for \bench{memcached}.}
    \kaan{I'm fairly certain that there's an issue with nested transactions in memcached. It looks like it tries to execute the whole benchmark while holding SGL, and I think it doesn't commit that transaction until the end.}
\end{figure*}
}

}{}

\end{document}